\documentclass[12pt]{iopart}

\usepackage{iopams}

\usepackage[utf8]{inputenc}
\usepackage{natbib}
\usepackage{setstack}
\usepackage[ngerman,english]{babel}
\usepackage{graphicx}
\usepackage[nooneline]{subfigure}
\usepackage{upgreek}
\usepackage{url}
\usepackage{xspace}
\usepackage{paralist}
\usepackage{verbatim}
\usepackage{icomma}

\newenvironment{align*}{\begin{eqnarray*}}{\end{eqnarray*}}

\newcommand{\bilder}{.}

\newcommand{\hb}{0.485\textwidth}

\newcommand{\db}{0.3133\textwidth}

\newcommand{\vib}{0.25\textwidth}
\newcommand{\veb}{0.342\textwidth}
\newcommand{\deb}{0.256\textwidth}

\newcommand{\ba}{\hspace{0.03\textwidth}}

\setcitestyle{comma,aysep={}}

\newcommand{\rs}{r_\mathrm{S}}
\newcommand{\rrmi}{r_\mathrm{m}}
\newcommand{\ud}{\mathrm{d}}

\begin{document}

\title[Sector models: II. Geodesics]{%
Sector models---A toolkit for teaching general relativity:
II. Geodesics
}

\author{C Zahn and U Kraus}

\bigskip
\address{Institut f\"ur Physik, Universität Hildesheim,
Universitätsplatz 1, 31141 Hildesheim, Germany}

\eads{
\mailto{corvin.zahn@uni-hildesheim.de},
\mailto{ute.kraus@uni-hildesheim.de}
}

\begin{abstract}

{\parindent=0pt

Sector models are tools
that make it possible to teach the basic principles of
the general theory of relativity
without going beyond elementary mathematics.
This contribution shows how sector models can be used to
determine geodesics.
We outline a workshop for high school and undergraduate students
that addresses gravitational light deflection
by means of the construction of geodesics on sector models.
Geodesics close to a black hole are used by way of example.
The contribution also describes a simplified calculation of sector
models that students can carry out on their own.
The accuracy of the geodesics constructed on sector models
is discussed in comparison with numerically computed solutions.
The teaching materials presented in this paper
are available online for teaching purposes at
{\tt www.spacetimetravel.org}.


}

\end{abstract}

\maketitle

\section{Introduction}

\label{sec.einleitung2}

To teach the basic principles of
the general theory of relativity
without going beyond elementary mathematics
remains a challenge
even a century after the completion
of the theory.
In view of this
objective,
we describe a novel approach
that is focussed on geometric insight
and makes do with
elementary mathematics as taught in school.
This approach is suitable
for learners who lack the qualifications
or the time
required to master the mathematical tools
that are needed for a standard introductory text,
i.e., advanced high school students
and undergraduate students,
especially those
in a physics minor programme
or in physics teacher education.
The approach can also be used
as a supplement to standard textbooks
(e.g., \citealp{har})
to strengthen geometric insight.

In a first paper (\citealp{teil1}, in the following referred to
as paper~I),
we have developed sector models
as a new type of physical model for curved spaces and spacetimes.
We have shown how they can be used to
convey the notion
of a space or a spacetime being curved.
In paper~I, sector models were first introduced
for two-dimensional curved surfaces of positive or negative curvature.
They were then developed for
three-dimensional curved spaces
and 1+1-dimensional curved spacetimes
using the Schwarzschild black hole as an example.

\begin{figure}
  \centering
  \includegraphics[width=0.8\textwidth]{\bilder/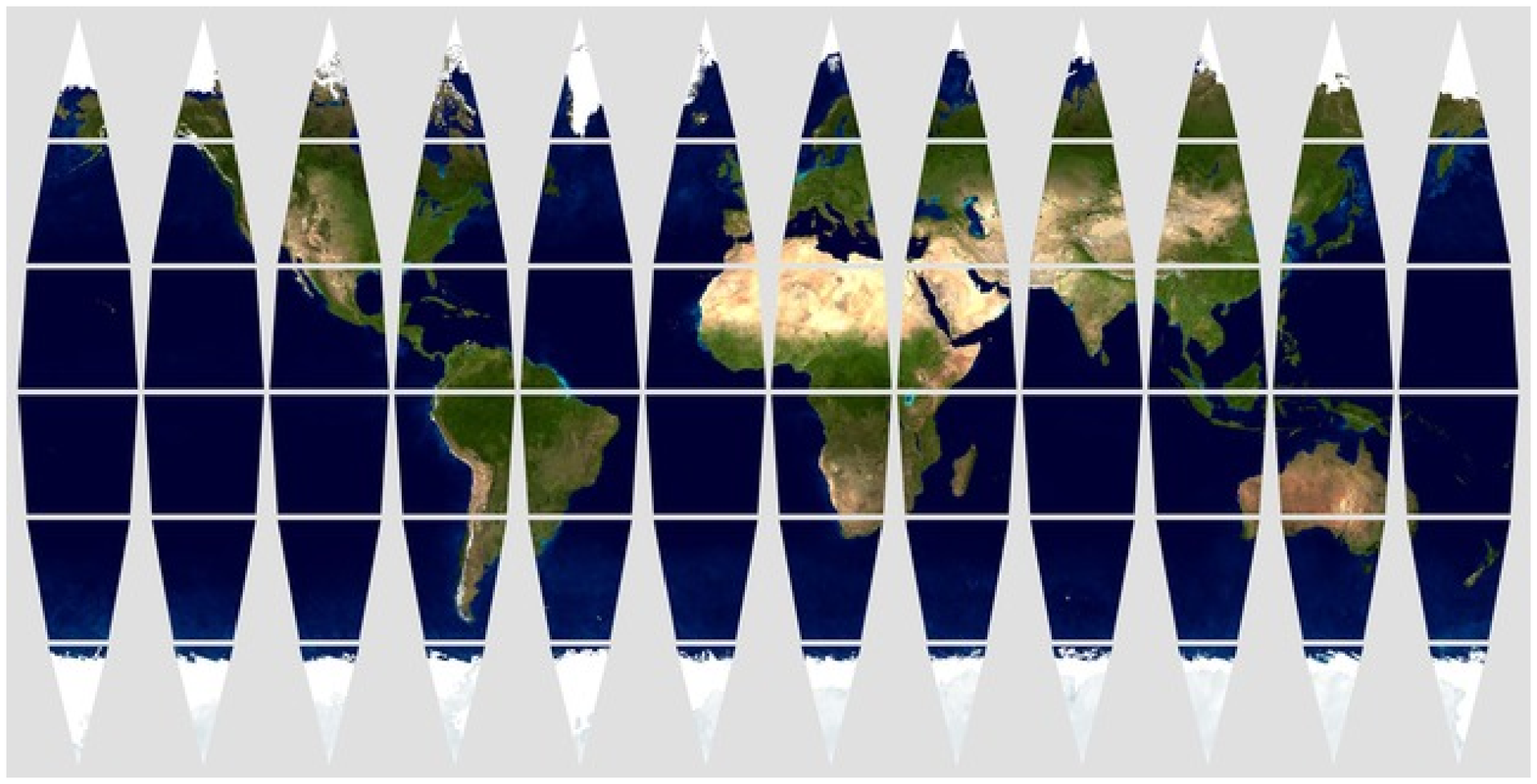}
  \caption{\protect\raggedright
    Sector model of the surface of the earth. Earth texture: NASA.
  }
  \label{fig.erdkarte}
\end{figure}

Sector models
implement
the description of curved spacetimes
used in the Regge calculus \citep{reg1961}
in the form of physical models.
Figure~\ref{fig.erdkarte} illustrates the basic principle
using the surface of the earth by way of example:
The surface is approximated by means of
small flat elements of area.
When these are laid out in the plane,
one obtains a world map;
this map is the sector model of the surface of the earth.
Two differences to ordinary world maps stand out:
The sector map is non-contiguous,
since
it is not possible to join all sectors
simultaneously to all their neighbours.
Also, the sector map
is undistorted
within the bounds of the discretization error,
preserving both lengths and angles.
It is, therefore, open to an intuitive geometric understanding.
The sector model of a curved three-dimensional space
is built along the same lines.
The flat elements of area
are replaced by blocks with euclidean geometry.
In the case of a curved spacetime,
the sectors are spacetime blocks with Minkowski geometry.

The general theory of relativity
describes the paths of light and of free particles
as geodesics of a curved spacetime.
The notion of geodesics, therefore, is an important point
in any introduction to general relativity.
This contribution shows
how sector models can be used
to introduce the concept of geodesics
and to determine geodesics by graphic construction.
Instead of
solving a system of ordinary differential equations,
geodesics are constructed with pencil and ruler.
The construction
implements the description of geodesics in the Regge calculus
\citep{wil1981}
and gives quantitatively correct results
(within the discretization error).

In this contribution, we outline a workshop
on gravitational light deflection
the way we teach it
for high school and
for undergraduate students (section~\ref{sec.workshopgeod}).
Section~\ref{sec.modell2}
is a discussion of
the approximations
involved in the construction of geodesics on sector models
and a study of the accuracy that can be achieved.
Conclusions and outlook follow in
section~\ref{sec.fazit2}.

\newcommand{\fotobreite}{1.0}

\section{Workshop on geodesics and light deflection}

\label{sec.workshopgeod}

The workshop starts by introducing the concept of geodesics,
using curved surfaces by way of example.
Geodesics are then constructed on the sector model of a black hole
in order to show how gravitational light deflection arises.
The black hole is used as an example
because close to it relativistic effects are large
and, therefore, clearly visible in the graphic constructions.
As an extension to the workshop,
we describe a simplified procedure
for the calculation of sector models
that students can use to create sector models on their own.
This enables them to study the geodesics of a spacetime
starting from a given metric.

\subsection{Geodesics on curved surfaces}

\label{sec.kugel}

The introduction to the workshop
includes
the explanation
that the general theory of relativity
describes the paths of light and of free particles as geodesics.
Depending on the background of the participants,
the significance of geodesics in general relativity may just be stated
or may be explained in more detail with reference to the equivalence
principle (e.g.,~\citealp{nat}, chapter 5).
In preparation for the determination of geodesics
in the vicinity of a black hole,
geodesics are first studied on curved surfaces.

The geodesic line on a curved surface
is introduced as a locally straight line.
Such a line keeps its direction at each point,
i.e., it neither bends nor kinks.
A criterion is described that permits to recognize a geodesic:
One imagines that
a narrow strip made of a non-elastic material
is glued along its centre line onto the line that is to be investigated.
If the line bends, the strip tears on the outside and buckles on the
inside---this shows that the line is not a geodesic.

\begin{figure}
    \centering
    \includegraphics[width=\db]{\bilder/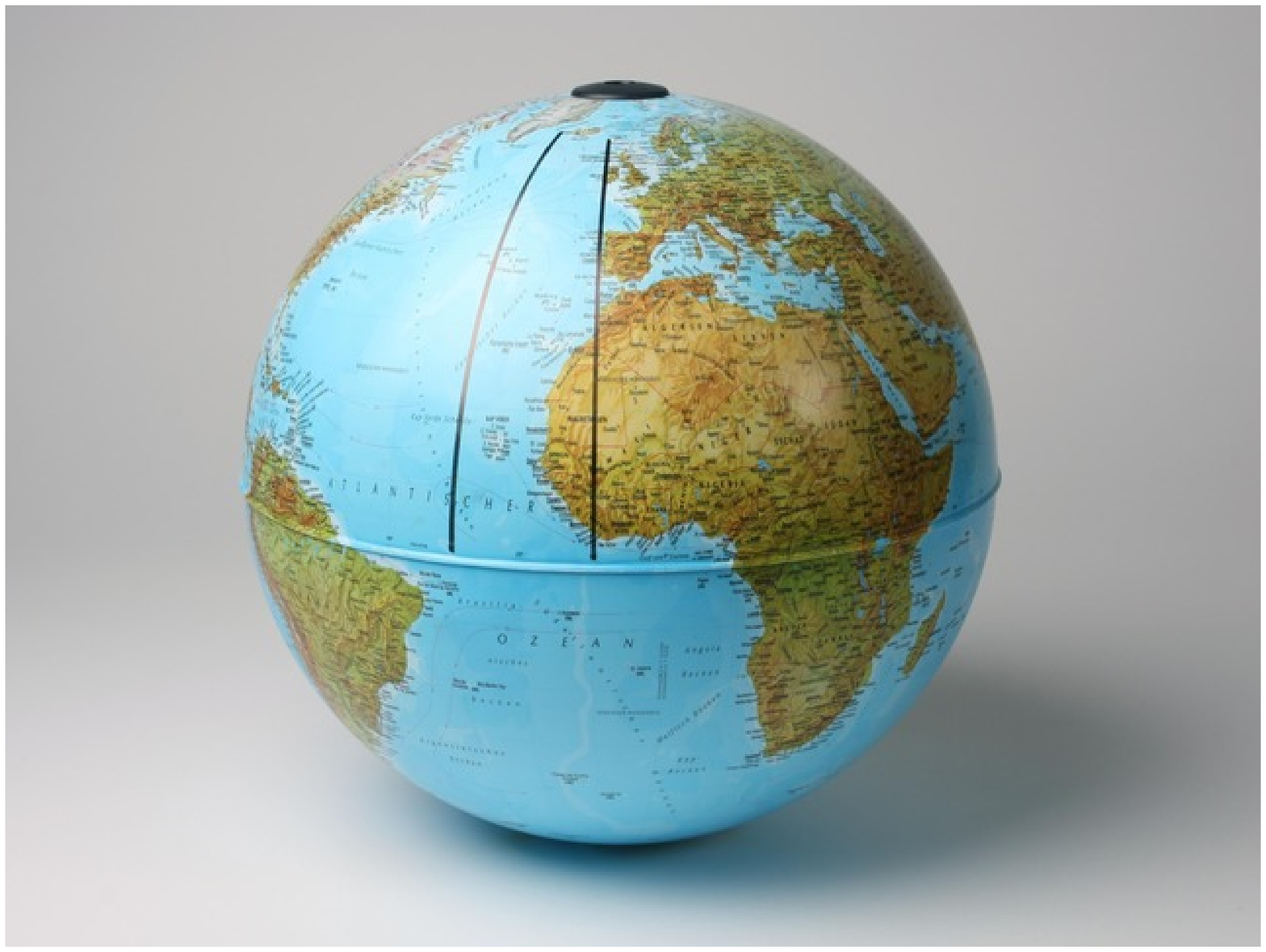}
    \caption{\label{fig.kugel}
   The geodesics on the sphere are the great circles.
   }
\end{figure}

\begin{figure}
  \centering
\subfigure[]{%
  \includegraphics[width=\db]{\bilder/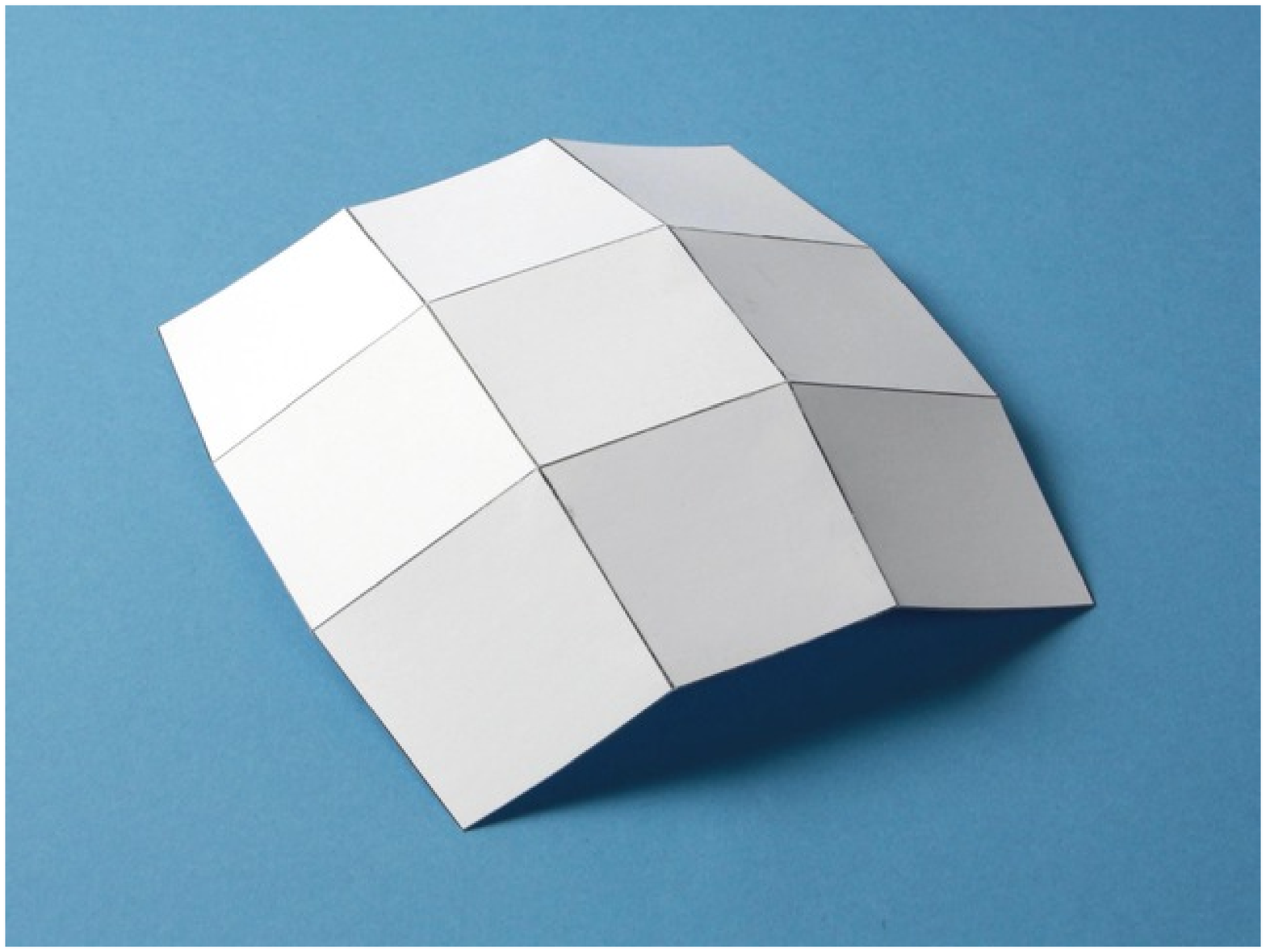}
}\ba%
\subfigure[]{%
  \includegraphics[width=\db]{\bilder/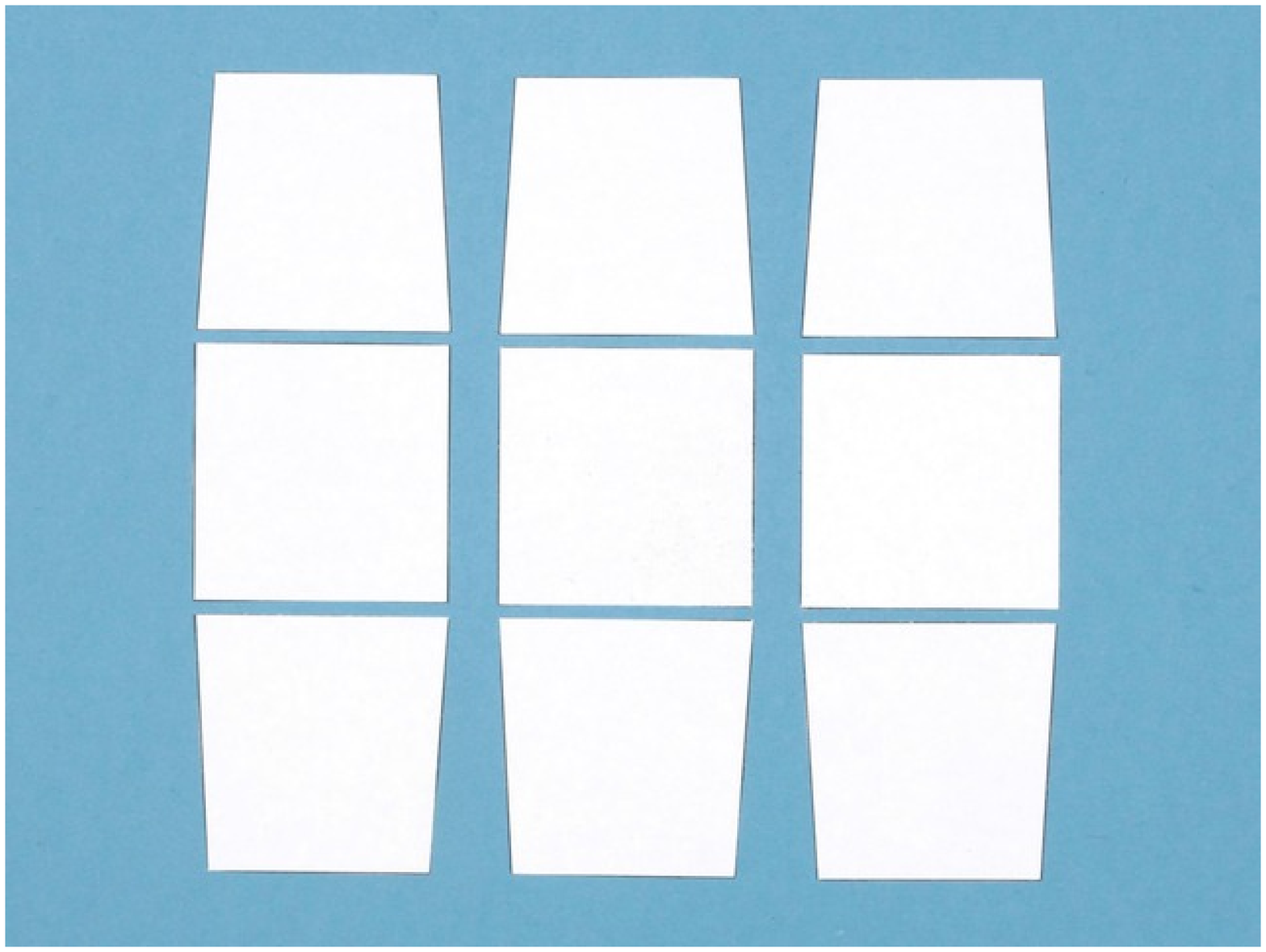}
}
\caption{\label{fig.kugelmodell}
     A spherical cap is approximated by facets (a)
     and represented as a sector model (b).
   }
\end{figure}

\begin{figure}
    \centering
    \includegraphics[width=\db]{\bilder/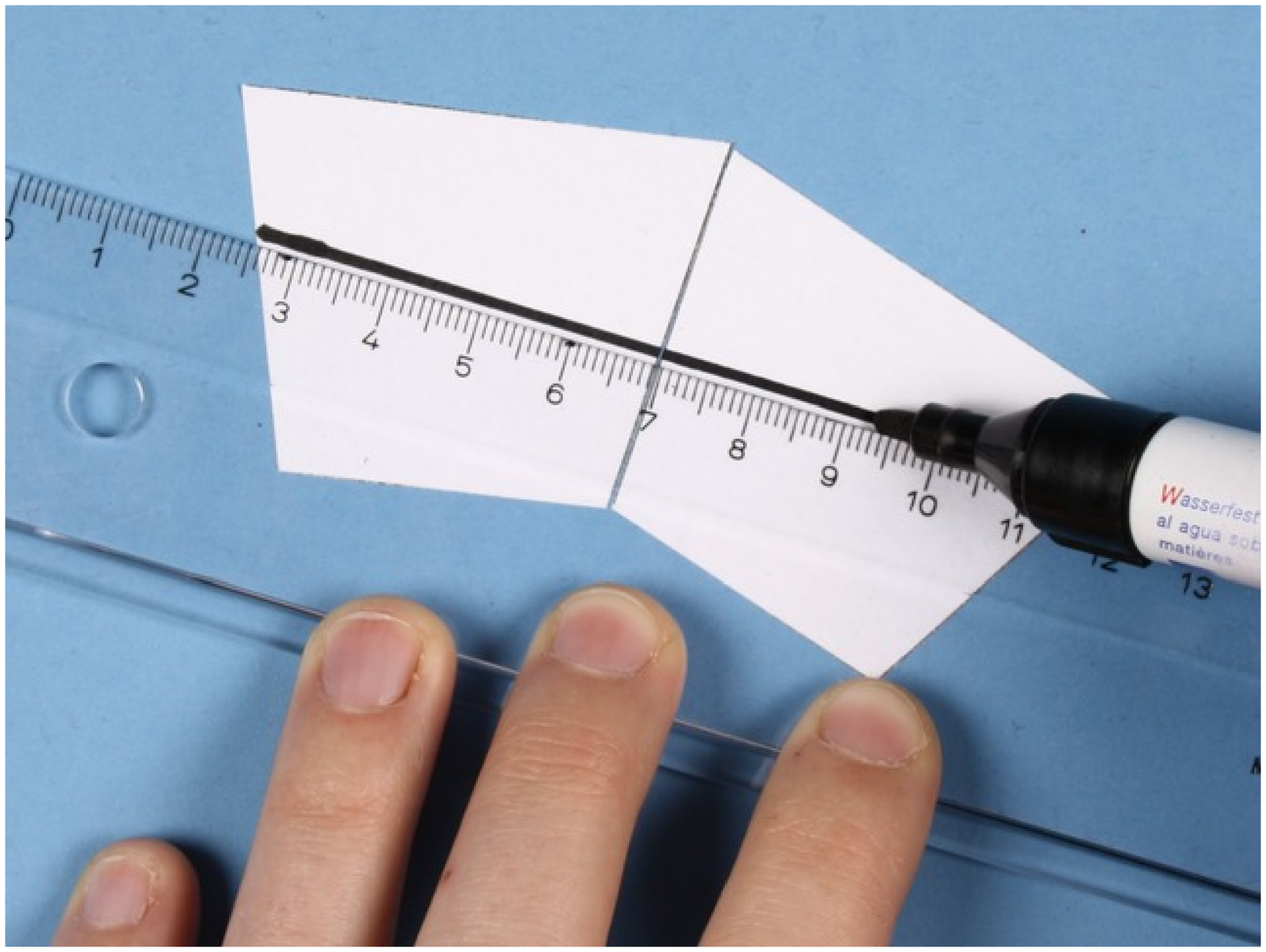}
    \caption{\label{fig.geodlineal}
     The construction of a geodesic on a sector model.
    }
\end{figure}

The sphere is used as the first example
(figure~\ref{fig.kugel}).
We consider
a line that starts from the equator due north
and
is locally straight.
It is obvious that the line is a line of longitude.
The lines of longitude, and
more generally all great circles,
are geodesics on the sphere.
Figure~\ref{fig.kugel} illustrates
a characteristic property of these geodesics:
Two lines of longitude are parallel at the equator;
towards the north pole they converge.
Generally speaking, geodesics
on the sphere
converge when starting in parallel.

\begin{figure}
  \centering
  \subfigure[]{%
    \includegraphics[width=\db]{\bilder/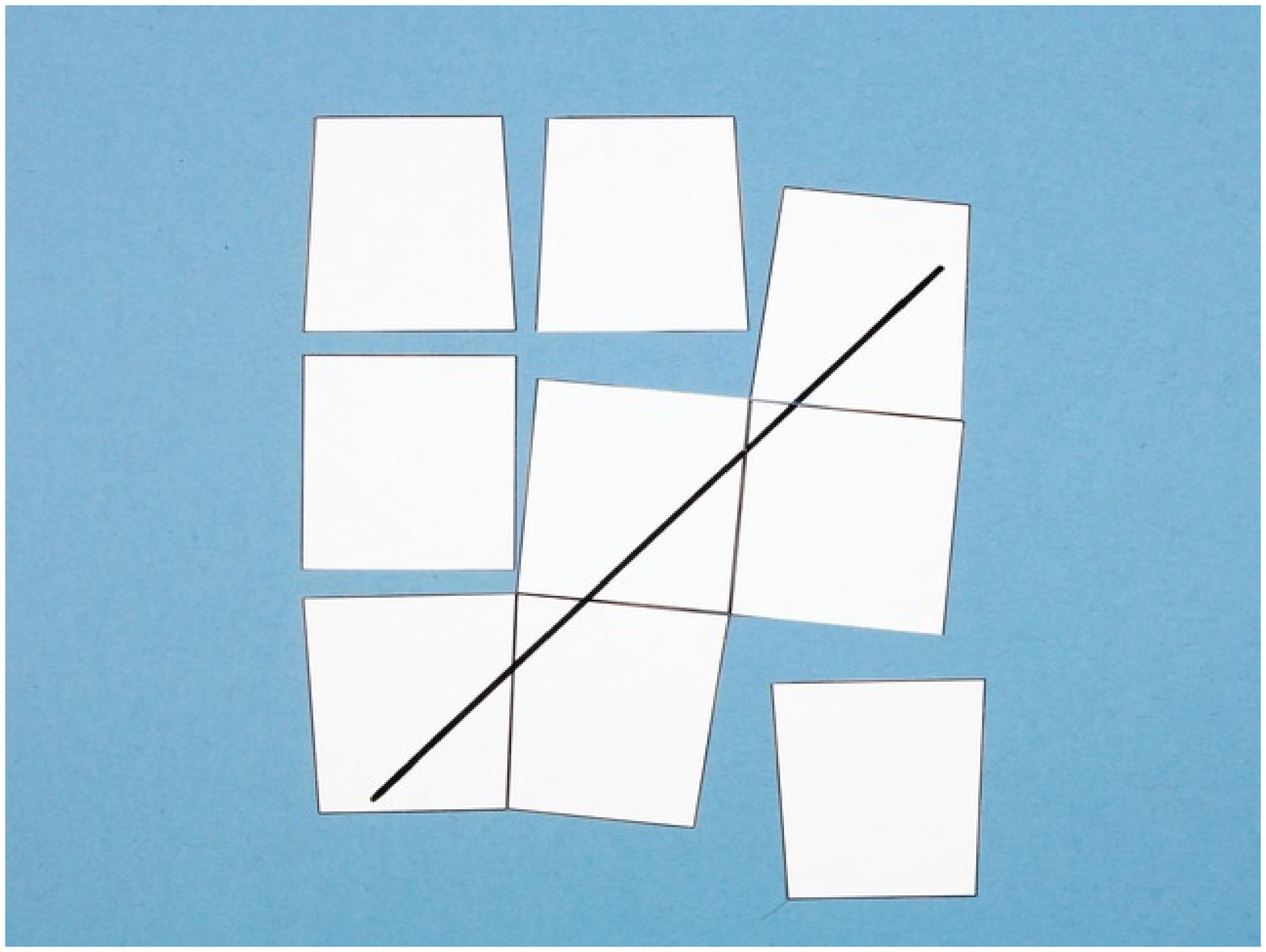}
  }\hfill%
  \subfigure[]{%
    \includegraphics[width=\db]{\bilder/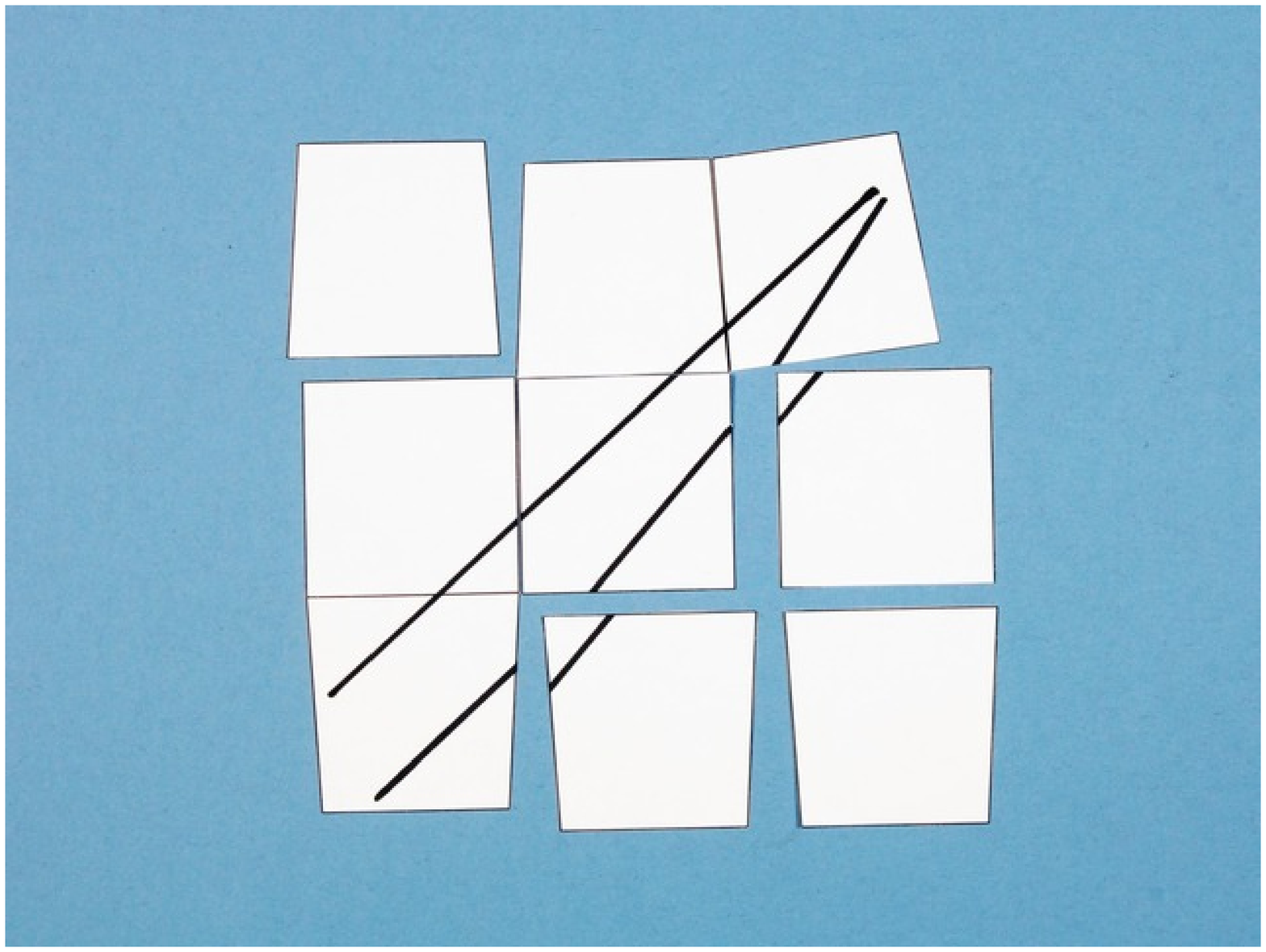}
  }\hfill%
  \subfigure[]{%
    \includegraphics[width=\db]{\bilder/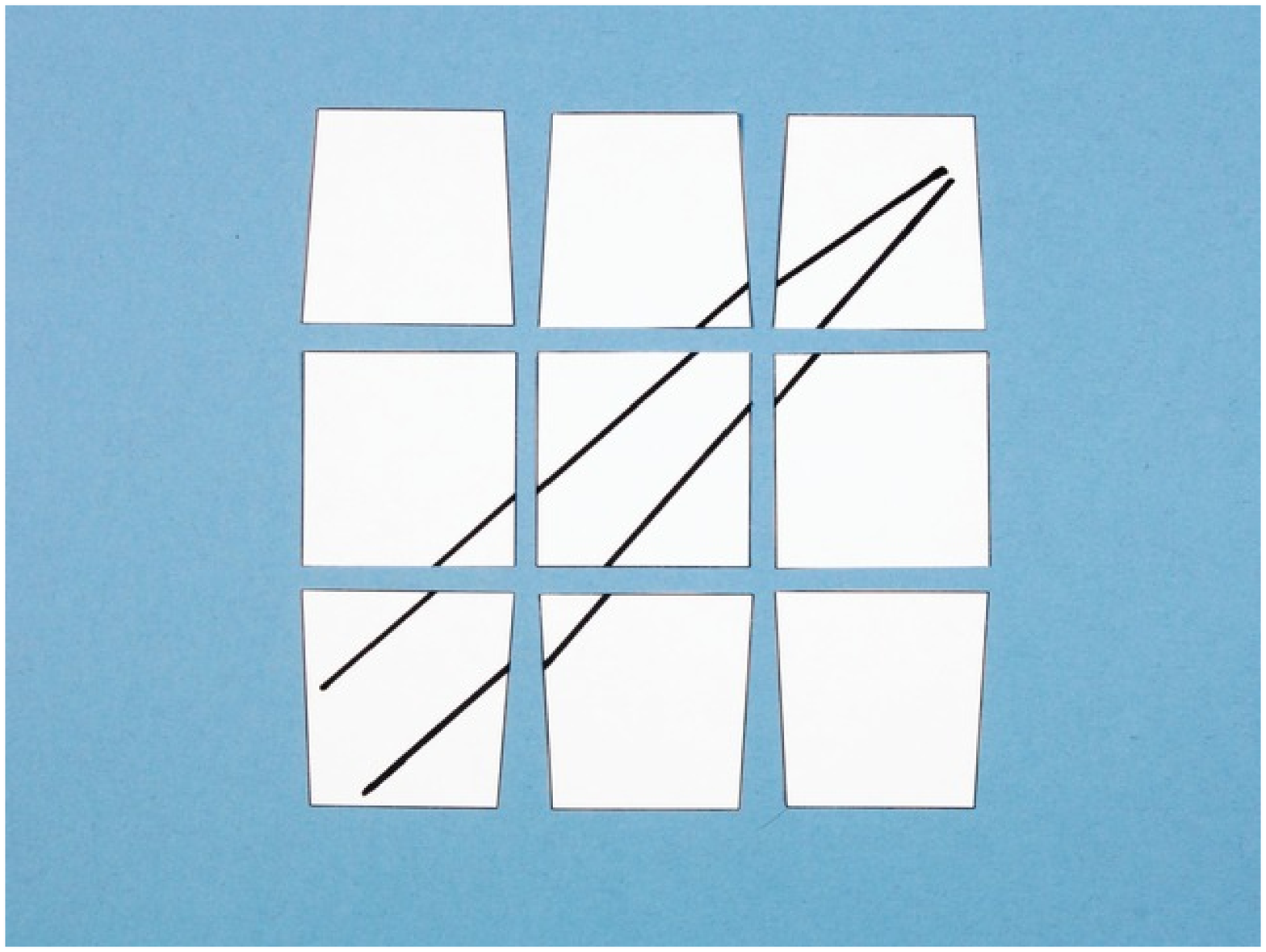}
  }
  \caption{\label{fig.kugelgeod}
    Geodesics on the sector model of a spherical cap. In (a) the sectors
    are joined along the bottom geodesic and in (b) along the top geodesic.
    The two geodesics are parallel in the bottom left sector and
    converge towards the right hand side ((b), (c)).
   }
\end{figure}

\begin{figure}
  \centering
  \subfigure[]{%
    \includegraphics[width=\db]{\bilder/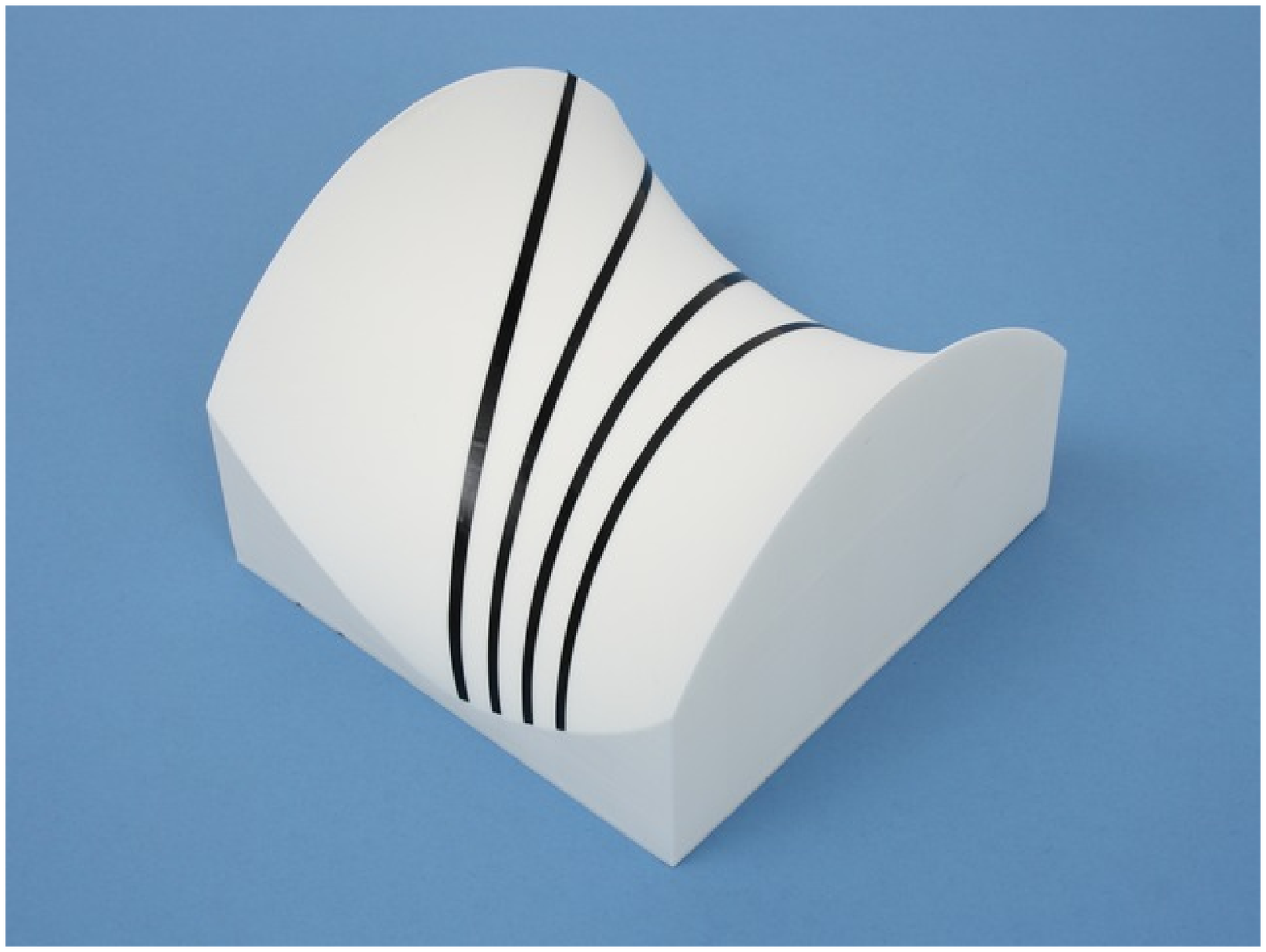}
  }\ba%
  \subfigure[]{%
    \includegraphics[width=\db]{\bilder/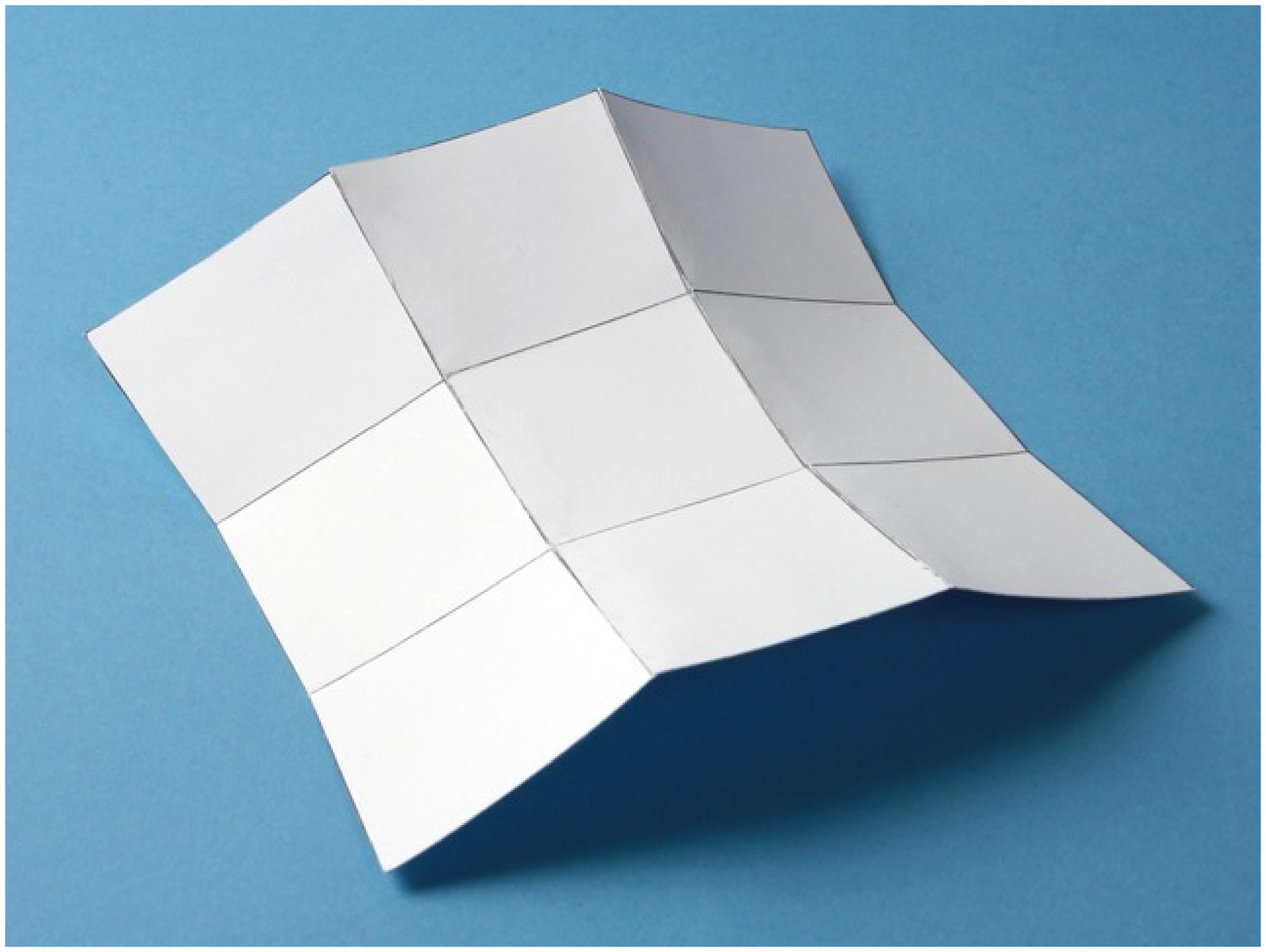}
  }\\*[2ex]%
  \subfigure[]{%
    \includegraphics[width=\db]{\bilder/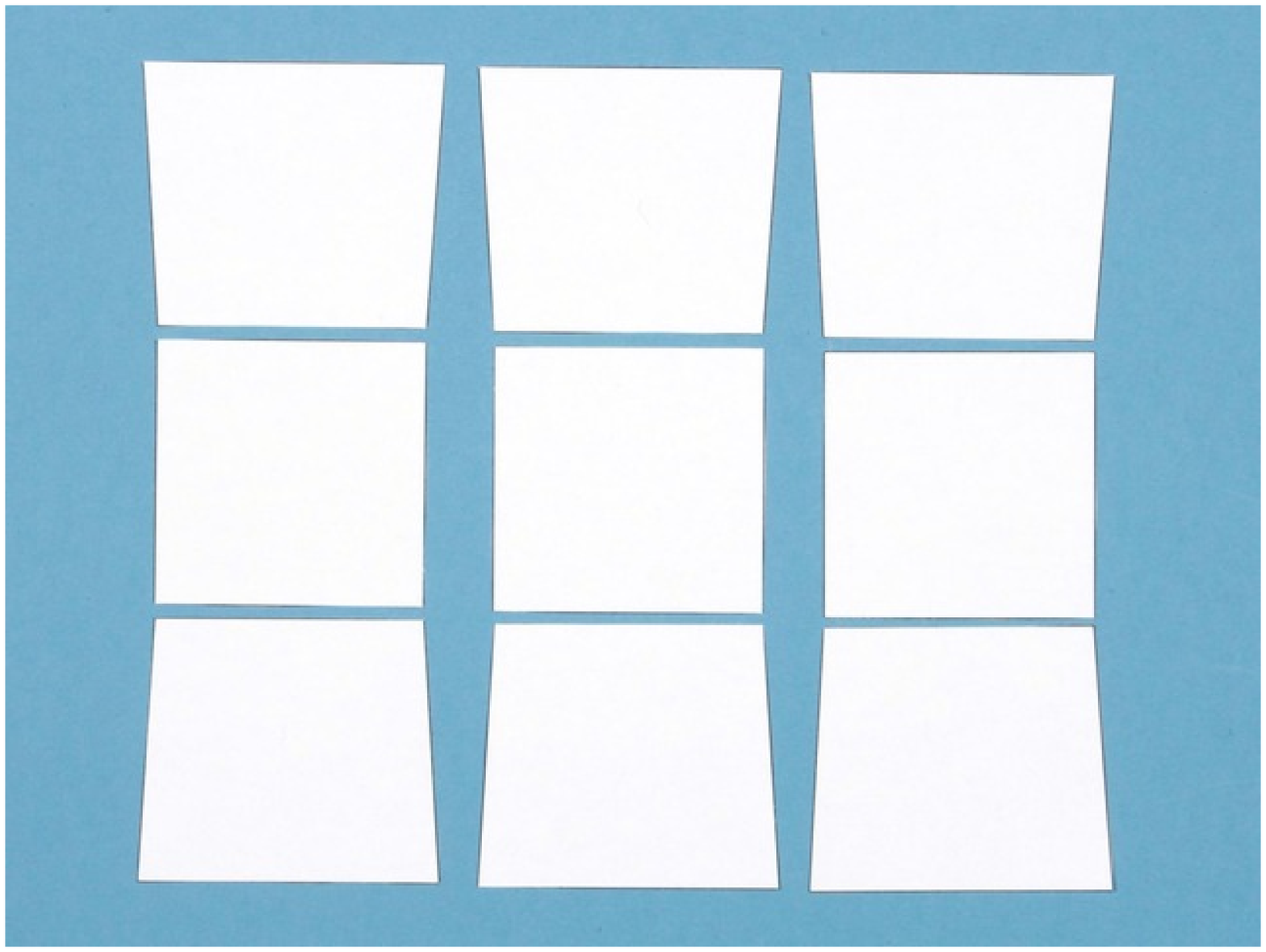}
  }\ba%
  \subfigure[]{%
    \includegraphics[width=\db]{\bilder/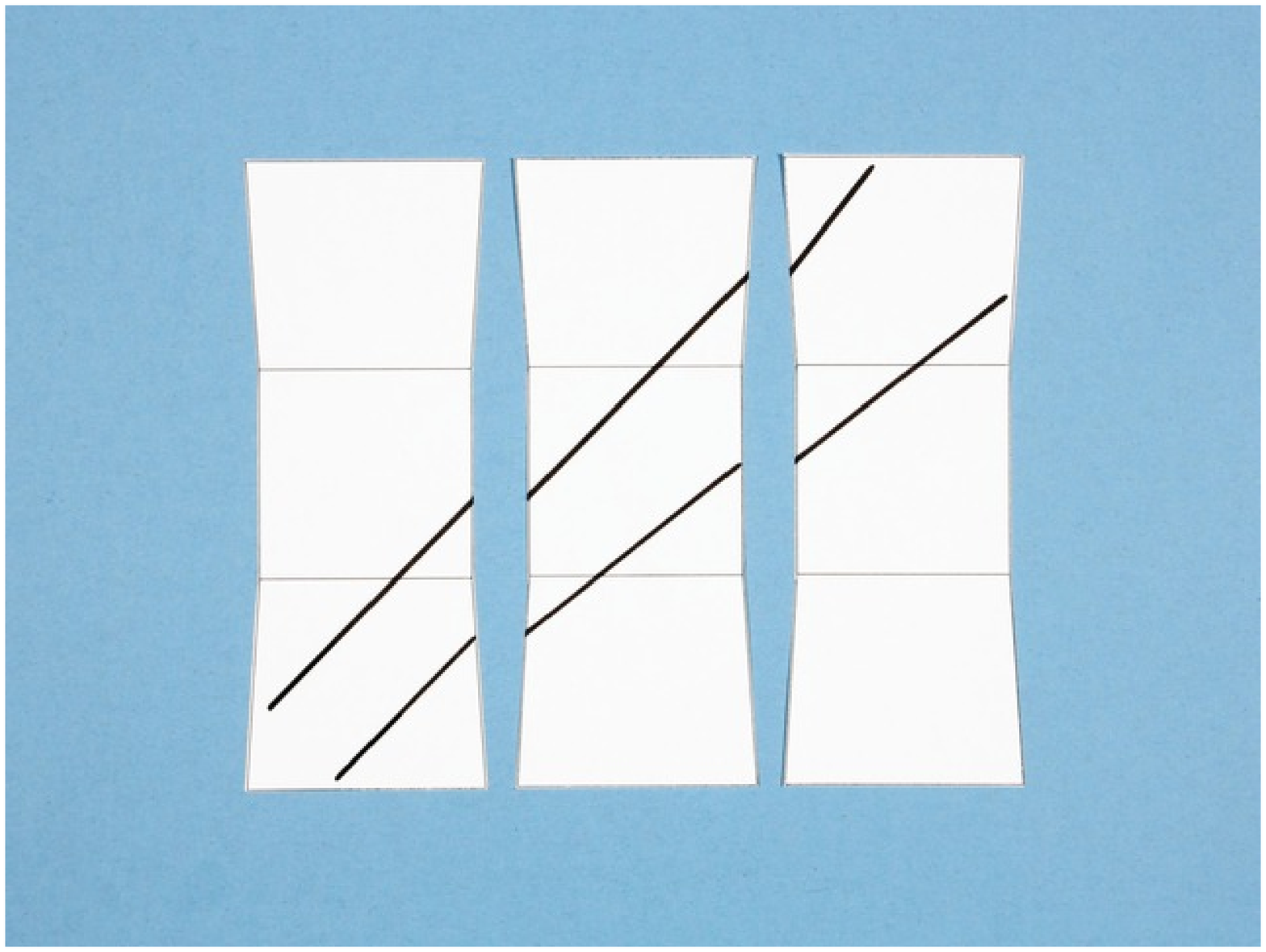}   
  }
  \caption{\label{fig.sattelgeod}
    A saddle (a) is in part approximated by facets (b) and
    represented as a sector model (c);
    geodesics starting parallel to each other diverge ((a), (d)).
   }
\end{figure}

In the next step we show
how this property of geodesics on the sphere
can be obtained from a sector model.
A spherical cap
is approximated by facets
(figure~\ref{fig.kugelmodell}(a)),
and the facets are laid out as a sector model
(figure~\ref{fig.kugelmodell}(b)).
Now a geodesic is drawn
onto the sector model.
Within a sector, i.e., on a flat element of area,
a geodesic is a straight line.
When the line reaches the border of a sector,
it is continued onto the neighbouring sector.
How to do this follows from the definition of the geodesic:
locally straight
(figure~\ref{fig.geodlineal}).
The two neighbouring sectors are joined at
their common edge
and the line is continued straight across the border.
In this way
the geodesic is continued across the sector model
(figure~\ref{fig.kugelgeod}(a)).
A second geodesic is then added
that is parallel to the first
in the bottom left sector
(figure~\ref{fig.kugelgeod}(b)).
One can see that
the two geodesics
starting in parallel on the lower left
converge towards the right hand side
(figures~\ref{fig.kugelgeod}(b), (c)).

A saddle is taken as a second example.
Adhesive strips glued to the surface show
that geodesics starting parallel to each other diverge
(figure~\ref{fig.sattelgeod}(a)).
The approximation of a part of the surface
by facets (figure~\ref{fig.sattelgeod}(b))
leads to the sector model (figure~\ref{fig.sattelgeod}(c)).
On the sector model,
two geodesics are constructed
that are parallel to each other in the lower left sector;
these geodesics diverge (figure~\ref{fig.sattelgeod}(d)).

The two examples show
that
sector models of curved surfaces can be used as tools to
study the properties of geodesics on these surfaces.

\begin{figure}
  \centering
  \subfigure[]{%
    \includegraphics[width=\hb]{\bilder/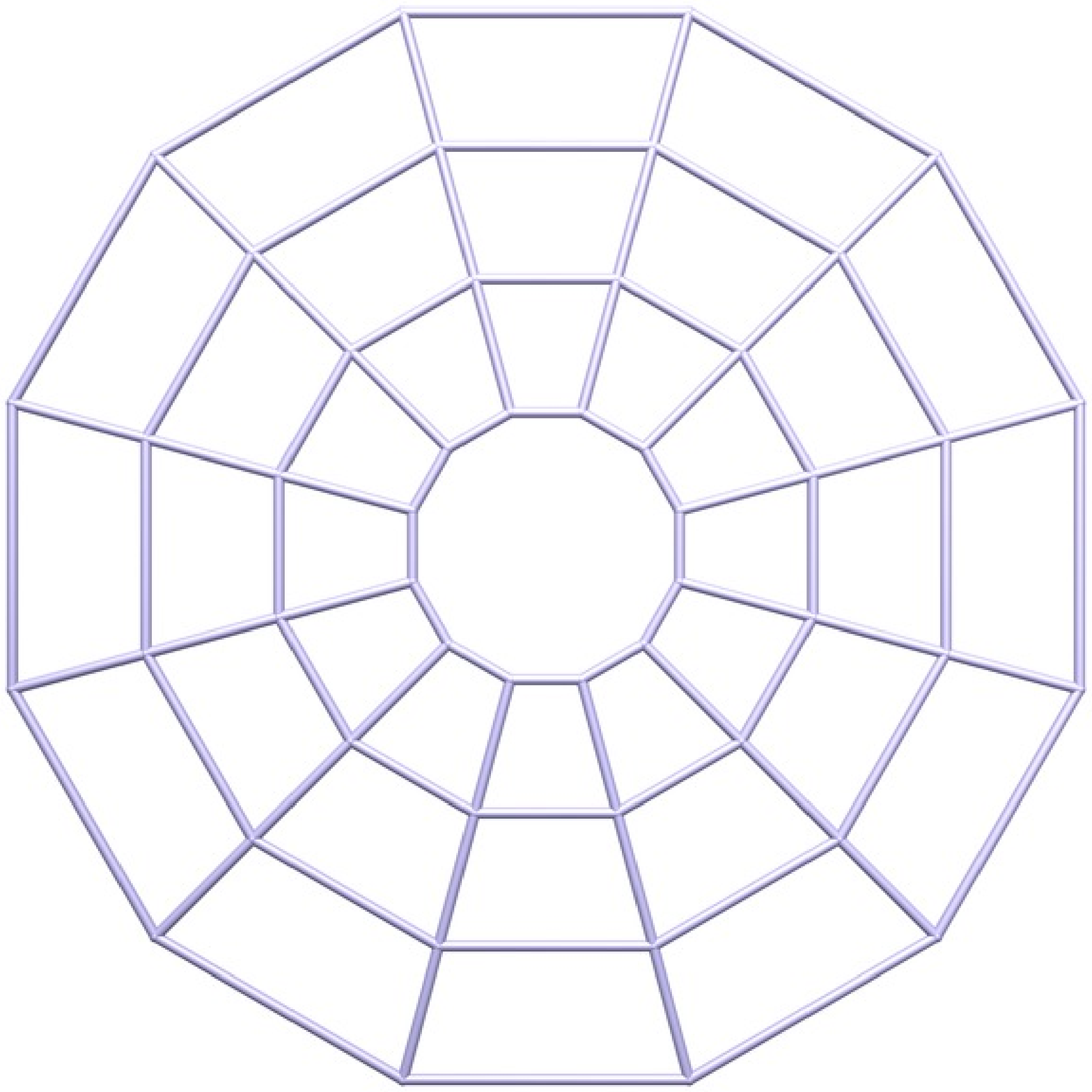}
  }\hfill%
  \subfigure[]{%
   \includegraphics[width=\hb]{\bilder/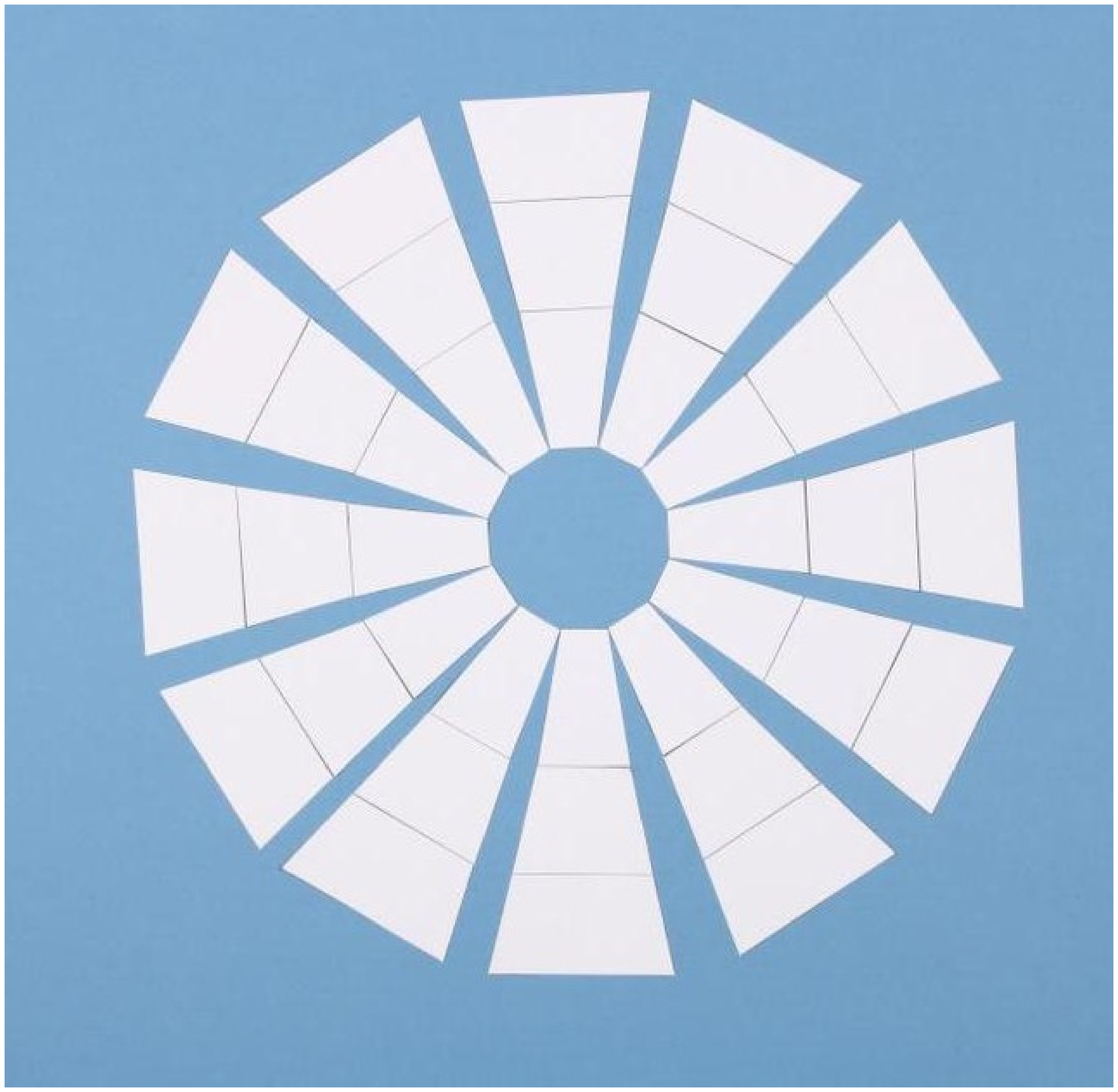}
  }
  \caption{\label{fig.gitter2d}
    A thought experiment on the construction of the sector model
    of the black hole equatorial plane:
    A lattice is erected around the black hole
    according to the pattern shown in (a).
    For each cell, the lengths of the four enclosing rods
    are measured and from these data paper sectors
    are constructed true to scale.
    The result is the sector model of a ring around the black hole (b).
    }
\end{figure}

\subsection{Geodesics close to a black hole}
\label{sec.ssmgeod}

The second part of the workshop
begins with
the presentation of a sector model
that permits to construct geodesics
in the vicinity of a black hole.
The sector model represents a plane of symmetry
of the black hole
that in the following we will call equatorial plane.%
\footnote{%
  We consider a non-rotating black hole.
  It has spherical symmetry, therefore, every geodesic
  lies in a plane that is a symmetry plane of the black hole.
  }

To introduce the sector model, its "construction" is described
in a thought experiment:
A spaceship is sent to a black hole
with the task of surveying
the space around it.
To this end, a lattice is erected around the black hole
according to the pattern shown in figure~\ref{fig.gitter2d}(a):
Rigid rods are arranged like an orb web
in the equatorial plane, centred on the black hole.
The whole lattice is located outside of the event horizon
because no such static structure is possible
in the inner region of the black hole.\footnote{%
The lattice covers the region
from $1.25$ to $5$ Schwarzschild radii
in the Schwarzschild radial coordinate,
see section~\ref{sec.slberechnung}.
}
Measurements are taken of the lattice:
Each cell is enclosed by four rods.
The lengths of the rods are measured
and the data sent to Earth.
There, each cell, reduced in size, is represented
by a sector.
The complete set of sectors forms
a true to scale model
of a ring around the black hole
(figure~\ref{fig.gitter2d}(b)).
Obviously,
the sectors cannot be arranged
to cover a ring without leaving gaps.
This indicates that the geometry
of the black hole equatorial plane
is different from the geometry of the plane surface
on which the sectors are laid out.
The equatorial plane of the black hole
is part of a curved space;
the flat support of the model
is a plane in euclidean space.
If one could place a black hole with the appropriate mass
in the centre of the model,
then all the flat pieces \emph{with the sizes and shapes as shown}
would fit without gaps.
The required mass amounts to about three earth masses
for the cut-out sheet
that is provided online
\citep{zah2018}.

\begin{figure}
  \centering
  \subfigure[]{%
    \includegraphics[width=\db]{\bilder/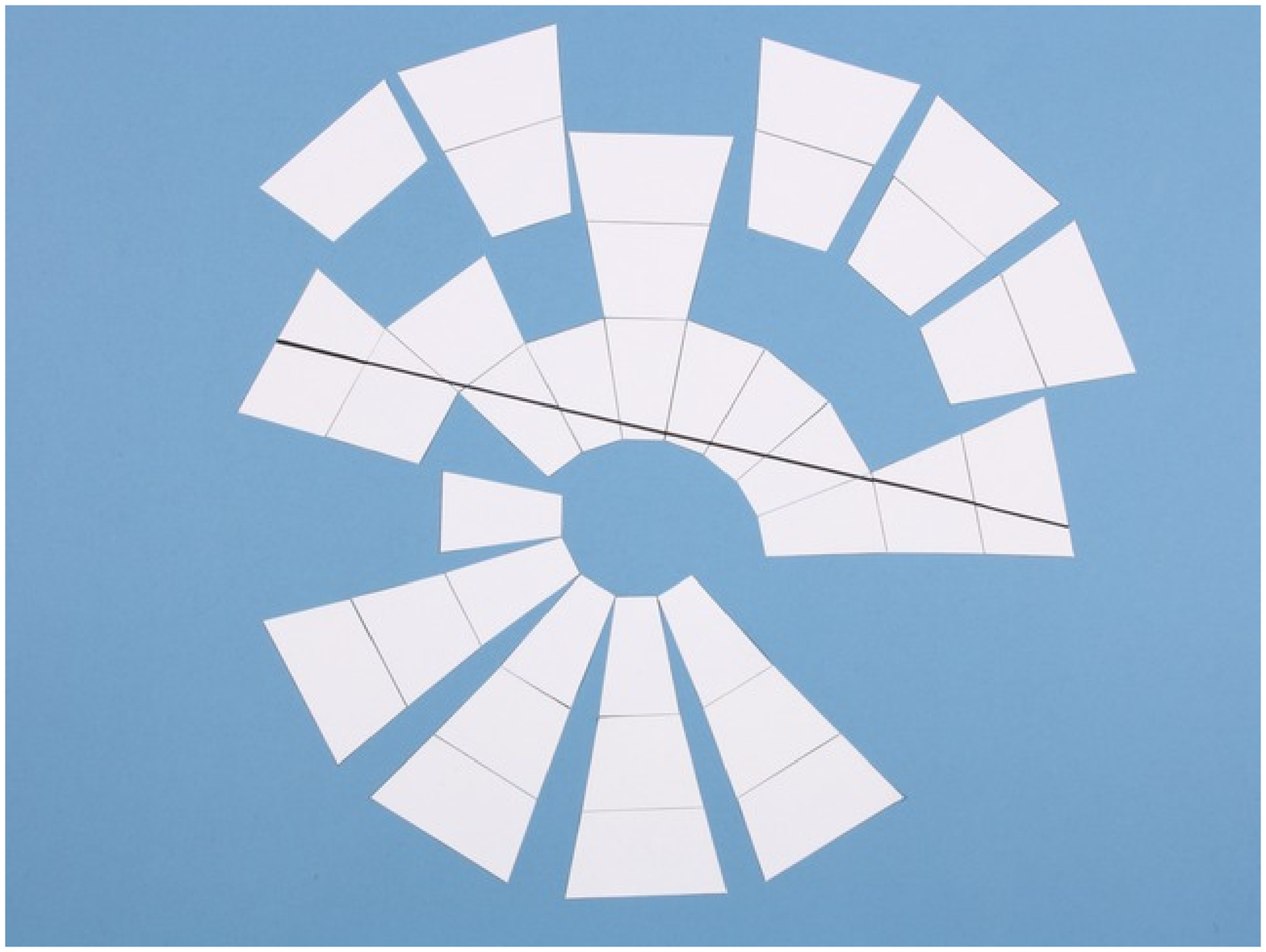}
  }\ba%
  \subfigure[]{%
    \includegraphics[width=\db]{\bilder/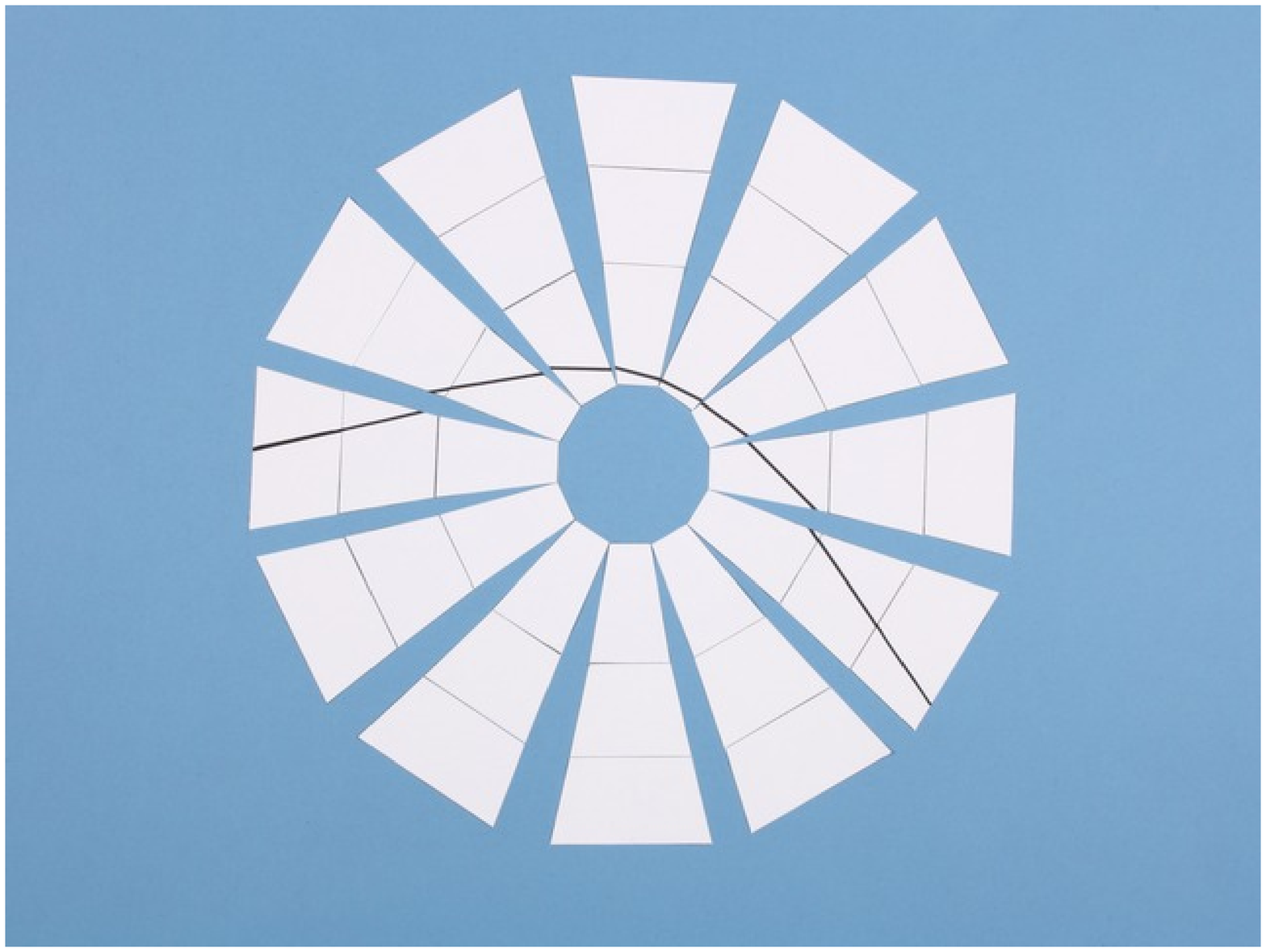}
  }
  \caption{\label{fig.slgeod}
A geodesic on the sector model of the equatorial plane of a black hole.
The line is locally straight;
the direction "far behind" the black hole differs from the direction
"far ahead".
In (a) the sectors are joined along the geodesic and in (b)
they are arranged symmetrically.
    }
\end{figure}

Alternatively,
the sector model can be introduced
via the workshop on curved space described in paper~I.
This workshop presents
a sector model of the curved three-dimensional space
around a black hole
(figure~5(b) in paper~I).
The equatorial plane of this model
(in figure~5(b) of paper~I: the green, nearly horizontal
sides of the blocks)
is identical with the sector model shown in
figure~\ref{fig.gitter2d}(b).

To build the sector model shown in
figure~\ref{fig.gitter2d}(b),
the sectors
are cut out of a sheet of paper
and are glued onto cardboard with
spray adhesive
(use repositionable spray adhesive
for repeated lifting and repositioning)%
\footnote{%
See section~\ref{sec.geodzeichnen}
for a method that does not require the use of adhesive.
}.
The sector model is then used to study geodesics close to a black hole.
First a single geodesic is drawn across the sector model.
As in the case of the curved surfaces
described above,
this is done by joining neighbouring sectors
and drawing a straight line
(figure~\ref{fig.slgeod}(a)).
It can be seen that the two end sections of the line
point in different directions
(Fig~\ref{fig.slgeod}(b)).
Thus, for a line
that passes close to a black hole
while keeping its direction at each point,
the direction "far behind" the black hole differs from the direction
"far ahead".
This construction illustrates the principle underlying
light deflection in a gravitational field:
Light propagates on a locally straight path;
when it passes a region of curved space,
the direction of propagation
afterwards is different from that before.

Two things must be borne in mind
when assessing the significance of this construction
on a sector model.
First, the geodesics constructed on sector models
are quantitatively correct.
When the condition of a locally straight line
is expressed mathematically,
the result is the geodesic equation
\citep[p.~70~ff]{wei1972}.
The graphically constructed geodesic
is a solution of this equation.
Since the sector model
is an approximate representation
of the curved space,
the geodesic drawn on it
is likewise an approximate solution.
Using an appropriately fine subdivision,
geodesics can
in principle
be graphically constructed
with high accuracy
(section~\ref{sec.modell2}).
Secondly one must bear in mind that
though the line constructed above
is a geodesic,
it is not a light ray.
This line is a geodesic in space.
But light propagates in space and time,
meaning that light rays are spacetime geodesics.
Nevertheless, the geodesic in space illustrates
by way of close analogy
the principle behind gravitational light deflection.

\begin{figure}
  \centering
  \subfigure[]{%
    \includegraphics[width=\db]{\bilder/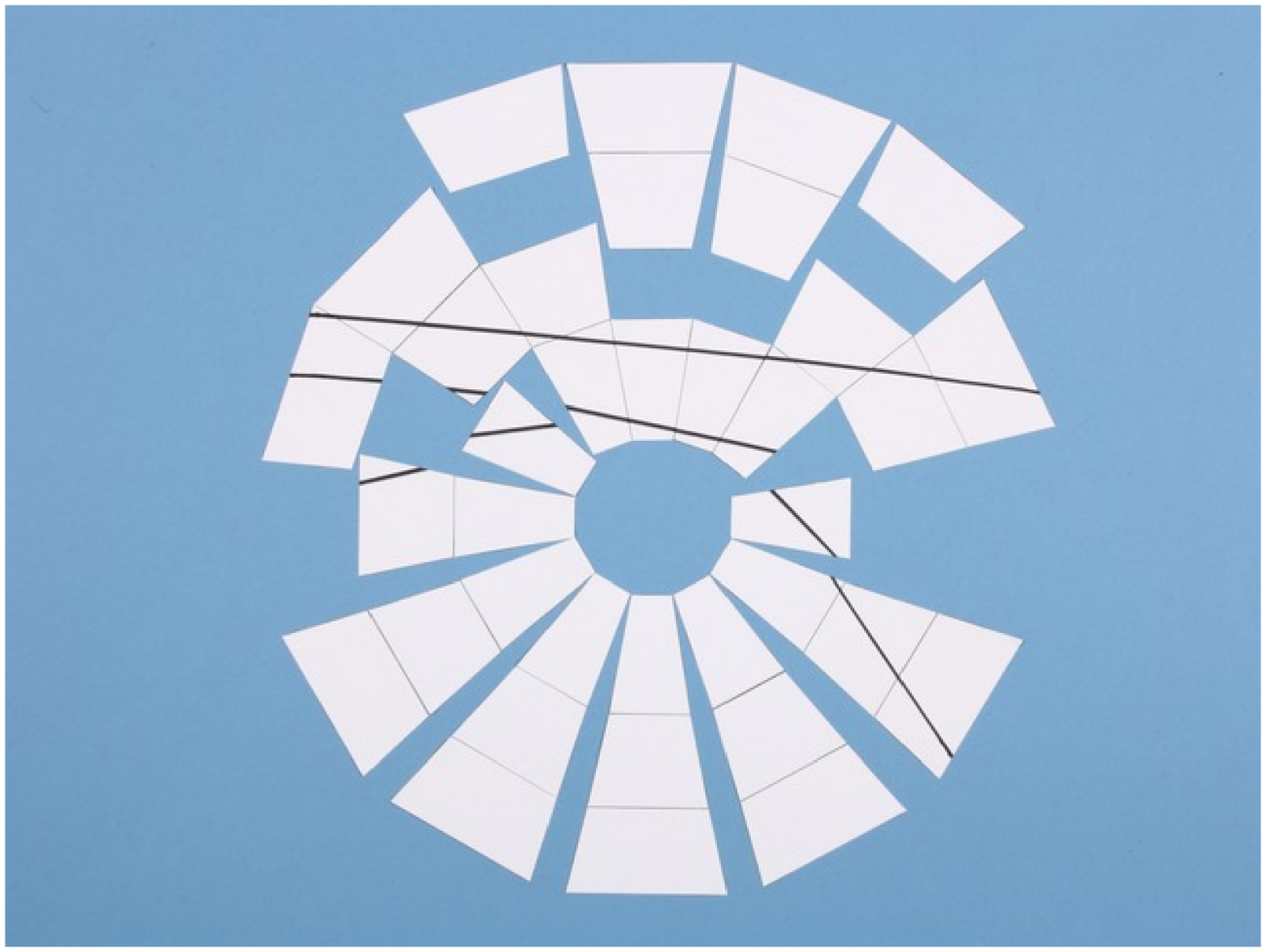}
  }\ba%
  \subfigure[]{%
    \includegraphics[width=\db]{\bilder/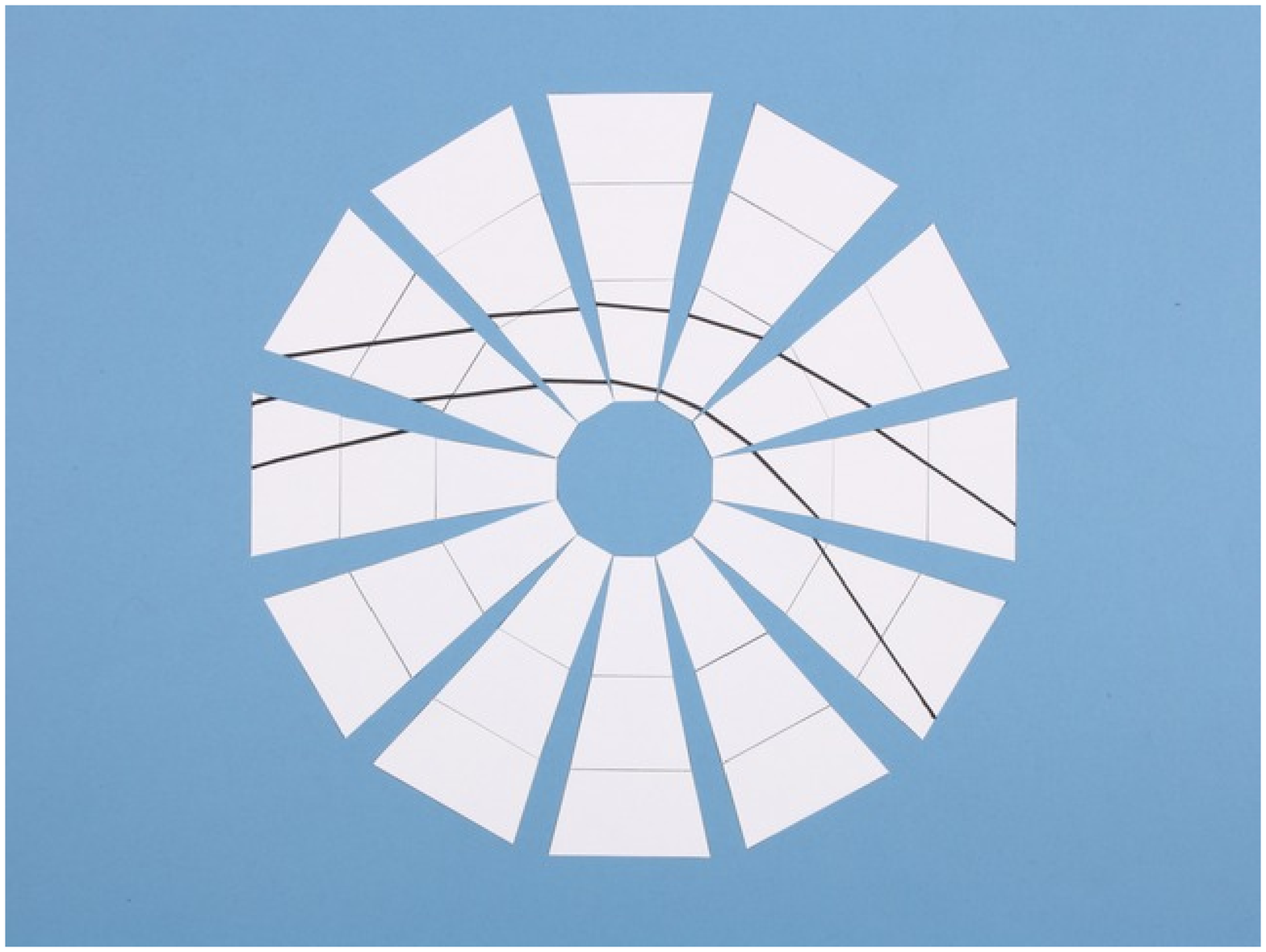}
  }
\caption{\label{fig.scharparallel}
To
the geodesic shown in figure \protect\ref{fig.slgeod},
a second geodesic is added that
on the far left is parallel to the first:
The inner geodesic is deflected more strongly,
and the two geodesics diverge.
In (a) the sectors are joined along the second geodesic,
in (b) they are arranged symmetrically.
}
\end{figure}

Even though geodesics in space
are not identical with light rays,
it is instructive to use them
for demonstrating
properties of geodesics.
One may for instance construct a second geodesic
that starts close to the first and in the same direction
(figure~\ref{fig.scharparallel}).
The geodesic that runs closer to the black hole
is deflected more strongly and the two geodesics diverge.
\begin{figure}
  \centering
  \subfigure[]{%
    \includegraphics[width=\db]{\bilder/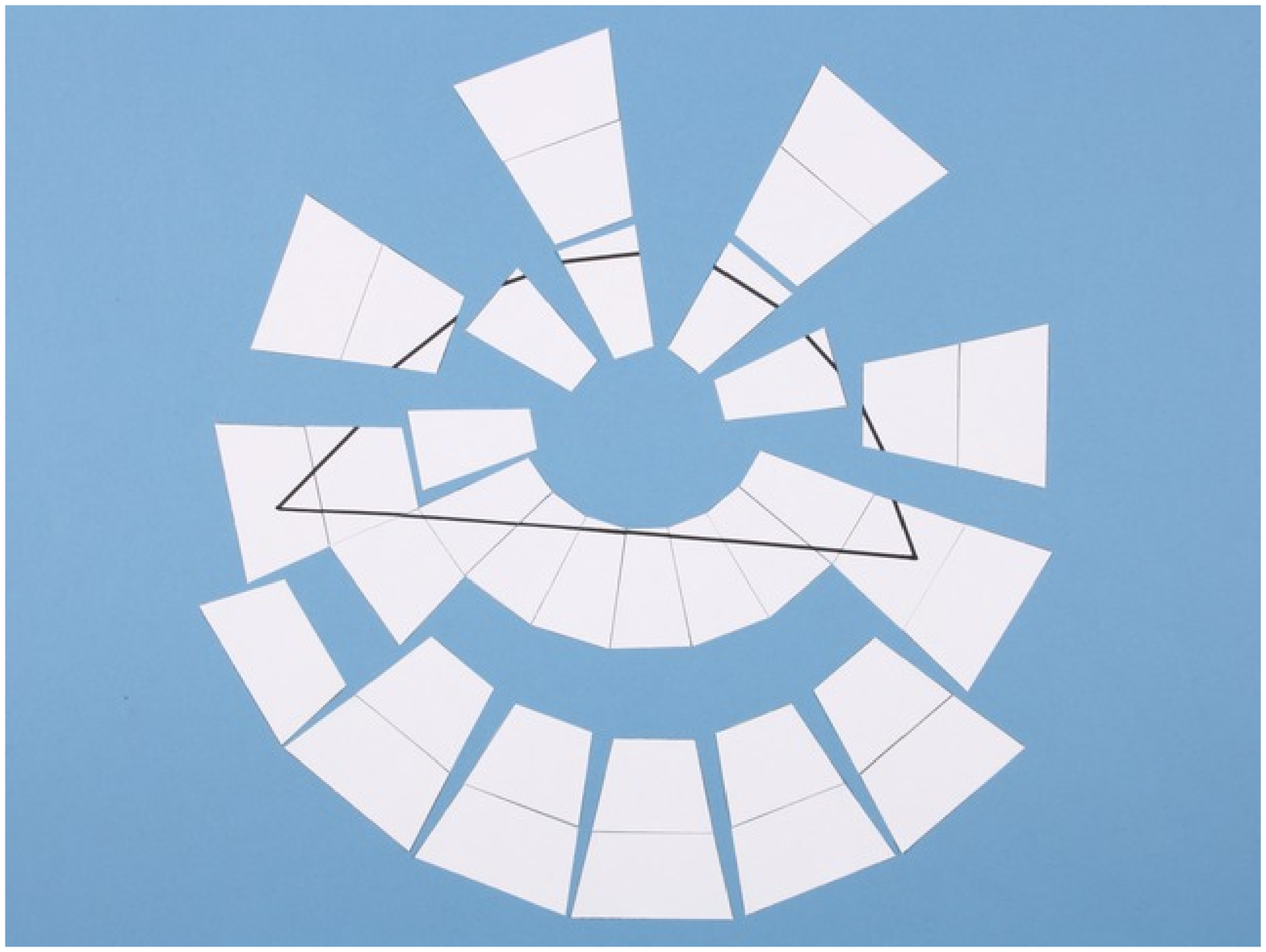}
  }\hfill%
  \subfigure[]{%
    \includegraphics[width=\db]{\bilder/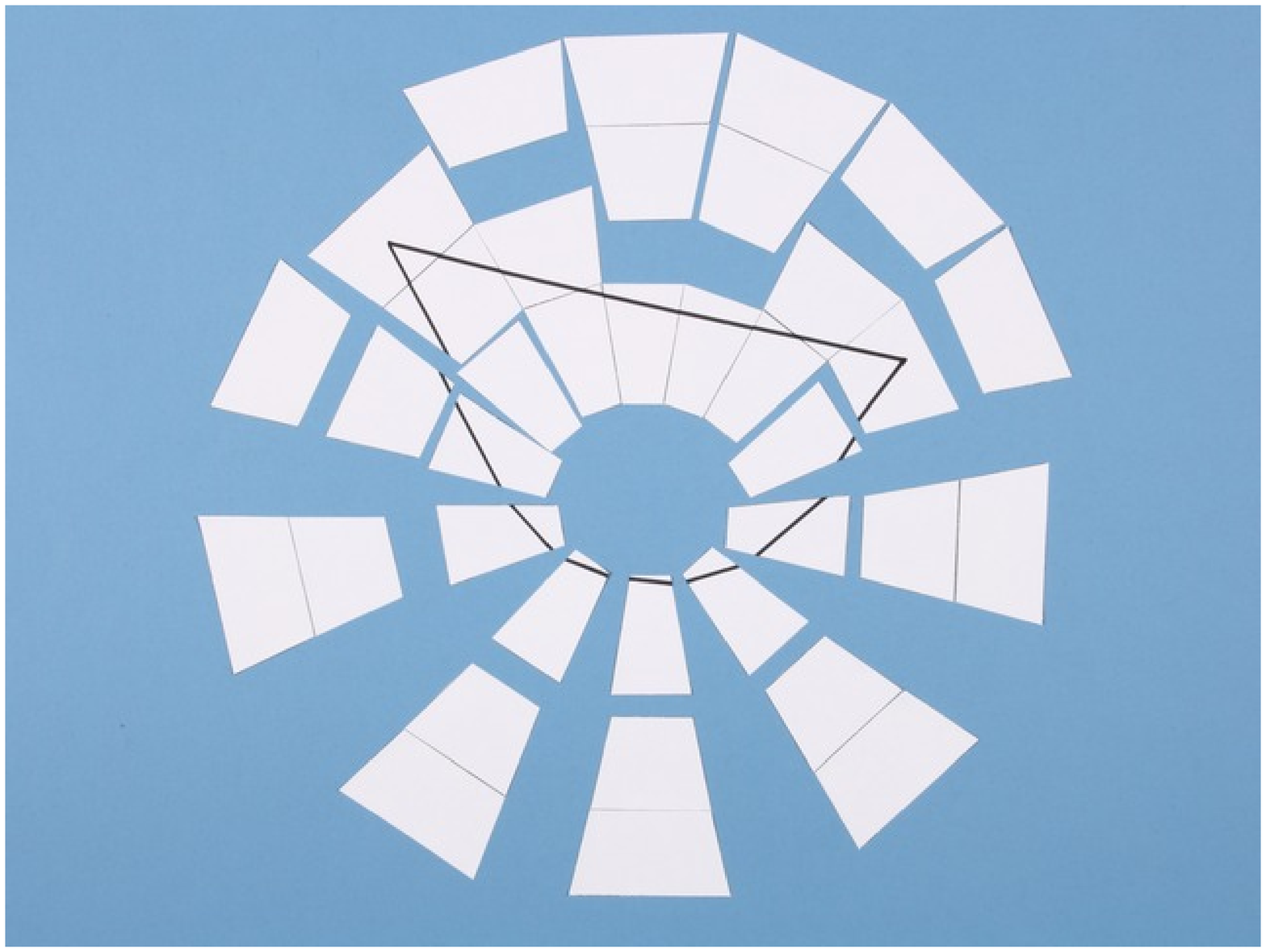}
  }\hfill%
  \subfigure[]{%
    \includegraphics[width=\db]{\bilder/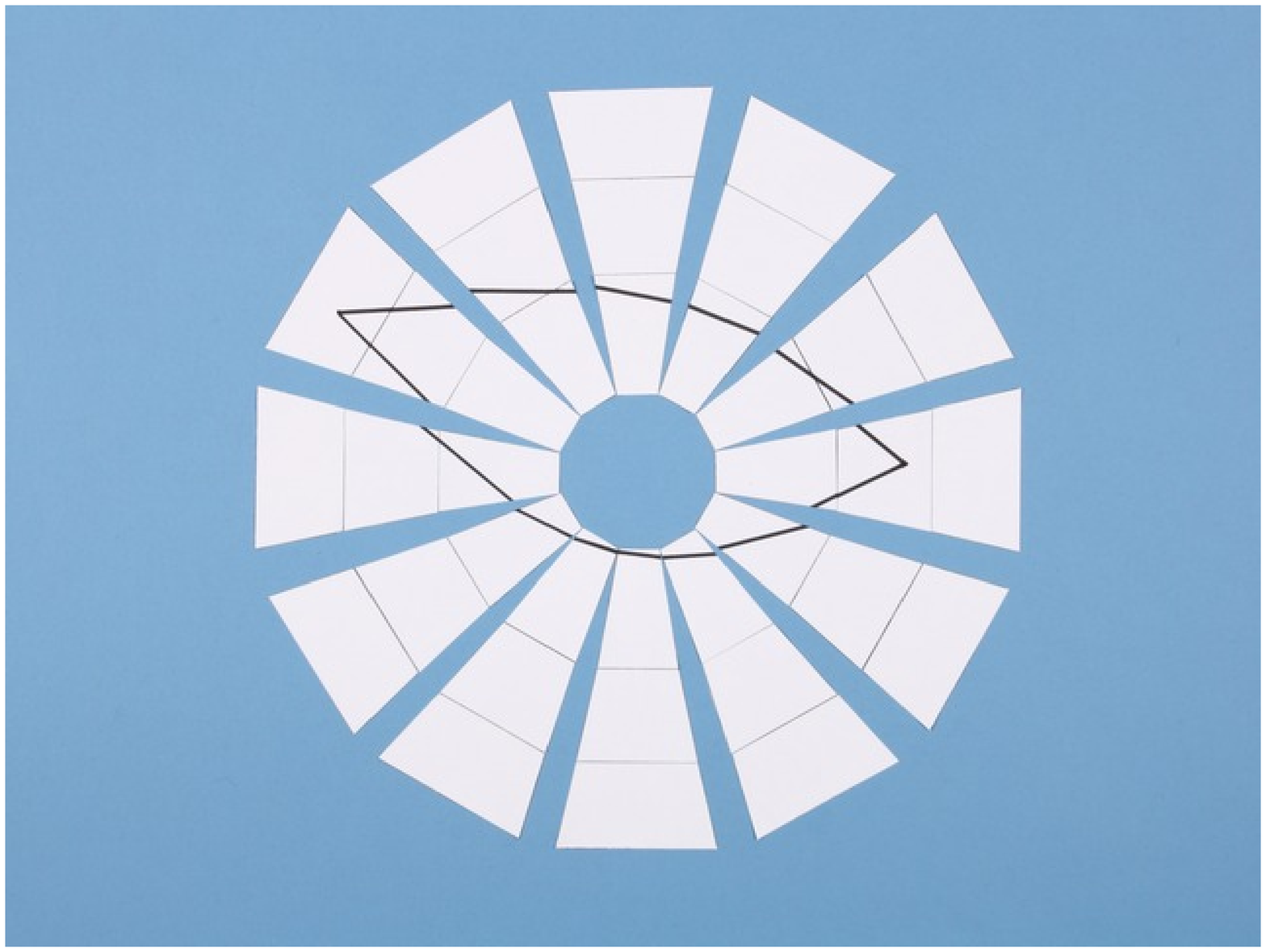}
  }
\caption{\label{fig.zweieck}
Two geodesics form a digon.
This illustrates the formation of double images
due to gravitational light deflection: Light emitted from
a source reaches the observer along two different paths.
In (a) and (b) the sectors are joined along the first
and the second geodesic, respectively,
in (c) they are arranged symmetrically.
}
\end{figure}
Or one may construct two geodesics
that come from the same point, pass the black hole
on opposite sides, and meet again
(figure~\ref{fig.zweieck}).
Thus, with geodesics it is possible to form a digon.
Extrapolated to light rays,
this construction shows how double images arise.

\subsection{The construction of geodesics using transfer sectors}
\label{sec.geodzeichnen}

\begin{figure}
  \centering
  \subfigure[]{%
    \includegraphics[scale=0.3]{\bilder/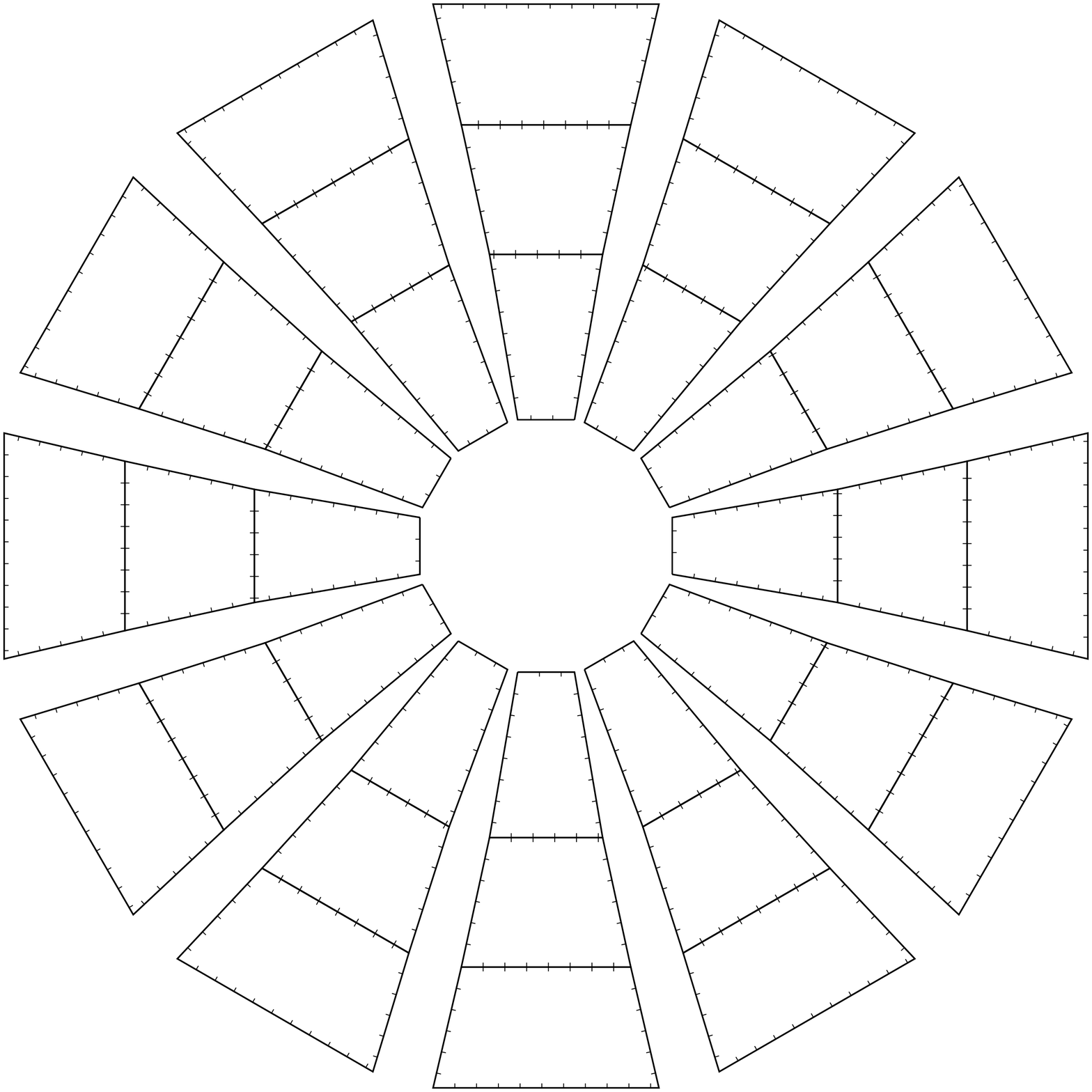}
  }\qquad%
  \subfigure[]{%
    \shortstack{%
      \mbox{\rule{2pt}{18bp} \raisebox{8pt}{$\rs$}}\\
      \vspace*{2cm}\\
      \includegraphics[scale=0.3]{\bilder/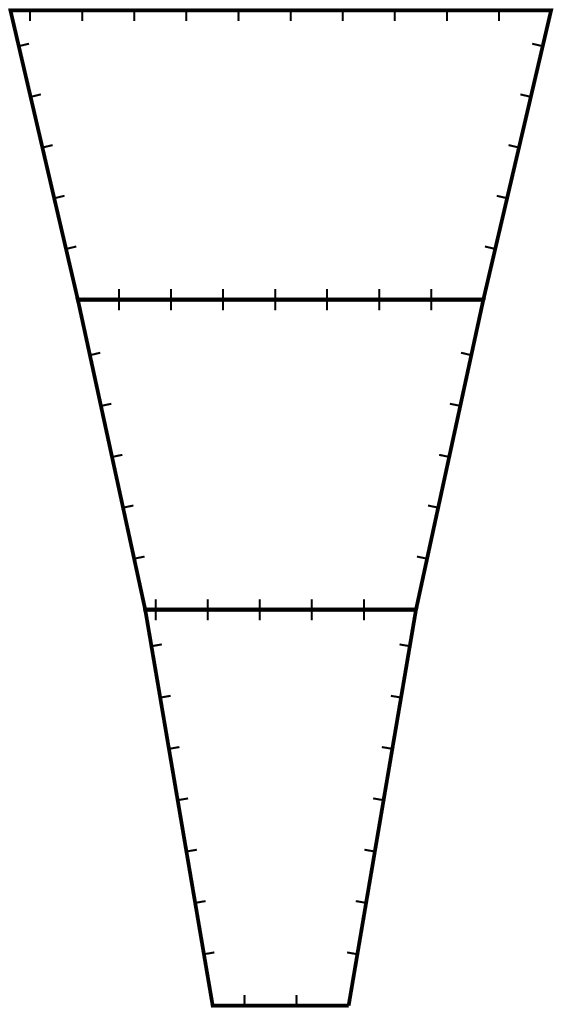}
    }
  }
  \caption{\label{fig.slmodell}
 Worksheet for the construction of geodesics
 close to a black hole. The worksheet includes
 the sector model of the equatorial plane
 arranged symmetrically
 and with tick marks (a)
 and a column of transfer sectors (b).
The length of the scale bar indicates the Schwarzschild radius
$\rs$ of the black hole.
}
\end{figure}

\begin{figure}
\centering
  \subfigure[]{%
    \includegraphics[width=\veb]{\bilder/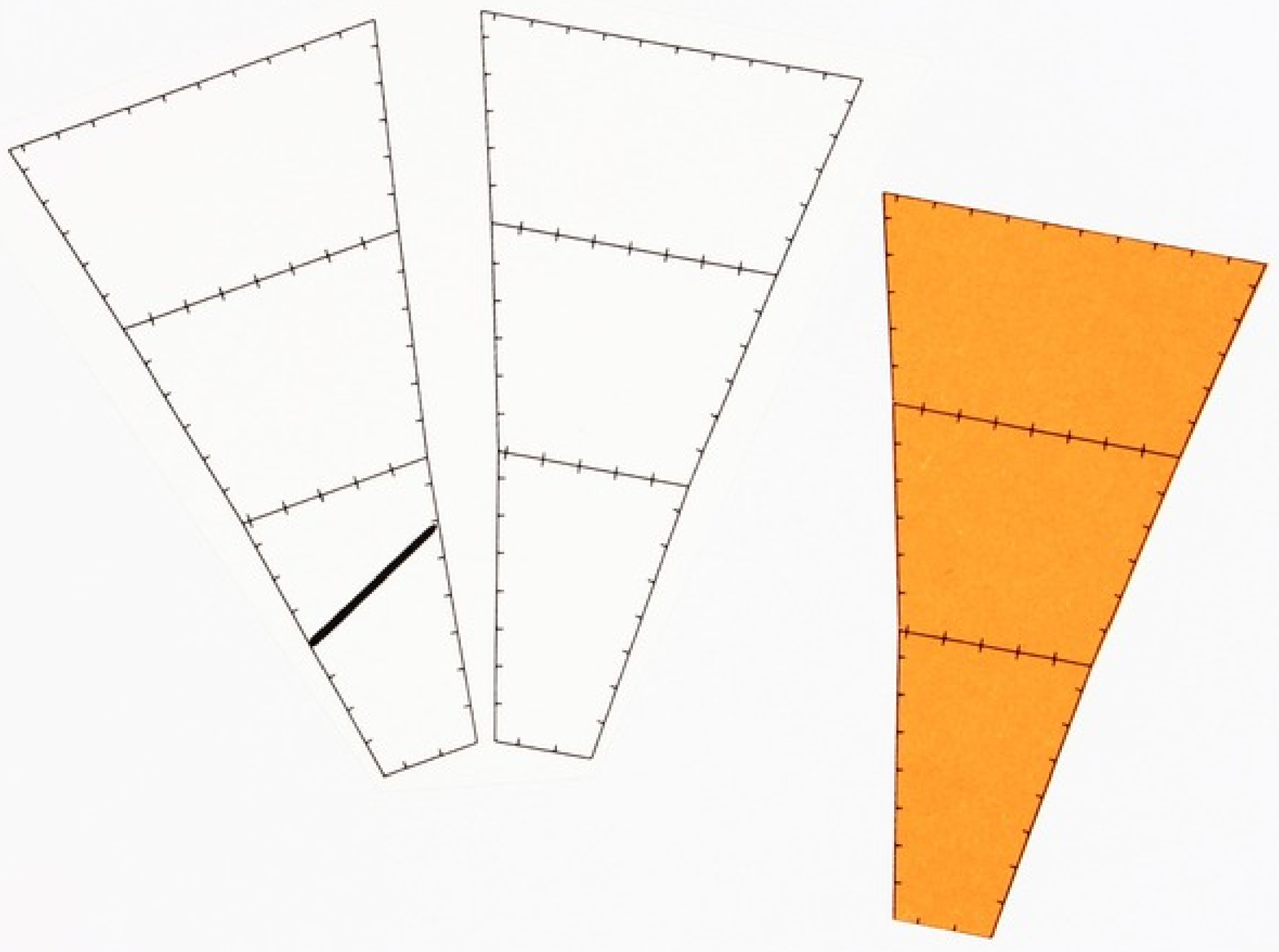}
  }\hfill%
  \subfigure[]{%
    \includegraphics[width=\deb]{\bilder/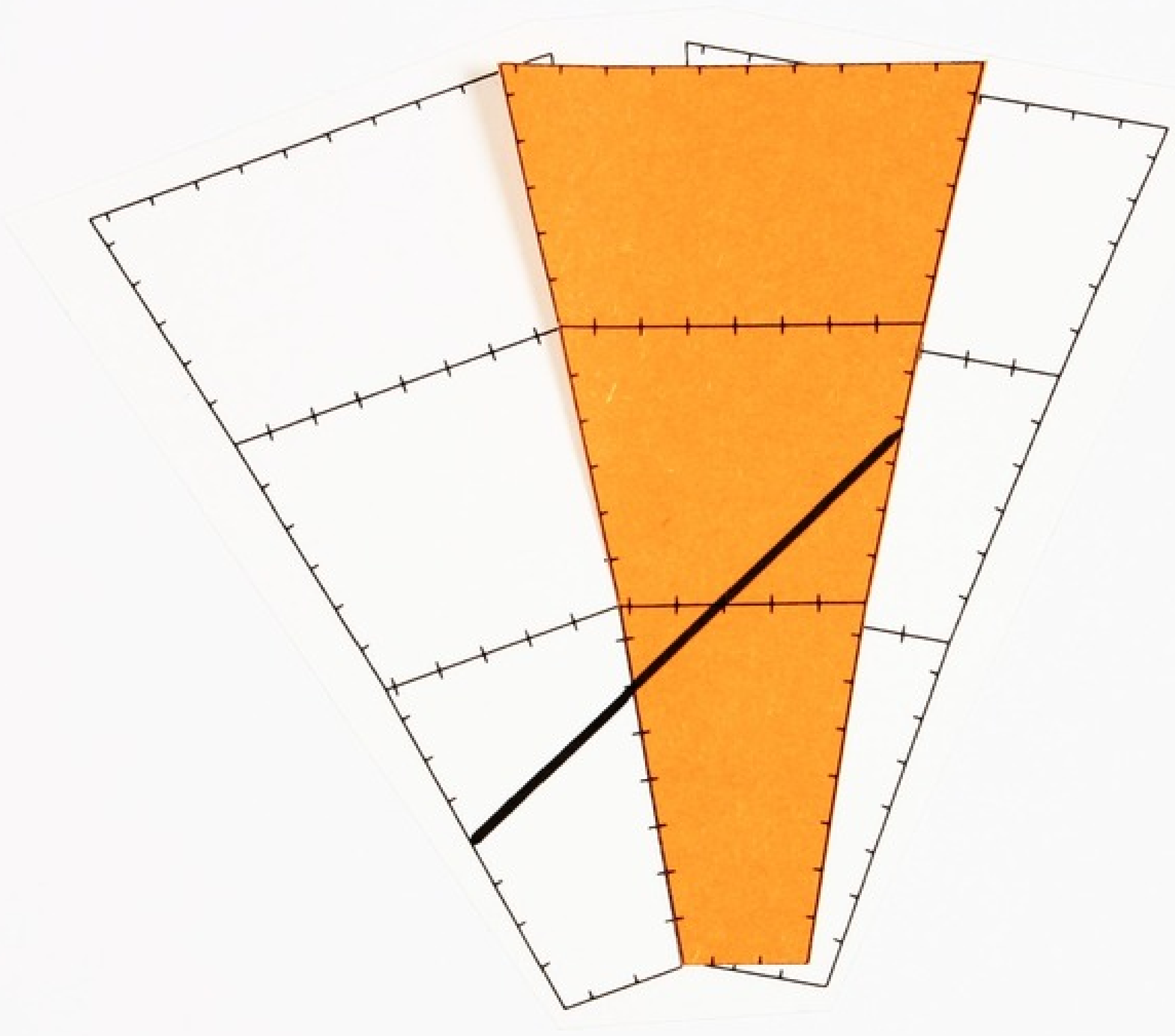}
  }\hfill%
  \subfigure[]{%
    \includegraphics[width=\veb]{\bilder/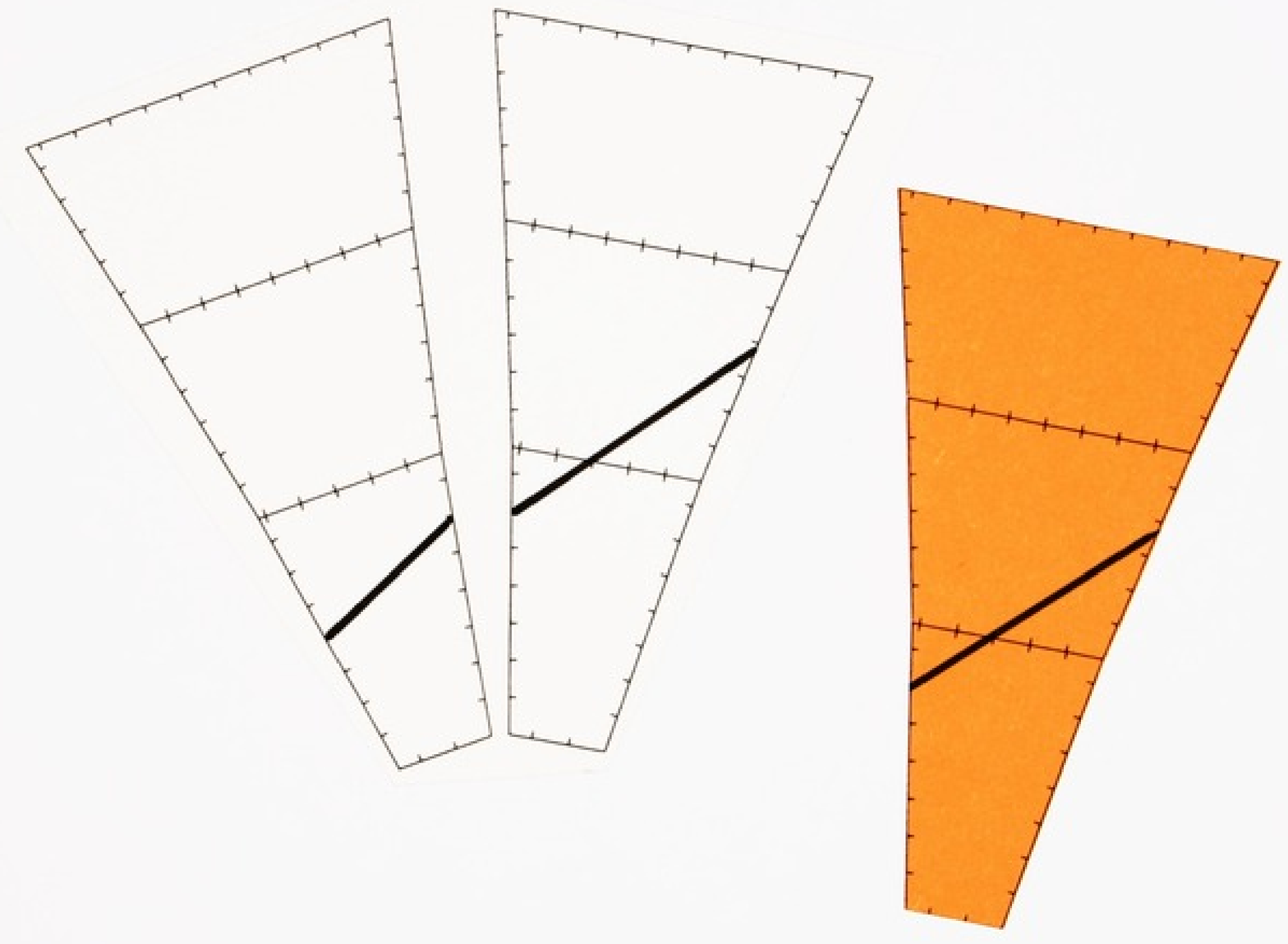}
  }
\caption{
\label{fig.transfer}
The construction of geodesics using transfer sectors (coloured).
A geodesic is drawn up to the border of the column (a),
continued across the transfer column (b),
and copied from the transfer column onto the neighbouring column (c).
}
\end{figure}

Figures~\ref{fig.slgeod},
\ref{fig.scharparallel},
and~\ref{fig.zweieck}
show sectors that have been cut out of a sheet of paper
and have been aligned along a geodesic or arranged symmetrically,
as required.
This procedure has the advantage that
each geodesic can separately be displayed as a straight line.
However, the construction of geodesics
can be carried out more easily and more quickly
without cutting out all the sectors.
To this end, one uses the complete model in symmetric layout
and with tick marks as shown in figure~\ref{fig.slmodell}(a).
A single additional column (figure~\ref{fig.slmodell}(b))
is cut out of a sheet of paper,
these are the so-called transfer sectors.
The construction of a geodesic starts on the symmetric model and
the line is first drawn up to the border of the column
(figure~\ref{fig.transfer}(a)).
The appropriate sector of the transfer column is then joined
and the line is continued across the column of transfer sectors
(figure~\ref{fig.transfer}(b)).
The line on the transfer column is copied
onto the neighbouring column of the symmetric model
(figure~\ref{fig.transfer}(c)).
This procedure is repeated up to the desired end point.
In the model shown in figure~\ref{fig.slmodell}(a),
equidistant tick marks have been added at the borders
to facilitate the copying of the lines.
The worksheet shown in figure~\ref{fig.slmodell}
is available
online
\citep{zah2018}.

\subsection{The construction of sector models}
\label{sec.modellberechnung}

A workshop can be implemented
as described above,
with sector models that are provided.
This is the most elementary and the shortest type of workshop.
But for sector models
to realize their full potential
as tools for studying curved spaces,
the participants should calculate and construct the models
on their own.
This enables them
to study other curved spaces
in the same way,
by, e.g., studying the geodesics
corresponding to a given metric.
Thus, setting up and solving the geodesic equation
is replaced by creating the sector model and
drawing the geodesics onto it.

The following section shows how one may introduce the
calculation of sector models by using the sphere
as an example.
This procedure is then extended to the calculation
of the sector model of the black hole equatorial plane.

\begin{figure}[t]
  \centering
  \subfigure[]{%
    \includegraphics{\bilder/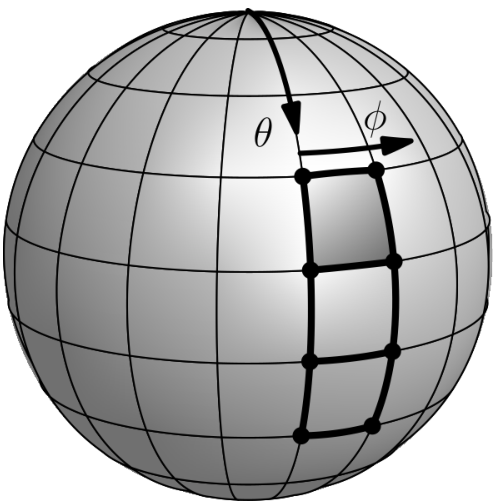}
  }\hfil%
  \subfigure[]{%
    \includegraphics[scale=1]{\bilder/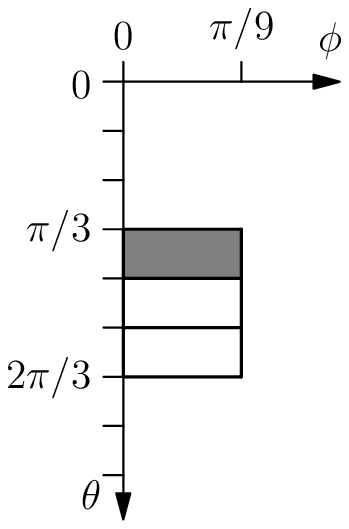}
  }\hfil%
  \subfigure[]{%
    \includegraphics[width=0.11\textwidth]{\bilder/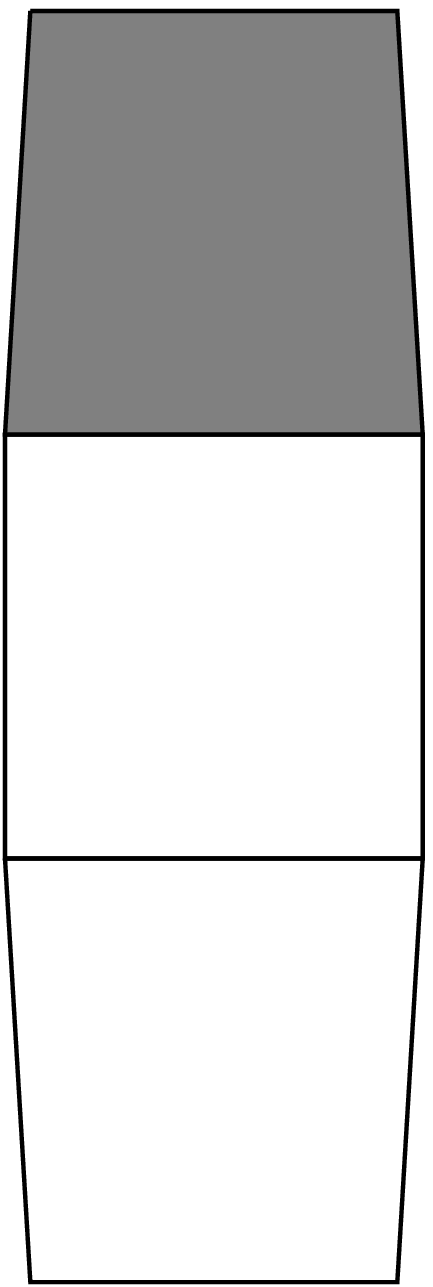}
  }
  \caption{\label{fig.kugelteilung}
   (a) The sphere described
   in polar coordinates $\theta$, $\phi$
   and subdivided into elements of area of 20 degrees by 20 degrees.
   (b) Three elements of area
   in $\phi$-$\theta$ coordinate space.
   (c) The corresponding sectors.
   They make up one column of the model shown in
   figure~\protect\ref{fig.kugelmodell}(b).
   For clarity, one and the same sector is
   highlighted in grey
   in all three component images.
}
\end{figure}

\subsubsection{Construction of the sector model of a spherical surface.}
\label{sec.kugelberechnung}

This example serves to introduce
the general procedure for
the construction of sector models.
The starting point is the concept of a metric
as a function that takes
the coordinates of two
nearby points as arguments
and returns their distance.
This can be introduced in an elementary
way by starting with curvilinear coordinates in the euclidean
plane (e.g., \citealp{kra2016}; \citealp{har}, p.~21~f; \citealp{nat}, p.~35~f).

For the sector model of the sphere,
the calculation
is based on the metric in the usual spherical coordinates $\theta$, $\phi$
(figure~\ref{fig.kugelteilung}(a)):
\begin{equation}
\ud s^2 = R^2\, \ud \theta^2 + R^2 \sin^2\theta \, \ud \phi^2,
\end{equation}
where $R$ is the radius of the sphere
(for an elementary derivation that can be used in the
workshop, see, e.g., \citealp{har}, p.~23~f;
\citealp{nat}, p.~37~ff).

The creation of the sector model
proceeds in three steps.
In the first step
the sphere is divided up into
elements of area, defined by their vertices.
In the example shown here, the elements of area are quadrilaterals
with vertices at 20~degree ($\pi/9$) intervals
in the angular coordinates $\theta$ and $\phi$,
respectively
(figures~\ref{fig.kugelteilung}(a), (b)).
In the second step
the edge lengths
of the quadrilaterals
are computed.
This is done in an approximate way
in order to keep the calculation simple.
For each edge,
one determines the length
by treating
the end points as
nearby points
in the sense of the metric.
For edges between vertices with the same longitude, one obtains
the length
\begin{equation}
\Delta s = R\,\Delta\theta \hspace*{1cm} (\Delta\phi=0),
\label{eq.kugel_laenge}
\end{equation}
in this example $\Delta s = R\pi/9$.
For edges between two vertices of the
same latitude, one finds
\begin{equation}
\Delta s = R\,\sin\theta\,\Delta\phi \hspace*{1cm} (\Delta\theta=0),
\label{eq.kugel_breite}
\end{equation}
depending on the angle $\theta$ of the circle of latitude.
For the sector models shown in the figures
and provided online,
the distance between vertices is determined
along geodesics
(see paper~I).
The difference between the approximate and the exact
edge lengths
amounts to $0.13\%$ at most in this example.

In step three flat pieces of area are constructed
from the edge lengths.
In the present example,
the elements of area on the sphere possess mirror symmetry.
The flat pieces of area are constructed with the same
symmetry property
in the shape of
symmetric trapezia
(figure~\ref{fig.kugelteilung}(c)).

\subsubsection{Construction of the sector model of the equatorial plane
of a black hole.}
\label{sec.slberechnung}

\begin{figure}
\centering
\subfigure[]{%
  \includegraphics[scale=1]{\bilder/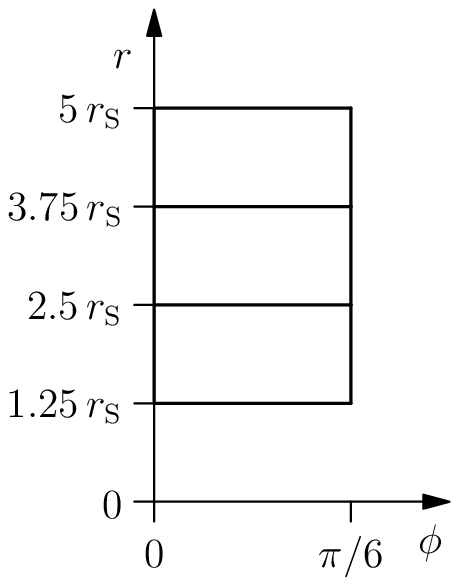}
}\qquad%
\subfigure[]{%
  \includegraphics[scale=0.8]{\bilder/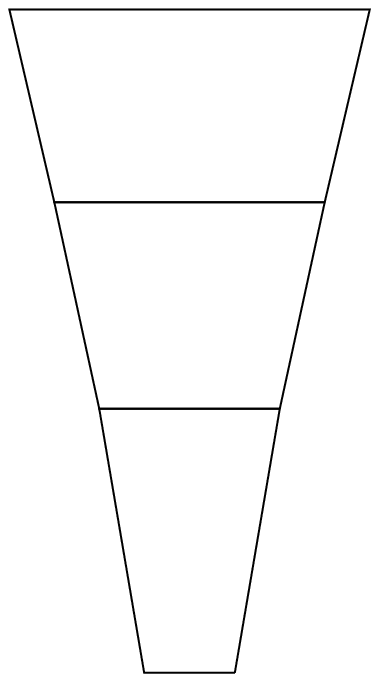}
}
\caption{\label{fig.sl}
Construction of the sector model of the equatorial plane of a black hole.
(a) The three elements of area of one column in $\phi$-$r$ coordinate space.
(b) The corresponding sectors.
}
\end{figure}

The starting point of the calculation is the metric of the
equatorial plane of a black hole
\begin{equation}
\ud s^2 = \frac{1}{1-\rs/r} \ud r^2 + r^2 \ud \phi^2
\label{eq.slmetrik}
\end{equation}
with the usual Schwarzschild coordinates $r$ and $\phi$.
Here, $\rs = 2 G M / c^2$ is the Schwarzschild radius
of the black hole with mass $M$, $G$ is the gravitational constant,
and $c$ the speed of light.
The sector model represents an annular part of the equatorial
plane, centred on the black hole.
The inner rim is located at $r=1.25\, \rs$
and the outer rim at $r=5\, \rs$.
The azimuthal angle $\phi$ has values between zero and $2\pi$.

First the ring is divided up into elements of area.
For that purpose, the $\phi$-range is subdivided into
twelve segments of coordinate length $\pi/6$ each.
Since the metric does not depend on the coordinate $\phi$,
only one of the twelve segments needs to be calculated.
The $r$-range is subdivided into three segments
of coordinate length $1.25\, \rs$ each
(figure~\ref{fig.sl}(a)).
Next, the edge lengths of the three quadrilaterals
shown in figure~\ref{fig.sl}(a) are calculated.
Using the metric,
the distance between vertices with the same value of $r$
is obtained as
\begin{equation}
\Delta s = r \,\Delta\phi \hspace*{1cm} (\Delta r = 0).
\end{equation}
When calculating the distance of vertices
with the same $\phi$-coordinate,
the first term of the metric comes into play.
Its metric coefficient $1\big/(1-\rs/r)$
depends on $r$ and, therefore, varies along
the edge.
Here we make another approximation
and
use the metric coefficient
at the mean $r$-coordinate $\rrmi$
of the edge:
\begin{equation}
\Delta s = \sqrt{\frac{1}{(1-\rs/\rrmi)}}\, \Delta r
     \hspace*{1cm} (\Delta\phi = 0),
\end{equation}
where $\rrmi = (r_1 + r_2)/2$
with the coordinates $r_1$ and $r_2$
of the associated vertices.
The sector models shown in the figures and provided
online are constructed with edge lengths
that are computed numerically for geodesics joining
the vertices.
The difference between the approximate and the exact edge
lengths
is largest for the innermost radial
edge and there amounts to 5.4\%.

Step three is the construction of the quadrilaterals.
The subdivision of the ring by radial cuts
creates elements of area that possess mirror symmetry
(figure~\ref{fig.gitter2d}(a)).
In accordance with this symmetry,
the sectors are constructed as symmetric trapezia.
Figure~\ref{fig.sl} shows the three sectors of a column
together with the corresponding three rectangles
in $\phi-r$ coordinate space.
The complete sector model
with twelve columns is shown in
figure~\ref{fig.slmodell}.

\subsection{Geodesics vs. curvature}
\label{sec.ergaenzunggeod}

\begin{figure}[t]
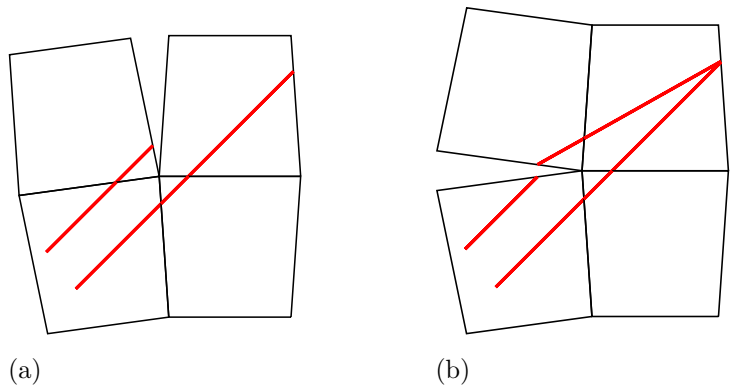

  \centering
  \subfigure[]{%
    \includegraphics[width=\vib]{\bilder/figure15a.eps}
  }\qquad\qquad%
  \subfigure[]{%
    \includegraphics[width=\vib]{\bilder/figure15b.eps}
  }
\caption{\label{fig.vertexgeod}
Two geodesics start in parallel (a), pass a vertex on opposite
sides, and are then inclined towards each other by the deficit
angle (b).
}
\end{figure}

When the workshop described above
is combined with the workshop on curvature
described in paper~I (section~2),
it is possible to address the connection between
the curvature of a surface and its geodesics.
In paper~I, the sphere and the saddle
are introduced as prototypes of surfaces with positive
and negative curvature, respectively.
The deficit angle at a vertex of the sector model
is shown to be a criterion for curvature: Positive curvature
is indicated by a positive deficit angle and vice versa.%
\footnote{%
The deficit angle of a vertex is positive if
a gap remains after joining
all adjacent sectors at this vertex
(figure~\ref{fig.vertexgeod} gives an example).
The deficit angle is negative
if, after joining all the sectors except one,
the remaining space is too small to accommodate the last sector.}

Using sector models, one can show
that the paths of neighbouring geodesics
also provide a criterion for determining curvature.
Figure~\ref{fig.vertexgeod}
shows neighbouring geodesics near a single vertex
with positive deficit angle.
Two geodesics that are parallel ahead of the vertex
and pass the vertex on opposite sides
(figure~\ref{fig.vertexgeod}(a))
converge behind it
(figure~\ref{fig.vertexgeod}(b)).
By construction,
the angle between the two directions behind the vertex
is the deficit angle.

Thus, geodesics starting in parallel
indicate positive curvature if they converge.
Conversely,
they indicate
negative curvature if they diverge.
Two examples for this criterion
are provided in section~\ref{sec.kugel}
in the form of
geodesics on the sphere
(figure~\ref{fig.kugelgeod}(c))
and on the saddle
(figure~\ref{fig.sattelgeod}(d)).
Applied to the equatorial plane of a black hole
where initially parallel geodesics diverge
(figure~\ref{fig.scharparallel}),
one concludes that the curvature is negative.

The argument given above is an illustration of the
equation of geodesic deviation
\begin{equation}
(\nabla_{\bf u}\nabla_{\bf u} {\bf D})^i =
  -  R^i{}_{jkl} u^j D^k u^l
\end{equation}
for two geodesics $x^i(\lambda)$ and
$x^i(\lambda) + D^i(\lambda)$
with
$u^i = d x^i / d\lambda$
and the Riemann curvature tensor
$R^i{}_{jkl}$.
In the sector model, the components of the Riemann curvature
tensor are represented by the deficit angles (paper~I,
section~3) and figure~\ref{fig.vertexgeod} illustrates
their impact on the change in distance between neighbouring geodesics.

\begin{figure}[p]
  \centering
  \subfigure[]{%
    \includegraphics[width=0.63\textwidth]{\bilder/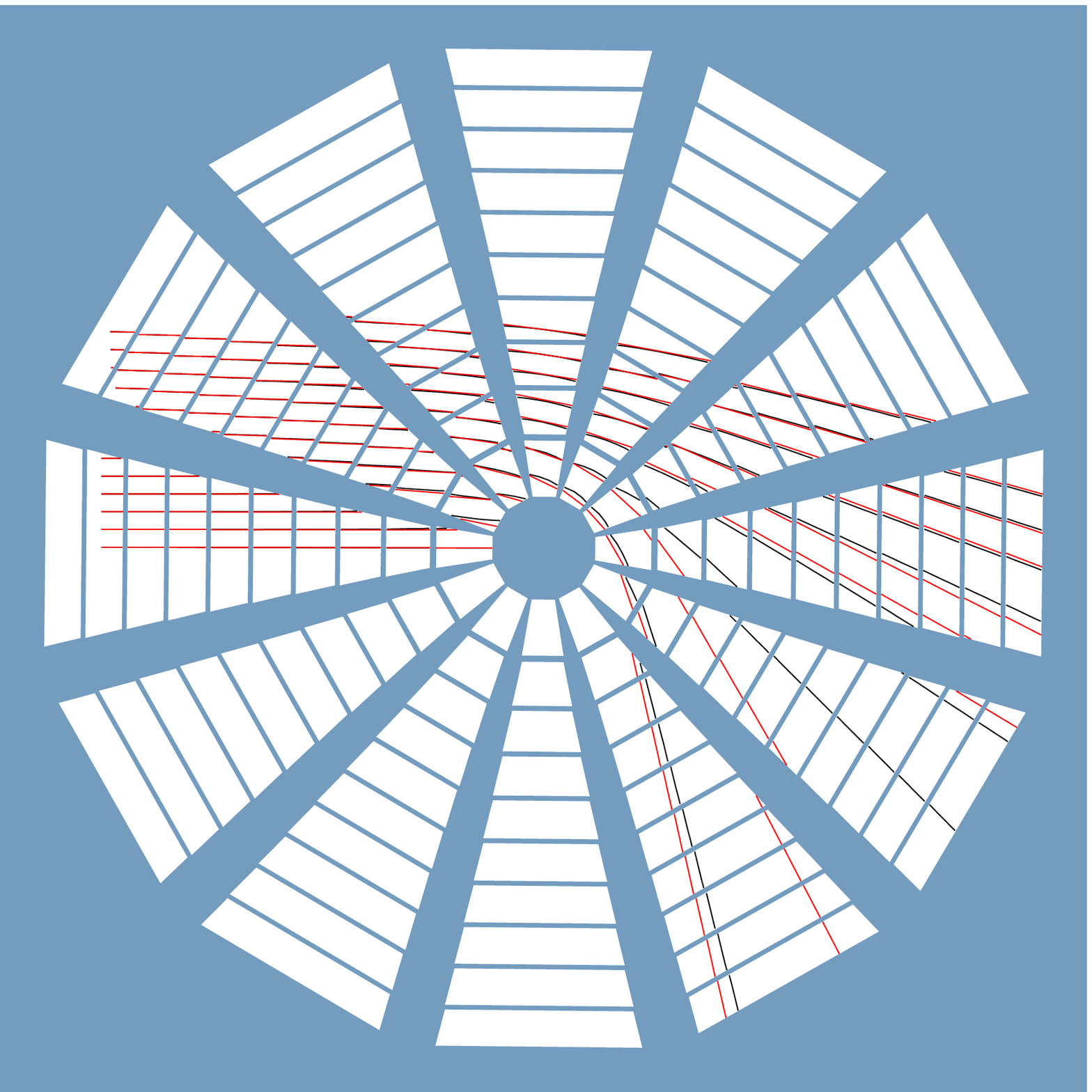}
  }
  \subfigure[]{%
    \includegraphics[width=0.63\textwidth]{\bilder/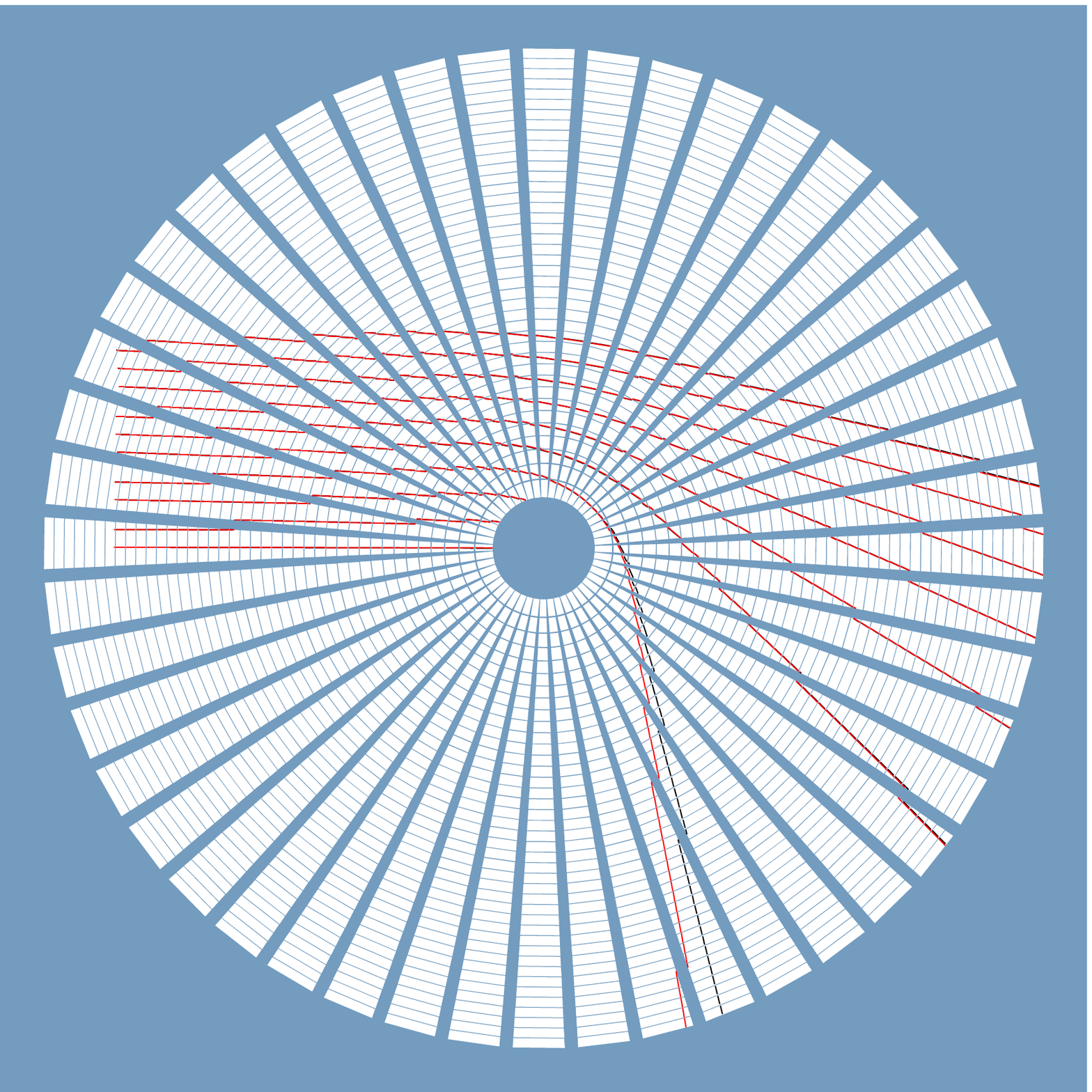}
  }
\caption{\label{fig.geodvergleich}
Geodesics constructed on sector models (red lines)
in comparison with numerical solutions of the geodesic equation
(black lines).
(a) Sector model with the resolution used in the workshop.
(b) Sector model with four times the resolution in each coordinate.
}
\end{figure}

\section{The accuracy of geodesics on sector models}
\label{sec.modell2}

\begin{figure}
  \centering
  \subfigure[]{%
    \includegraphics[width=\hb]{\bilder/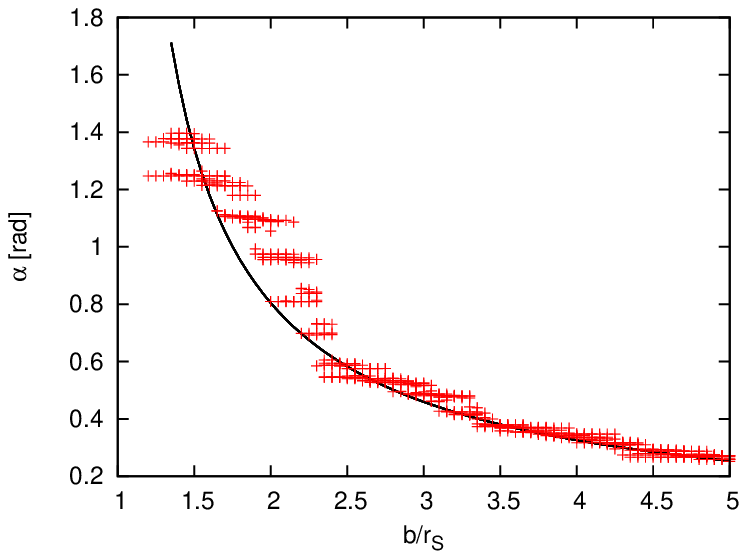}
  }\hfill%
  \subfigure[]{%
    \includegraphics[width=\hb]{\bilder/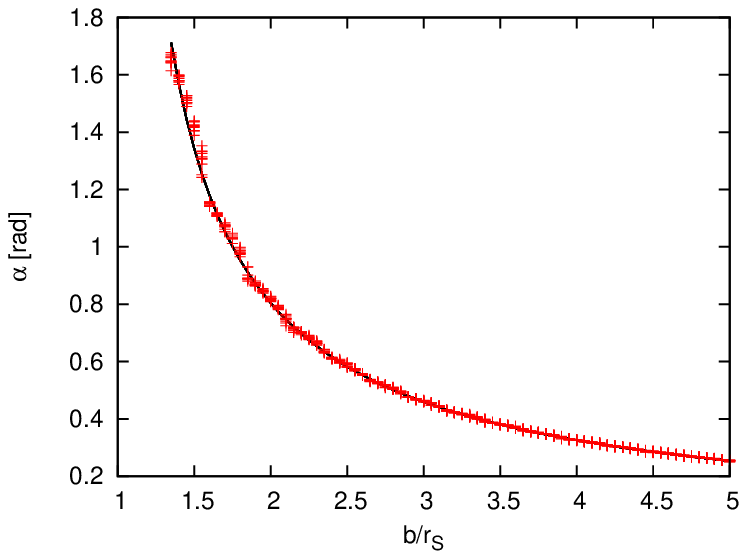}
  }
\caption{\label{fig.ablenkwinkel}
Relation between the angle of deflection $\alpha$ and
the impact parameter $b$ of a geodesic,
determined by constructing geodesics on a sector model
(symbols) and
by integrating the geodesic equation (line).
(a) Sector model with the resolution used in the workshop.
(b) Sector model with four times the resolution in each coordinate.
}
\end{figure}

In the Regge calculus,
geodesics are described as straight lines
in the flat sectors \citep{wil1981,wil1984,bre1993}.
On the sector models,
this description is implemented
by graphic construction.
Thus, the geodesics constructed in this way
are in principle quantitatively correct.
Their accuracy, however, depends on the resolution
of the sector model.
For use in the workshops,
the resolution is deliberately chosen to be coarse,
in order for the models to be easy to handle.
In this section, we study the accuracy of the graphic method
by comparing its results with numerically computed geodesics.
Two sector models of the equatorial plane of a black hole are used
in the comparison.
Both cover the region between $r=1.25\,\rs$ and
$r=13.75\,\rs$. The first one has the resolution used in the workshop
($\Delta r = 1.25\,\rs$, $\Delta\phi = \pi/6$),
it consists of 10~rings of 12~sectors each
(figure~\ref{fig.geodvergleich}(a)).
The second one has four times this resolution in each coordinate
($\Delta r = 0.3125\,\rs$,
$\Delta\phi = \pi/24$),
thus it consists of 40~rings of 48~sectors each
(figure~\ref{fig.geodvergleich}(b)).
Figure~\ref{fig.geodvergleich}
shows the geodesics obtained in the Regge calculus
in comparison with
the numerical solutions of the geodesic equation.
For this comparison, the paths computed from the geodesic equation
are plotted onto the sector models.
The mapping of Schwarzschild coordinates onto sector points is carried out
by interpolation \citep{hor2005}.
For a quantitative comparison,
the angle of deflection
as a function of the impact parameter
was determined
on the same two sector models
and compared with the values obtained by integration
(figure~\ref{fig.ablenkwinkel}).
On the sector models,
ten geodesics were constructed
for each value of the impact parameter.
They are rotated in $\phi$-direction with respect to each
other and thus have different locations with respect to the
sector boundaries.
For the coarser resolution,
the agreement is good if the deflection is small,
otherwise the deviation may be significant
(figures~\ref{fig.geodvergleich}(a), \ref{fig.ablenkwinkel}(a)).
For the higher resolution sector model,
the agreement is generally good
(figures~\ref{fig.geodvergleich}(b), \ref{fig.ablenkwinkel}(b)).
For qualitative considerations,
the accuracy on the sector model used in the workshop
is
satisfactory.

\section{Conclusions and outlook}

\label{sec.fazit2}

\subsection{Summary and pedagogical comments}

We have shown how sector models can be used as tools
to determine geodesics.
On the one hand this provides geometric insight
and on the other hand it is a possibility to determine geodesics
graphically.
The concept of a geodesic as a locally straight line
is illustrated by implementing this definition directly on
a sector model using pencil and ruler (section~\ref{sec.kugel}).
The graphic construction of geodesics shows
that their direction
after crossing a region of curved space
differs from the direction they had before
(section~\ref{sec.ssmgeod}),
thus showing clearly
how gravitational light deflection
arises.
Since the graphically constructed geodesics represent
solutions
of the geodesic equation,
one obtains quantitatively correct results.
Due to the rather coarse resolution of the sector models
used in the workshop,
the accuracy of these geodesics is not high.
From a pedagogical point of view, though,
a coarse resolution is an advantage.
The deficit angles are then large enough
to allow the illustration of the
effects of curvature
by considering a single vertex.
This provides
a clear picture of the connection between curvature
and the run of neighbouring geodesics
(section~\ref{sec.ergaenzunggeod}).
In this paper we consider spatial geodesics only.
An extension to geodesics in spacetime is described in the
sequel to this contribution \citep{teil3}.

The workshop on geodesics and gravitational light deflection
outlined in this contribution
has been developed
in several cycles of testing and revision
\citep{kra2005,zah2010,zah2013,kra2016}.
It was mainly
tested with
classes grades 10 to 13
(age 16 to 19 years)
and with pre-service teachers.

There are different possible uses for
the material presented above,
depending on the teaching goals and on the time available.
If the goal is
to provide a short and direct approach to the phenomenon
of gravitational light deflection,
e.g.\ in an astronomy class,
then the workshop can be held as described in
sections~\ref{sec.kugel} to~\ref{sec.geodzeichnen}
with the sector models provided as worksheets.
No previous knowledge of the concept of metric
is required;
the graphic construction is easily carried out
and conveys an appropriate understanding of
light paths
as geodesics.
In a course that aims at introducing the basic concepts
of general relativity,
one can let the participants compute
the sector models
of the sphere and
of the black hole equatorial plane.
The participants then acquire the necessary skills
for studying
the geometry of a surface
when they are given the metric.
Answers are here obtained graphically
that in a standard university course would be found by calculations.
Since
sector models and the graphic construction of geodesics
directly correspond to
the mathematical description by means of the metric and
the geodesic equation,
this material can also be used as a supplement
to a standard course
in order to strengthen geometric insight.

\subsection{Comparison with other graphic approaches}

In comparison with other graphic representations of geodesics,
constructions on sector models stand out by the fact that they
clearly show geodesics to be locally straight as well as
by their straightforward construction.

Explanations of optical phenomena
due to gravitational light deflection,
e.g.\ double images,
typically use drawings
that depict light rays as curved lines.
In this context, light rays are described as "bent".
These drawings and explanations
do not express the fact that light paths
are geodesics, i.e.\ (locally) straight lines,
and may thereby encourage misconceptions.
The construction of geodesics on sector models
clearly shows that there is no contradiction between
the line being locally straight
and the occurrence of light deflection (figure~\ref{fig.slgeod}).
The construction
can also relate geodesics to light rays being drawn as curved lines:
On a world map, the surface of the earth is projected onto the plane
and the geodesics of the sphere appear as curved lines.
These are distortions due to the projection.
Analogously, in a projection that maps the sector model
of figure~\ref{fig.slgeod} onto a
plane circular ring,
there will be distortions and
the geodesics in the equatorial plane of the black hole
will appear curved.

A commonly used visualization
shows geodesics on the embedding surface
of the equatorial plane of a star or a black hole
in order to illustrate
the deflection of light
(\citealp{dinv}, p.~209).
This is equivalent to the geodesics constructed
in section~\ref{sec.ssmgeod}.
When the sectors of the model shown there
are joined at the common edges,
one obtains a faceted surface
that is an approximation of the embedding surface.
The same caveat applies to the geodesics on the embedding surface
as to the geodesics on the sector model:
The subspace is purely spatial
so that gravitational light deflection
is illustrated by way of analogy
with spacelike geodesics.
Experience shows
that the concept of an embedding surface
is a difficult one
for the audience at which this workshop is targeted
and that the embedding surface is quite likely to be
misinterpreted as the geometric shape of the black hole
\citep{zah2010}.
With respect to embedding surfaces,
sector models have the advantage that
their calculation is elementary,
especially with the simplified method of
section~\ref{sec.modellberechnung}.
Also, sector models are easily built as physical paper models
and are readily duplicated,
so that all participants of a workshop can
carry out the construction of geodesics
on models of their own.

A description of geodesics
that is related to the representation on sector models
is diSessa's construction on so-called wedge maps \citep{dis1981}.
To create a wedge map,
the symmetry plane of a spherically symmetric spacetime
is divided up into strips by a number of radial cuts;
the strips (the wedges) are regarded as flat segments.
On the wedges, geodesics are determined numerically
as in the Regge calculus.
Thus, the construction of geodesics on the wedge map
follows the same prescription
that is used here for sector models.
The numerical method, however, is more involved
than the graphic construction used here,
concerning
both the mathematical description
and the requirement of programming skills.

\subsection{Outlook}

In paper~I,
three fundamental questions were raised
that should be answered by an introduction
to general relativity:
What is a curved spacetime? How does matter move in a curved
spacetime? How is the distribution of matter linked to the
curvature of the spacetime?
The concept of curved spaces and spacetimes
was discussed in paper~I.
To address the second question,
the present part~II
describes
geodesics in space
and its sequel
describes geodesics in spacetime
\citep{teil3}.
The relation
between curvature and the distribution of matter
will be treated in a fourth part of this series.

\end{document}


\title[Sektormodelle: II. Geodäten]{%
Sektormodelle -- Ein Werkzeugkasten zur Vermittlung der
               Allgemeinen Relativitätstheorie:\\
II. Geodäten
}

\author{C Zahn und U Kraus}

\bigskip
\address{Institut f\"ur Physik, Universität Hildesheim,
Universitätsplatz 1, 31141 Hildesheim}

\eads{
\mailto{corvin.zahn@uni-hildesheim.de},
\mailto{ute.kraus@uni-hildesheim.de}
}

\begin{abstract}

{\parindent=0pt

Sektormodelle sind Werkzeuge,
mit denen
die Grundprinzipien der Allgemeinen Relativitätstheorie
vermittelt werden können,
ohne bei der Formulierung über Schulmathematik hinauszugehen.
Dieser Beitrag zeigt, wie Sektormodelle dazu verwendet
werden können Geodäten zu bestimmen.
Wir stellen einen Workshop
für Schüler/innen und Studierende vor,
in dem die gravitative Lichtablenkung
mittels der Konstruktion von Geodäten auf Sektormodellen
behandelt wird;
als Beispiel dienen Geodäten in der Nähe eines Schwarzen
Lochs.
Der Beitrag beschreibt ferner eine vereinfachte Berechnung von Sektormodellen,
die Schüler/innen und Studierende
selbst durchführen können.
Die Genauigkeit der auf Sektormodellen konstruierten Geodäten wird
im Vergleich mit numerisch berechneten Lösungen diskutiert.
Die vorgestellten Materialien stehen online unter
{\tt www.tempolimit-lichtgeschwindigkeit.de}
für den Unterricht zur Verfügung.
}

\end{abstract}

\maketitle

\section{Einleitung}

\label{sec.einleitung2}

Die Allgemeine Relativitätstheorie
ohne den
mathematischen Apparat
in ihren Grundzügen zu vermitteln
ist ein Anliegen,
das auch hundert Jahre nach der Entwicklung
der Theorie
nichts an Aktualität verloren hat.
Im Hinblick auf dieses Ziel
beschreiben wir
einen neuen Zugang, der mit
Schulmathematik auskommt.
Zielgruppe sind Schüler/innen der Sekundarstufe
sowie Bachelor-, Lehramts- und Nebenfachstudierende, d.\,h.
Interessent/innen, denen entweder das Vorwissen
oder die Zeit fehlen,
um sich den für die übliche Darstellung benötigten
mathematischen Apparat anzueignen.
Dieser Zugang kann aber auch als Ergänzung zu
Standardlehrbüchern  (z.\,B. \citealp{har}) eingesetzt
werden, um ein anschauliches, geometrisches Verständnis
zu fördern.

In einer ersten Arbeit (\citealp{teil1}, im Folgenden
als Teil~I bezeichnet) haben wir Sektormodelle als
neuen Typ von Anschauungsmodellen für gekrümmte Räume
und Raumzeiten
entwickelt.
Wir haben aufgezeigt, wie mit ihrer Hilfe
eine Vorstellung von gekrümmten Räumen und Raumzeiten
vermittelt werden kann.
Sektormodelle werden in Teil~I anhand
von zweidimensionalen, positiv bzw. negativ gekrümmten Flächen
eingeführt und dann am Beispiel
des Schwarzen Lochs auf dreidimensionale gekrümmte Räume
und 1+1-dimensionale gekrümmte Raumzeiten erweitert.

\begin{figure}
  \centering
  \includegraphics[width=0.8\textwidth]{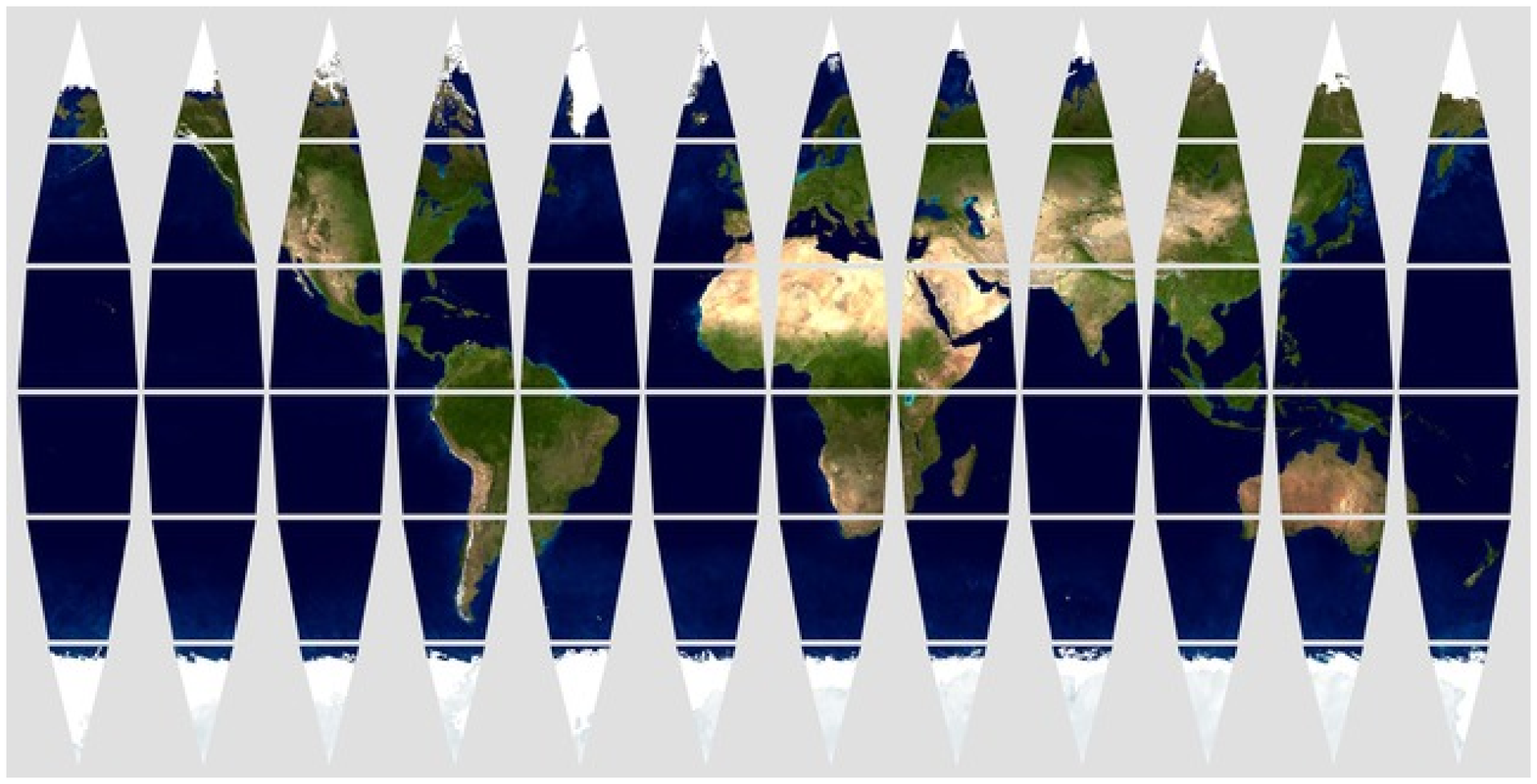}
  \caption{\protect\raggedright
    Sektormodell der Erdoberfläche.
    Erdtextur: NASA.
  }
  \label{fig.erdkarte}
\end{figure}

Sektormodelle
realisieren die im Regge-Kalkül eingesetzte
Darstellung gekrümmter Raumzeiten \citep{reg1961}
in Form von gegenständlichen Modellen.
Abb.~\ref{fig.erdkarte} illustriert das Prinzip
am Beispiel der Erdoberfläche:
Die Oberfläche der Erdkugel wird
durch kleine, ebene Flächenstücke angenähert.
Legt man die Flächenstücke in der Ebene aus,
erhält man eine Weltkarte;
diese stellt das Sektormodell der Erdoberfläche dar.
Zwei Unterschiede zu üblichen Weltkarten
fallen ins Auge: Die aus Sektoren bestehende Weltkarte hat kein
geschlossenes Kartenbild, da
sich die Sektoren nicht sämtlich
lückenlos aneinanderfügen lassen.
Und die Sektorkarte
ist im Rahmen der Diskretisierungsfehler
unverzerrt, d.\,h. sowohl längen- als auch winkeltreu
und daher unmittelbar der geometrischen Anschauung
zugänglich.
Das Sektormodell eines gekrümmten dreidimensionalen Raums
ist analog aufgebaut;
an die Stelle der ebenen Flächenstücke
treten Klötzchen, deren innere Geometrie euklidisch ist.
Im Falle einer
gekrümmten Raumzeit
sind die Sektoren Raumzeitelemente mit Minkowskigeometrie.

Die Allgemeine Relativitätstheorie
beschreibt die Bahnen von Licht und freien Teilchen
als Geodäten einer im Allgemeinen gekrümmten Raumzeit.
Das Konzept der Geodäte ist deshalb ein
wichtiger Punkt jeder Einführung in die Relativitätstheorie.
Der vorliegende Beitrag zeigt, wie anhand von Sektormodellen
der Begriff der Geodäte eingeführt und
Geodäten zeichnerisch ermittelt werden können.
An die Stelle der
Lösung eines Systems gewöhnlicher Differentialgleichungen
tritt die Konstruktion mit dem Lineal.
Die Konstruktion der Geodäten entspricht
der Beschreibung von Geodäten im Regge-Kalkül
\citep{wil1981}
und liefert im Rahmen des Diskretisierungsfehlers
quantitativ richtige Resultate.

In diesem Beitrag stellen wir
einen Workshop zur gravitativen Lichtablenkung vor,
den wir in dieser Form mit Schüler/innen
sowie mit Studierenden durchführen (Abschnitt~\ref{sec.workshopgeod}).
In Abschnitt~\ref{sec.modell2}
werden die Näherungen diskutiert,
die mit der Konstruktion von Geodäten auf Sektormodellen
verbunden sind und es wird die
erzielbare Genauigkeit
untersucht.
Fazit und Ausblick folgen in
Abschnitt~\ref{sec.fazit2}.

\newcommand{\fotobreite}{1.0}

\section{Workshop Geodäten und Lichtablenkung}

\label{sec.workshopgeod}

In diesem Workshop wird zunächst
am Beispiel von gekrümmten Flächen der Begriff der Geodäte eingeführt.
Es wird dann
das Zustandekommen der gravitativen Lichtablenkung verdeutlicht,
indem
auf einem Sektormodell eines Schwarzen Lochs Geodäten konstruiert
werden.
Das Schwarze Loch wird als Beispiel gewählt,
weil in seiner Nähe
relativistische Effekte gro\3 und deshalb in den Zeichnungen
klar zu erkennen sind.
Als Erweiterung des Workshops wird beschrieben,
wie Schüler/innen und Studierende in einem vereinfachten Verfahren
selbstständig zweidimensionale Sektormodelle erstellen können.
Dies versetzt sie in die Lage, Geodäten einer gegebenen Raumzeit
ausgehend von deren Metrik zu untersuchen.

\subsection{Geodäten auf gekrümmten Flächen}

\label{sec.kugel}

In der Einführung zum Workshop wird erläutert,
dass die Allgemeine Relativitätstheorie
Lichtwege und Bahnen frei fallender Teilchen als Geodäten beschreibt.
Je nach Teilnehmerkreis kann die Bedeutung von Geodäten
in der Relativitätstheorie lediglich mitgeteilt
oder aber mit Bezug auf das Äquivalenzprinzip näher erläutert werden
(z.\,B.~\citealp{nat}, Kap. 5).
Als Vorbereitung für die Bestimmung von Geodäten in der Nähe
eines Schwarzen Lochs
werden zunächst
Geodäten auf gekrümmten Flächen veranschaulicht.

Eine Geodäte auf einer gekrümmten Fläche wird als lokal gerade Linie
eingeführt. Eine solche Linie behält
an jedem Punkt ihre Richtung bei,
macht also keinen Bogen und keinen Knick.
Es wird ein anschauliches Kriterium dafür angegeben, ob eine
Linie eine Geodäte ist:
Man denkt sich einen schmalen Streifen
aus einem nicht dehnbaren Material
längs seiner Mittellinie auf die zu untersuchende Linie geklebt.
Falls die Linie einen Bogen macht, reißt der Streifen auf der Außenseite
ein und wirft auf der Innenseite Falten -- dies zeigt an, dass es sich
nicht um eine Geodäte handelt.

\begin{figure}
    \centering
    \includegraphics[width=\db]{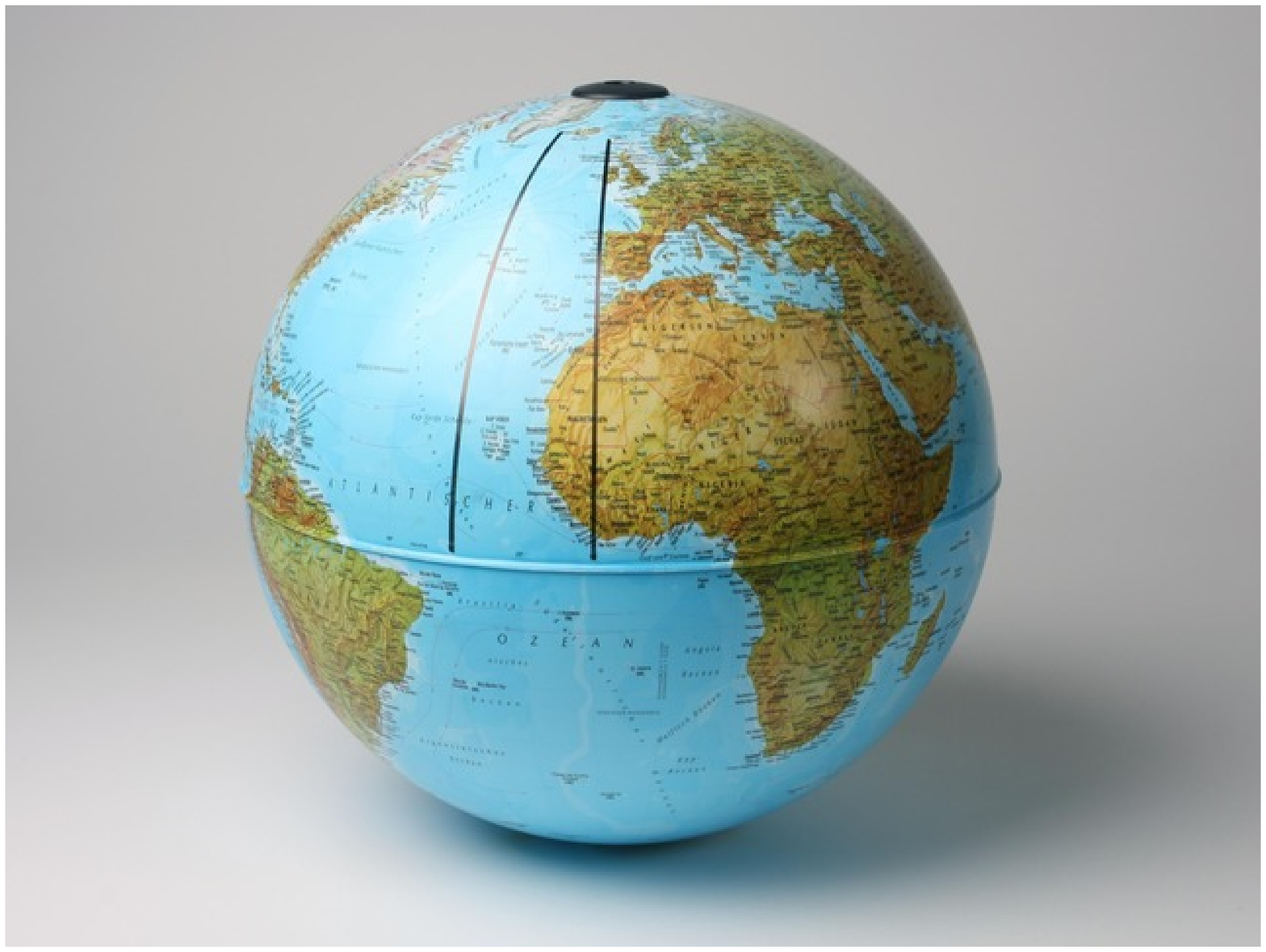}
    \caption{\label{fig.kugel}
     Die Geodäten auf der Kugeloberfläche sind die Großkreise.
   }
\end{figure}

\begin{figure}
  \centering
\subfigure[]{%
  \includegraphics[width=\db]{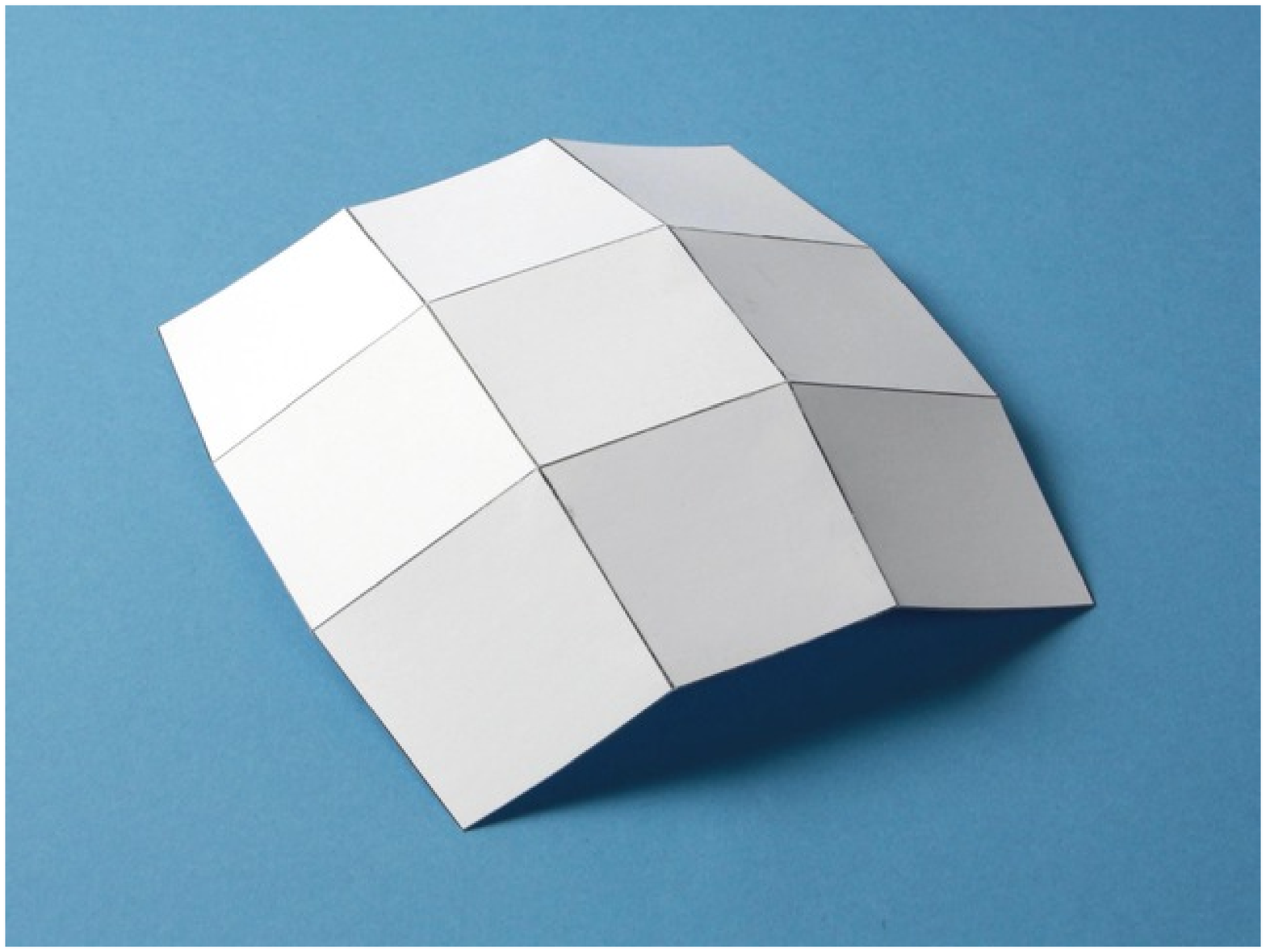}
}\ba%
\subfigure[]{%
  \includegraphics[width=\db]{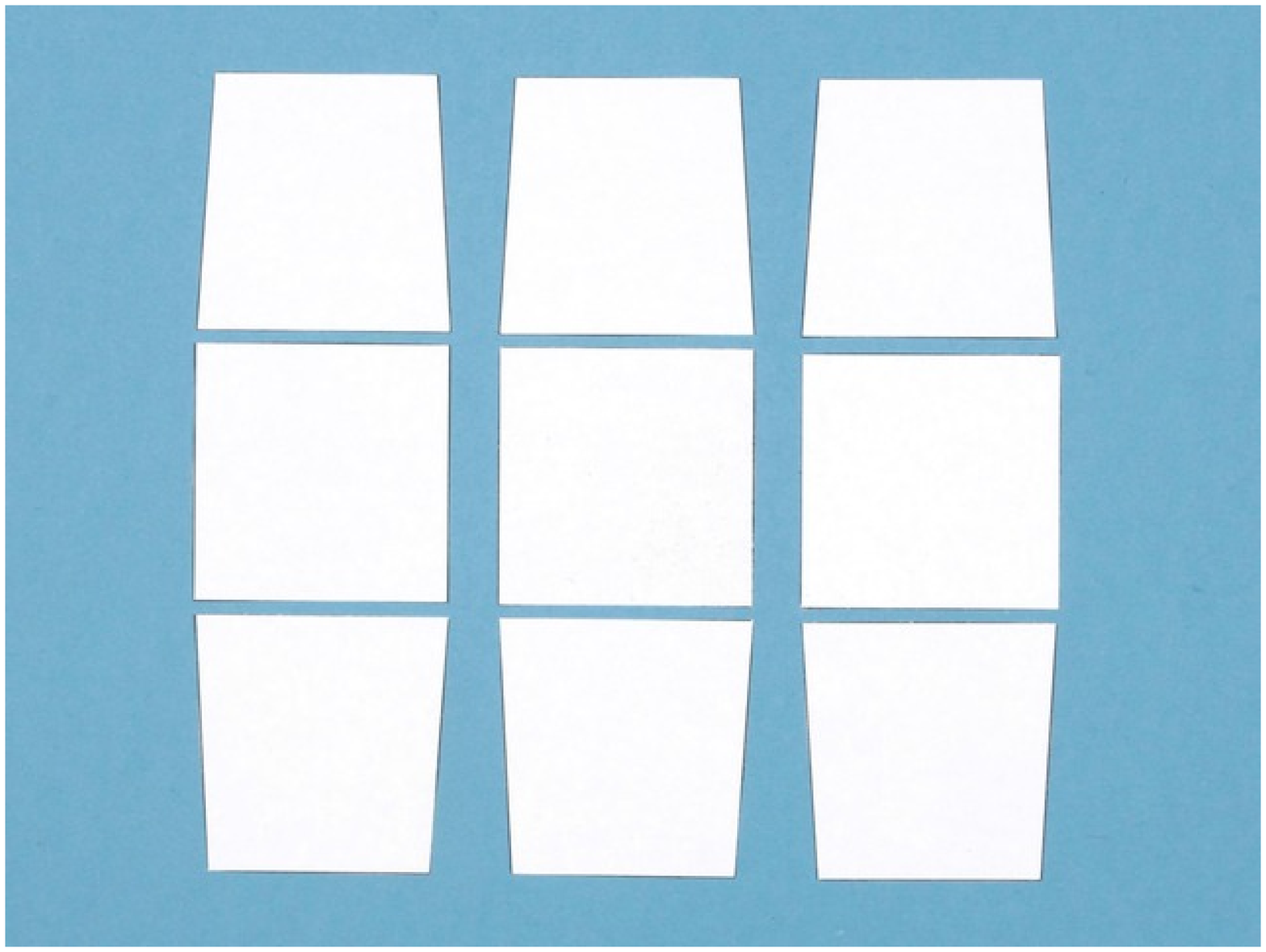}
}
\caption{\label{fig.kugelmodell}
     Eine Kugelkalotte wird durch eine Facettenfläche angenähert (a)
     und als Sektormodell dargestellt (b).
   }
\end{figure}

\begin{figure}
    \centering
    \includegraphics[width=\db]{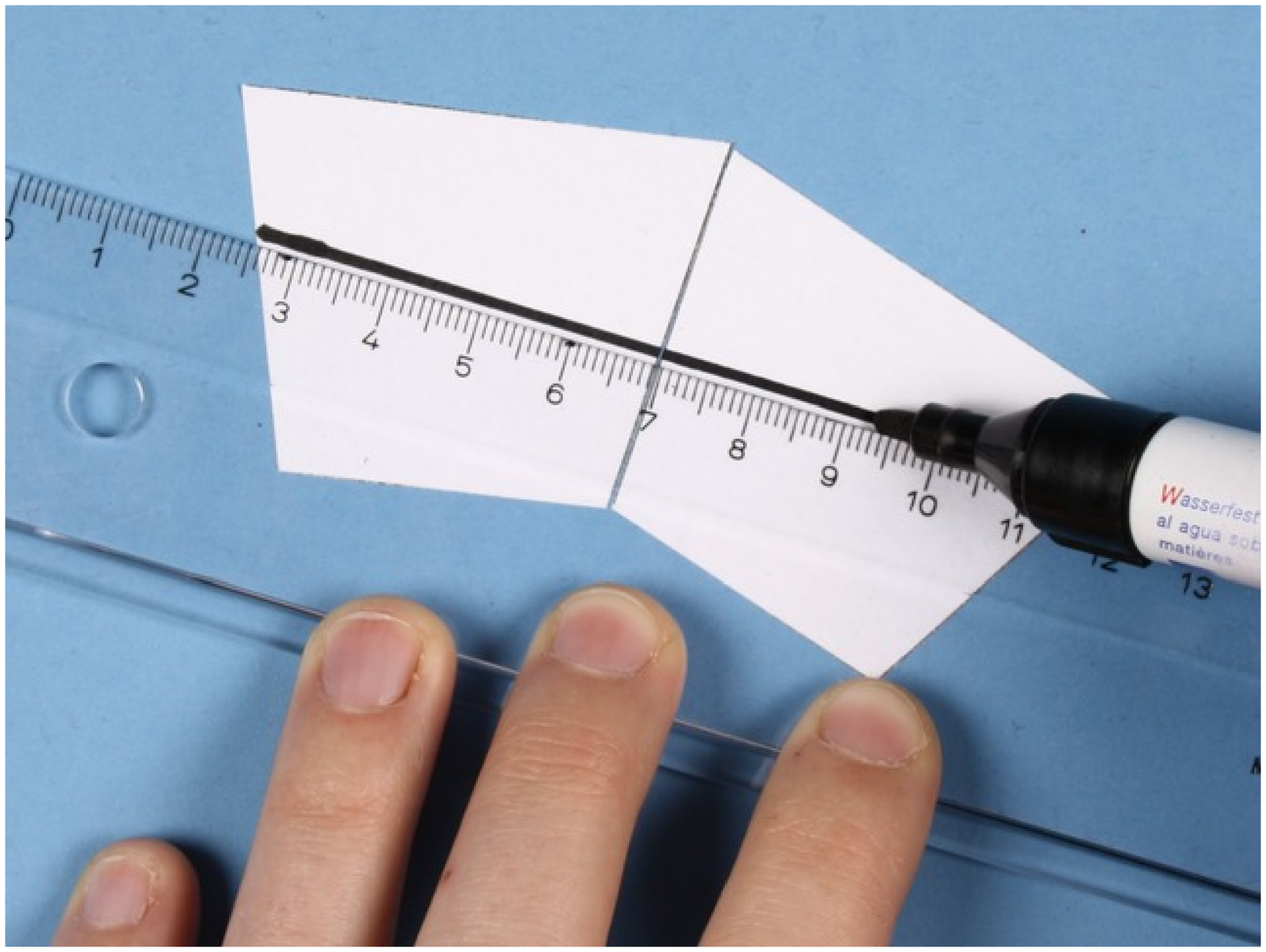}
    \caption{\label{fig.geodlineal}
     Konstruktion einer Geodäte auf einem Sektormodell.
    }
\end{figure}

Als erstes
Beispiel dient die Oberfläche einer Kugel
(Abb.~\ref{fig.kugel}).
Es wird eine Linie betrachtet,
die am Äquator in Richtung Nordpol startet
und lokal gerade verläuft.
Sie liegt offensichtlich auf einem Längenkreis.
Längenkreise, allgemeiner alle Großkreise,
sind also
Geodäten auf der Kugeloberfläche.
Abb.~\ref{fig.kugel} illustriert
eine charakteristische Eigenschaft
dieser Geodäten:
Zwei Längenkreise sind am Äquator parallel;
in ihrem weiteren Verlauf zum Pol hin nähern sie sich einander an.
Allgemein formuliert heißt das für die Kugeloberfläche,
dass parallel startende Geodäten
aufeinander zu laufen.

\begin{figure}
  \centering
  \subfigure[]{%
    \includegraphics[width=\db]{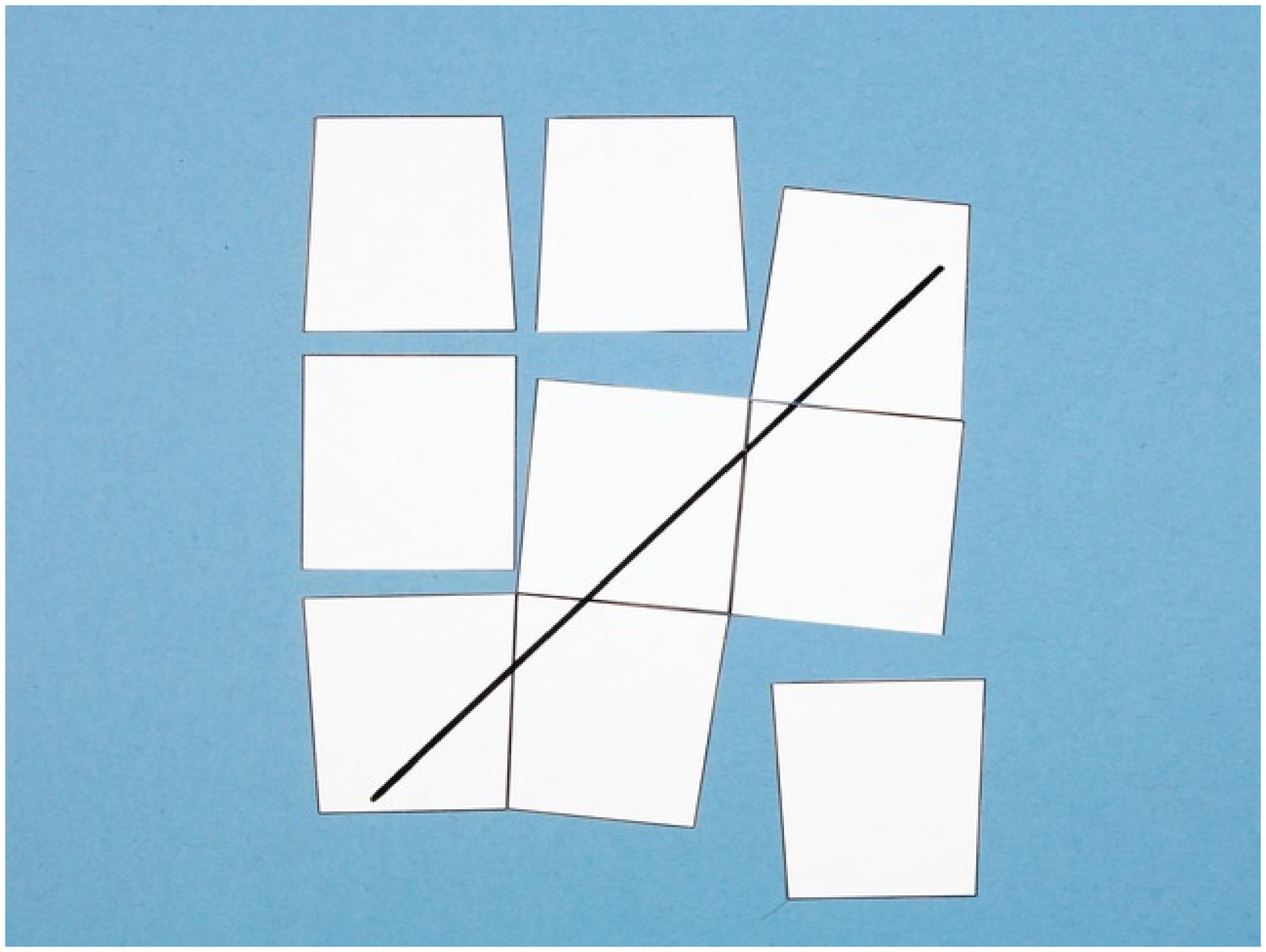}
  }\hfill%
  \subfigure[]{%
    \includegraphics[width=\db]{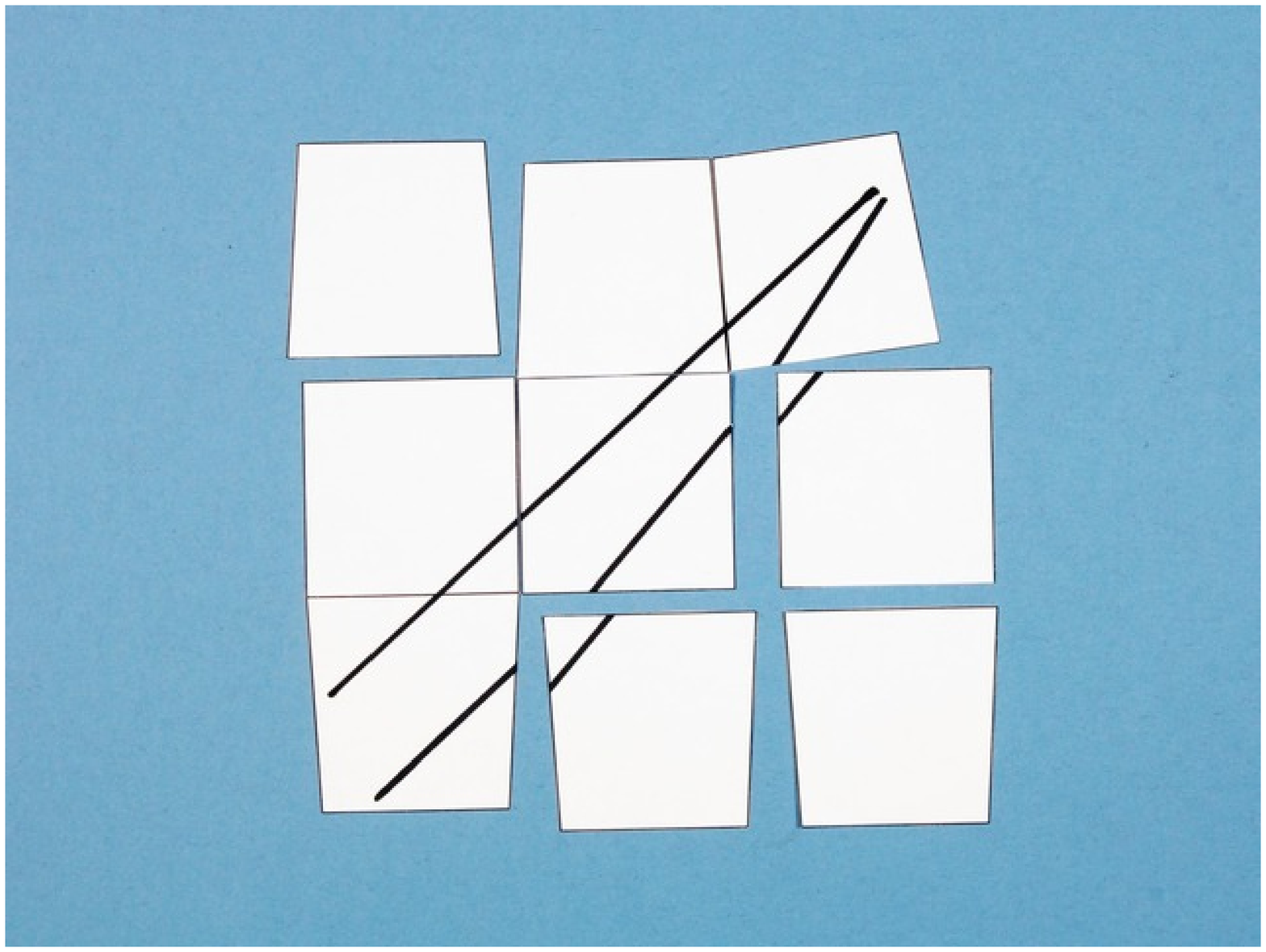}
  }\hfill%
  \subfigure[]{%
    \includegraphics[width=\db]{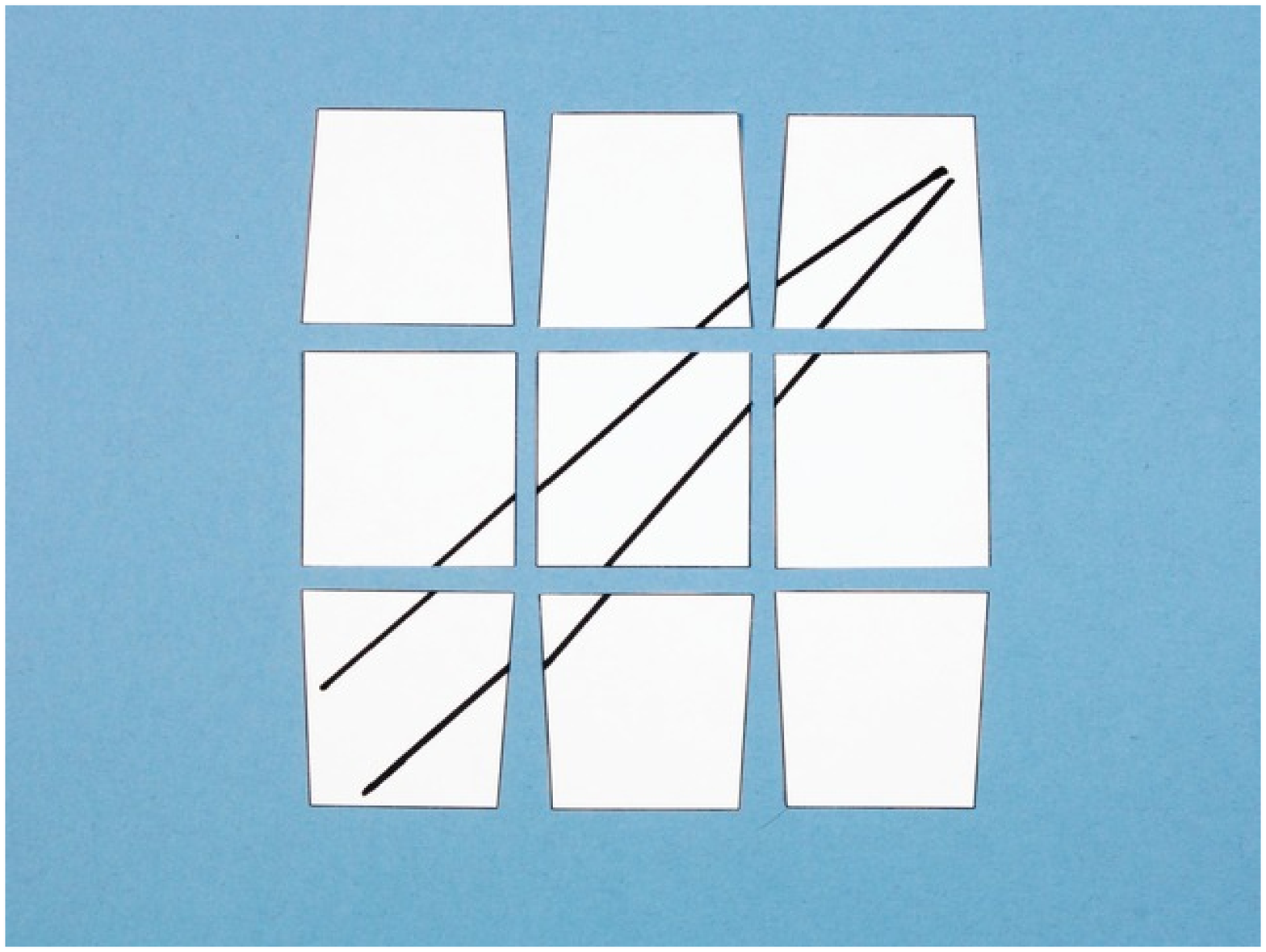}
  }
  \caption{\label{fig.kugelgeod}
    Geodäten auf dem Sektormodell einer Kugelkalotte.
    Die Sektoren sind in (a) längs der unteren Geodäte aneinandergelegt
    und in (b) längs der oberen. Im linken unteren Sektor verlaufen die
    beiden Geodäten parallel, nach rechts nähern sie sich einander
    an ((b), (c)).
   }
\end{figure}

\begin{figure}
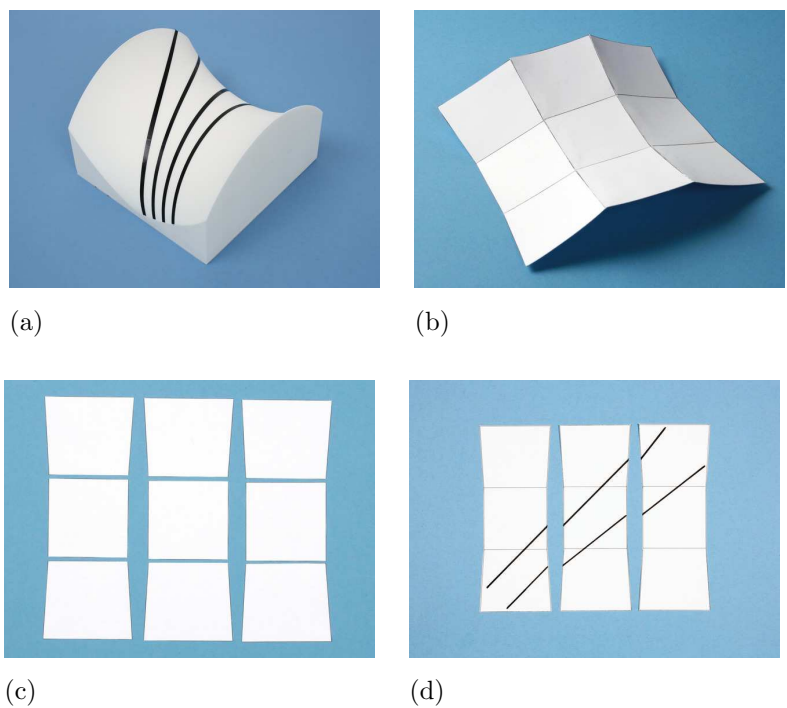

  \centering
  \subfigure[]{%
    \includegraphics[width=\db]{\bilder/figure6a.eps}
  }\ba%
  \subfigure[]{%
    \includegraphics[width=\db]{\bilder/figure6b.eps}
  }\\*[2ex]%
  \subfigure[]{%
    \includegraphics[width=\db]{\bilder/figure6c.eps}
  }\ba%
  \subfigure[]{%
    \includegraphics[width=\db]{\bilder/figure6d.eps}   
  }
  \caption{\label{fig.sattelgeod}
    Aus einer Sattelfläche (a) wird
    ein Ausschnitt über die Annäherung als Facettenfläche (b)
    als Sektormodell dargestellt (c);
    parallel startende Geodäten laufen auseinander ((a), (d)).
   }
\end{figure}

Im nächsten Schritt wird gezeigt, wie man
diese Eigenschaft von Geodäten der Kugeloberfläche
mithilfe eines Sektormodells gewinnen kann.
Eine Kugelkalotte wird durch eine Facettenfläche angenähert
(Abb.~\ref{fig.kugelmodell}(a));
die Facetten werden als Sektormodell ausgelegt
(Abb.~\ref{fig.kugelmodell}(b)).
Nun soll eine Geodäte
quer durch das Sektormodell gezeichnet werden.
Innerhalb eines Sektors, der ja eben ist,
ist die Geodäte eine gerade Linie.
Wenn die Linie den Rand eines Sektors erreicht,
wird sie in den Nachbarsektor fortgesetzt.
Wie das zu geschehen hat, folgt aus der Definition
der Geodäte: lokal gerade
(Abb.~\ref{fig.geodlineal}).
Die beiden benachbarten Sektoren werden an ihrer
gemeinsamen Kante zusammengelegt und die Linie wird
über die Sektorgrenze hinweg geradlinig fortgesetzt.
Auf diese Weise wird die Geodäte quer durch das
Sektormodell gezeichnet (Abb.~\ref{fig.kugelgeod}(a)).
Es wird dann
eine zweite Geodäte hinzugefügt,
die im
linken unteren Sektor
parallel zur ersten verläuft
(Abb.~\ref{fig.kugelgeod}(b)).
Man erkennt,
dass sich die beiden
parallel startenden Geodäten annähern
(Abb.~\ref{fig.kugelgeod}(b), (c)).

Als zweites Beispiel dient eine Sattelfläche.
Aufgeklebte Klebestreifen zeigen, dass parallel startende Geodäten
sich voneinander entfernen (Abb.~\ref{fig.sattelgeod}(a)).
Die
Annäherung eines Ausschnitts der Fläche
durch eine Facettenfläche (Abb.~\ref{fig.sattelgeod}(b))
führt zu einem Sektormodell
(Abb.~\ref{fig.sattelgeod}(c)).
Es werden zwei Geodäten eingezeichnet, die in der linken
unteren Ecke des Modells parallel verlaufen; diese entfernen sich
voneinander (Abb.~\ref{fig.sattelgeod}(d)).

An den beiden Beispielen wird deutlich,
dass die Sektormodelle gekrümmter Flächen dazu geeignet sind,
die Eigenschaften von Geodäten auf den Flächen zu untersuchen.

\begin{figure}
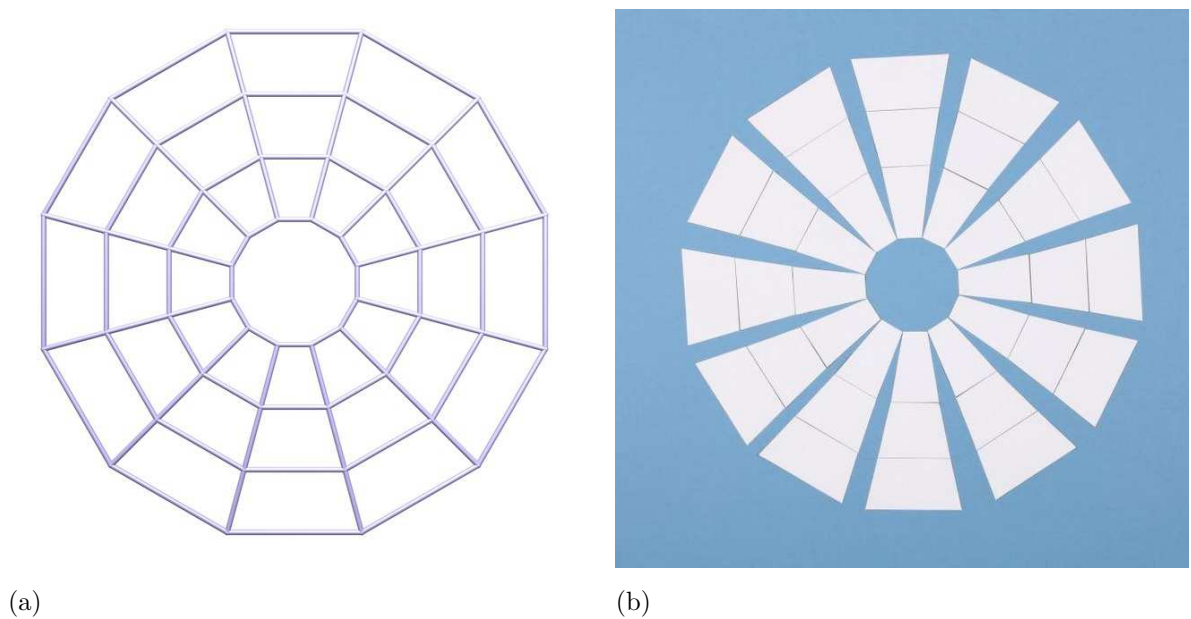

  \centering
  \subfigure[]{%
    \includegraphics[width=\hb]{\bilder/figure7a.eps}
  }\hfill%
  \subfigure[]{%
   \includegraphics[width=\hb]{\bilder/figure7b.eps}
  }
  \caption{\label{fig.gitter2d}
    Ein Gedankenexperiment zum
    Sektormodell für die Äquatorfl\"ache eines Schwarzen Lochs:
    Um das Schwarze Loch wird ein Gittergerüst
    nach dem in (a) gezeigten Schema errichtet.
    Für jede Zelle werden die Längen der sie umschließenden vier Stäbe
    abgemessen; aus diesen Daten werden maßstabsgerechte
    Sektoren aus Papier konstruiert.
    Es resultiert das Sektormodell des Rings um das Schwarze Loch (b).
    }
\end{figure}

\subsection{Geodäten in der Nähe eines Schwarzen Lochs}
\label{sec.ssmgeod}

Im zweiten Teil des Workshops wird
zunächst
ein Sektormodell vorgestellt,
das es erlaubt Geodäten in der Nähe eines Schwarzen Lochs
zu konstruieren.
Das Sektormodell stellt eine
Symmetrieebene des Schwarzen Lochs dar; diese wird
im Folgenden als Äquatorebene bezeichnet.%
\footnote{%
Wir betrachten ein nichtrotierendes Schwarzes Loch.
Es ist kugelsymmetrisch, weshalb jede Geodäte in einer
Ebene verläuft, die eine Symmetrieebene des Schwarzen Lochs
darstellt.
}

Das Sektormodell wird eingeführt, indem im Gedankenexperiment
seine \glqq Entstehung\grqq\ beschrieben wird:
Ein Raumschiff wird in die Nähe eines Schwarzen Lochs geschickt,
um die Geometrie des Raums zu vermessen. Dazu wird um das Schwarze Loch
nach dem in Abb.~\ref{fig.gitter2d}(a)
dargestellten Schema ein Gittergerüst errichtet.
In Art eines Radnetzes, das auf das Schwarze Loch zentriert ist,
werden starre Stäbe in der Äquatorebene
des Schwarzen Lochs angeordnet. Das ganze Gerüst
befindet sich
außerhalb des Ereignishorizonts, da im Inneren des Schwarzen Lochs
eine solche statische Struktur nicht möglich ist.\footnote{%
Das Gitter überdeckt den Bereich
von $1,25$ bis $5$ Schwarzschildradien in der
Schwarzschildschen Radialkoordinate,
s.~Abschnitt~\ref{sec.slberechnung}.
}
Das Gittergerüst wird vermessen:
Jede einzelne Gitterzelle ist von vier Stäben umschlossen.
Deren Längen werden bestimmt und zur Erde übermittelt.
Dort wird die Gitterzelle
verkleinert
als Sektor dargestellt.
Alle Sektoren zusammen bilden
das maßstabsgetreue Modell eines
Rings um das Schwarze Loch
(Abb.~\ref{fig.gitter2d}(b)).
Sie lassen sich allerdings nicht
lückenlos zu einem Ring zusammenschieben.
Dies zeigt an, dass die Äquatorebene des Schwarzen Lochs
eine andere Geometrie hat
als die ebene Fläche, auf der die Sektoren ausgelegt sind.
Die Äquatorebene des Schwarzen Lochs
ist Teil eines gekrümmten Raums;
die Unterlage des Modells ist
eine Ebene im euklidischen Raum.
Könnte man ein Schwarzes Loch der passenden Größe in die Mitte des
Modells setzen, dann würden die Flächenstücke \emph{so wie sie sind}
lückenlos zusammenpassen.
Für
die online verfügbare Vorlage
\citep{zah2018}
hat das zu dem Modell passende
Schwarze Loch ungefähr die dreifache Erdmasse.

\begin{figure}
  \centering
  \subfigure[]{%
    \includegraphics[width=\db]{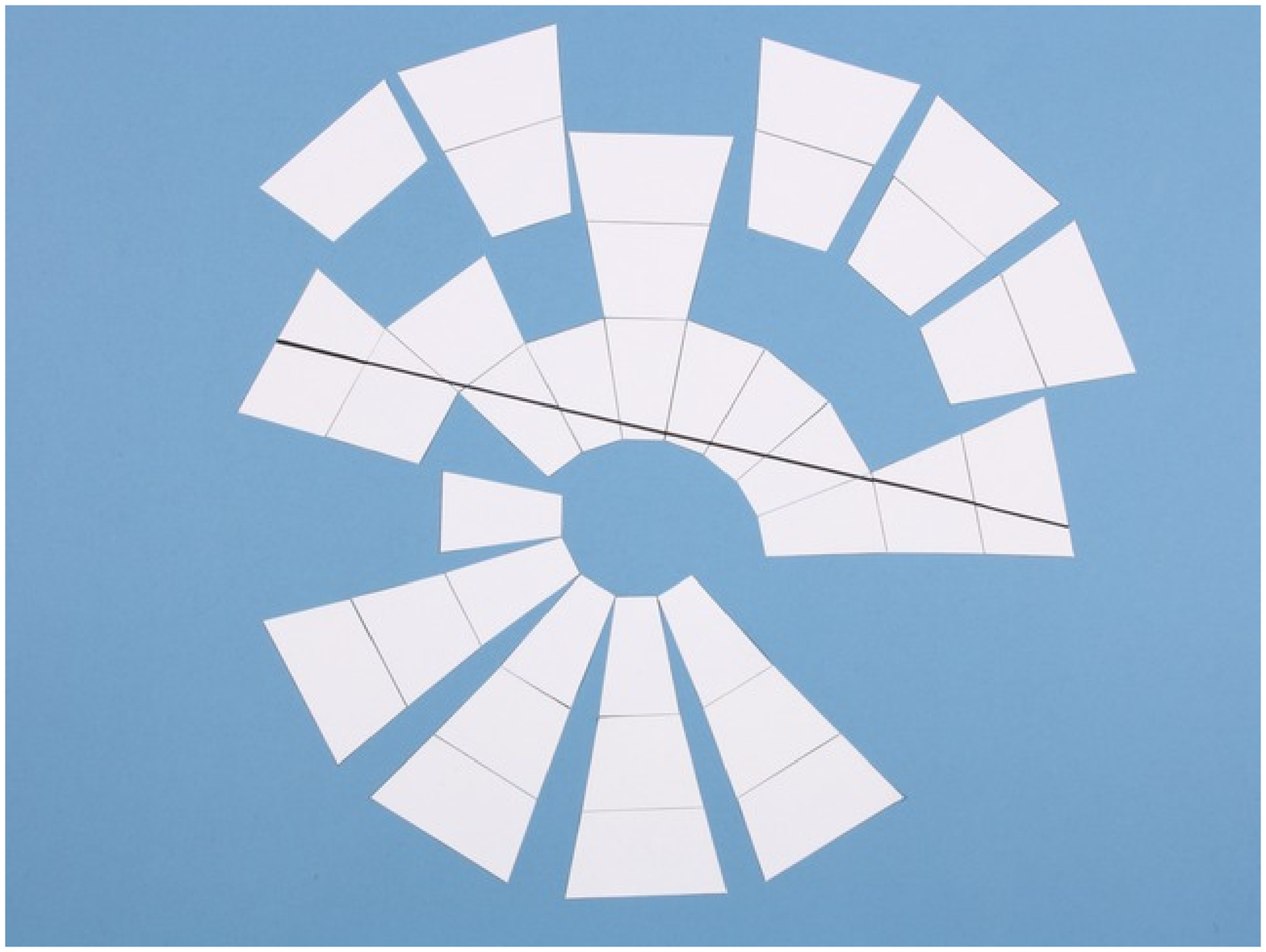}
  }\ba%
  \subfigure[]{%
    \includegraphics[width=\db]{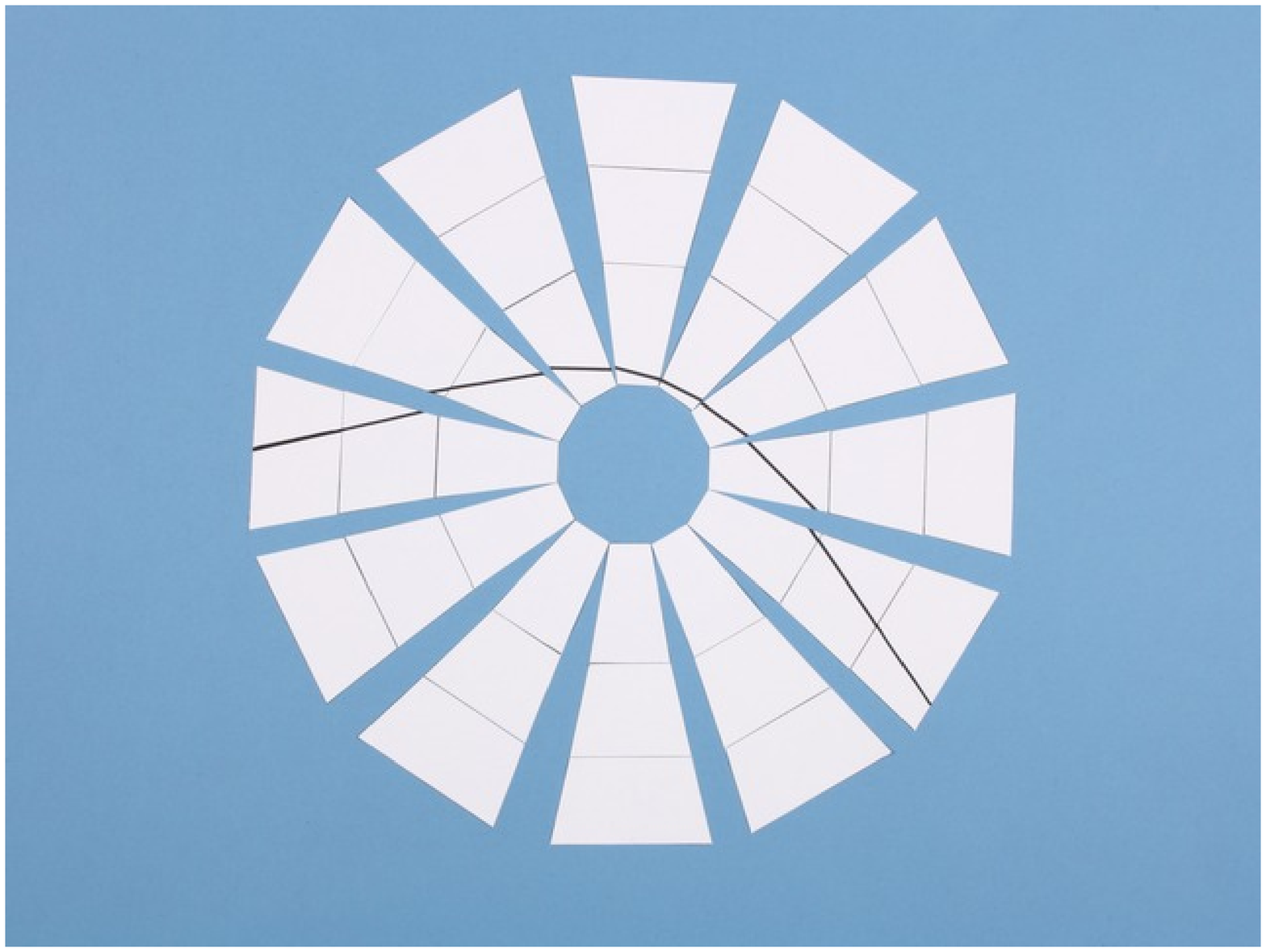}
  }
  \caption{\label{fig.slgeod}
     Eine Geodäte auf dem Sektormodell der Äquatorfläche eines
     Schwarzen Lochs. Die Linie ist lokal gerade;
     die beiden Enden zeigen in unterschiedliche Richtungen.
     Die Sektoren sind in (a)
     längs der Geodäte aneinandergelegt und in (b)
     symmetrisch angeordnet.
    }
\end{figure}

Alternativ
kann das Sektormodell mit dem
Workshop über gekrümmte Räume
aus Teil~I eingeführt werden.
Dort wird ein Sektormodell des dreidimensionalen gekrümmten
Raums um ein Schwarzes Loch vorgestellt
(Abb.~5(b) in Teil~I),
dessen Äquatorfläche
genau das in
Abb.~\ref{fig.gitter2d}(b) gezeigte Modell ist
(in Abb.~5(b) von Teil~I: die grünen, annähernd
horizontal ausgerichteten Seiten der Klötzchen).

Das Sektormodell von Abb.~\ref{fig.gitter2d}(b)
wird
für den Workshop vorbereitet,
indem die Sektoren
aus Papier ausgeschnitten
und mit Sprühkleber
(Kleber für wiederlösbare und wiederverklebbare Verbindungen)
auf Karton fixiert werden%
\footnote{%
S. Abschnitt~\ref{sec.geodzeichnen} für
ein Verfahren, das ohne Aufkleben von Sektoren auskommt.
}.
Dann wird das Sektormodell
dazu verwendet,
Geodäten
in der Nähe eines Schwarzen Lochs zu untersuchen.
Zunächst wird
eine einzelne Geodäte quer durch das Sektormodell gezeichnet.
Wie oben
am Beispiel gekrümmter Flächen beschrieben,
geschieht dies
durch
Aneinanderlegen benachbarter Sektoren
und Zeichnen einer geraden Linie
(Abb~\ref{fig.slgeod}(a)).
Man erkennt,
dass die beiden Enden der Linie in unterschiedliche Richtungen zeigen
(Abb~\ref{fig.slgeod}(b)).
Eine Linie, die nah an einem Schwarzen Loch vorbeiführt
und dabei in jedem Punkt ihre Richtung beibehält,
hat also
\glqq weit nach\grqq\ dem Schwarzen Loch eine andere Richtung als
\glqq weit davor\grqq.
Die Konstruktion verdeutlicht
das Prinzip der Lichtablenkung im Schwerefeld:
Licht breitet sich lokal geradlinig aus;
wenn es einen Bereich gekrümmter Raumzeit durchquert,
ist die Ausbreitungsrichtung hinterher anders als vorher.

Um die Aussagekraft dieser Konstruktion auf dem Sektormodell
einzuschätzen, sind zwei Dinge zu bedenken.
Zum einen sind die auf Sektormodellen konstruierten Geodäten
quantitativ richtig.
Wenn man die Bedingung des lokal geraden Verlaufs
mathematisch formuliert, erhält man die Geodätengleichung
\citep[S. 70 ff]{wei1972}. Die konstruierte Geodäte ist eine
Lösung dieser Gleichung.
Da das Sektormodell den gekrümmten Raum
näherungsweise darstellt,
ist auch
die gezeichnete Geodäte eine Näherungslösung.
Durch eine entsprechend feine Unterteilung können
Geodäten
aber
prinzipiell auch mit hoher Genauigkeit
konstruiert werden (s.~Abschnitt~\ref{sec.modell2}).
Zum anderen muss bedacht werden,
dass die konstruierte Linie
zwar eine Geodäte ist, aber dennoch keinen Lichtstrahl darstellt.
Die gezeichnete Geodäte ist rein räumlich.
Licht jedoch breitet sich in Raum und Zeit aus,
was bedeutet, dass Lichtwege raumzeitliche Geodäten sind.
Die rein räumliche Geodäte illustriert also im Sinne einer Analogie,
wie Lichtablenkung prinzipiell zustandekommt.

\begin{figure}
  \centering
  \subfigure[]{%
    \includegraphics[width=\db]{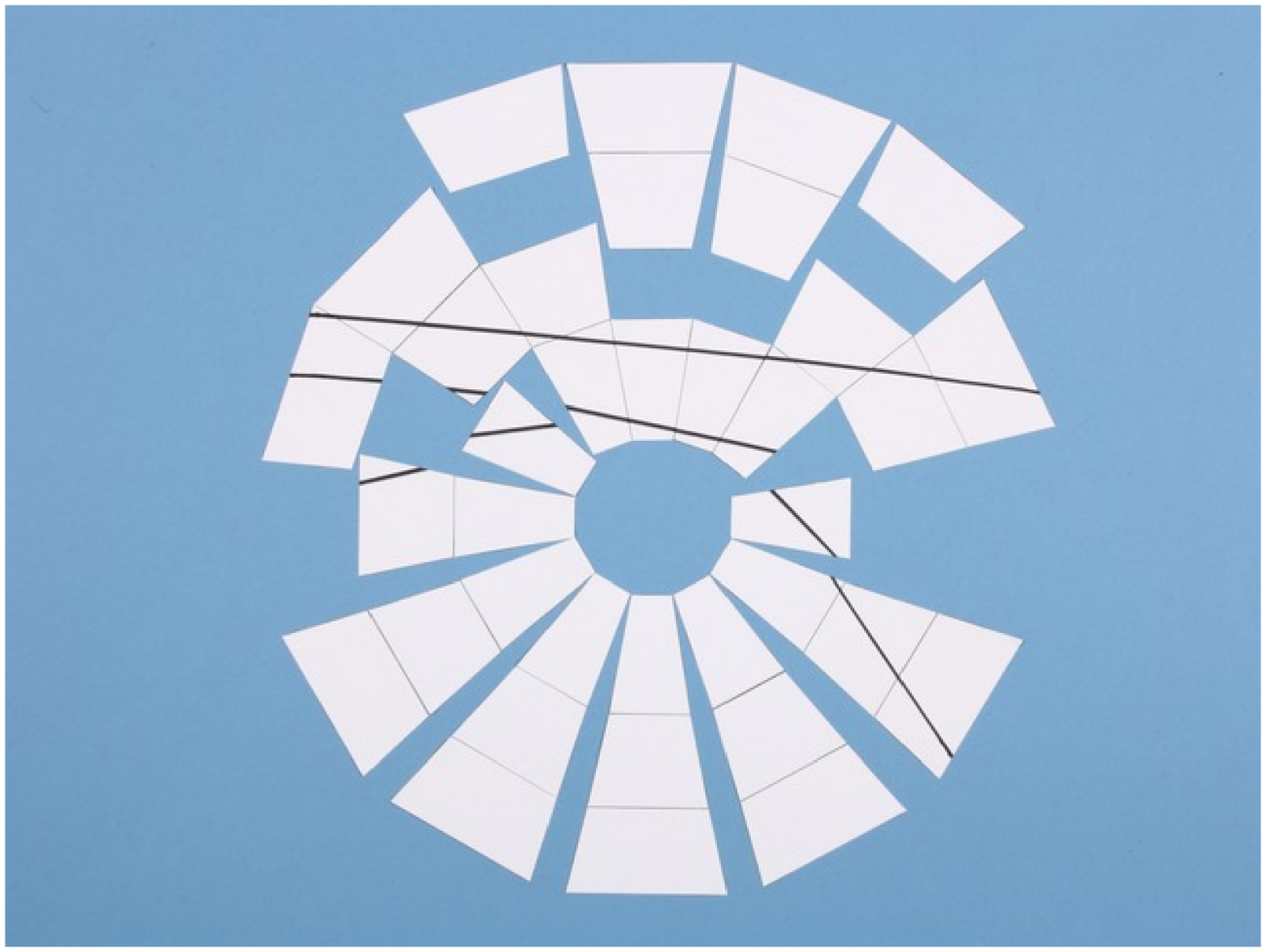}
  }\ba%
  \subfigure[]{%
    \includegraphics[width=\db]{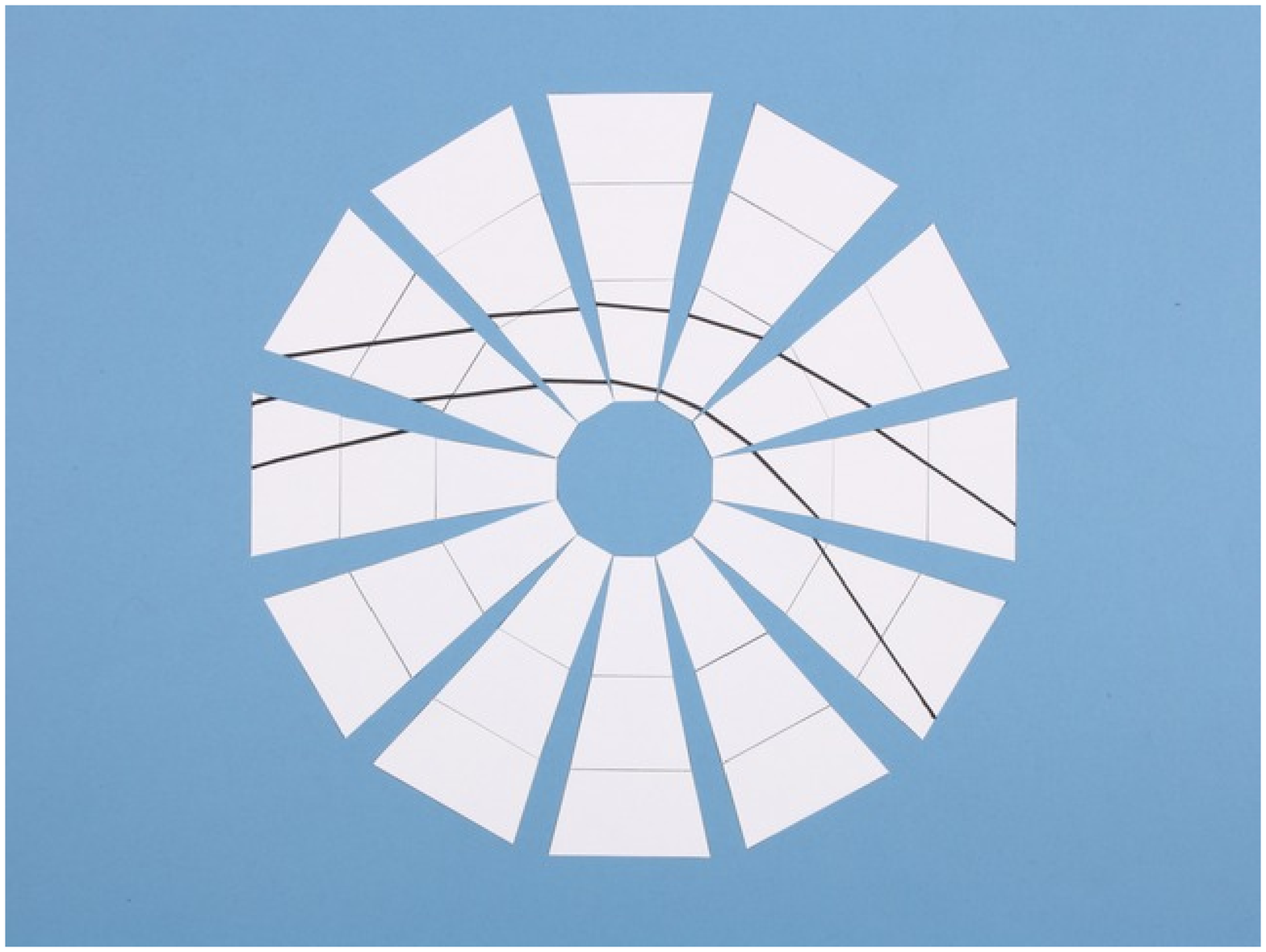}
  }
\caption{\label{fig.scharparallel}
Zu der in Abb.~\protect\ref{fig.slgeod} gezeigten Geodäte
wird eine zweite hinzugefügt,
die links außen parallel zur ersten
verläuft:
Die innere Geodäte wird stärker abgelenkt; die beiden Geodäten
laufen auseinander.
Die Sektoren sind in (a)
längs der zweiten Geodäte aneinandergelegt und in (b)
symmetrisch angeordnet.
}
\end{figure}

Auch wenn r\"aumliche Geod\"aten nicht
identisch mit Lichtstrahlen
sind,
ist es instruktiv,
an ihnen
Eigenschaften von Geod\"aten
aufzuzeigen.
So kann man eine zweite Geodäte konstruieren,
die
in der Nähe der ersten und parallel zu ihr beginnt
(Abb.~\ref{fig.scharparallel}).
Diejenige der beiden Geodäten, die dem Schwarzen Loch näher kommt,
wird stärker abgelenkt; die beiden Geodäten laufen
auseinander.
\begin{figure}
  \centering
  \subfigure[]{%
    \includegraphics[width=\db]{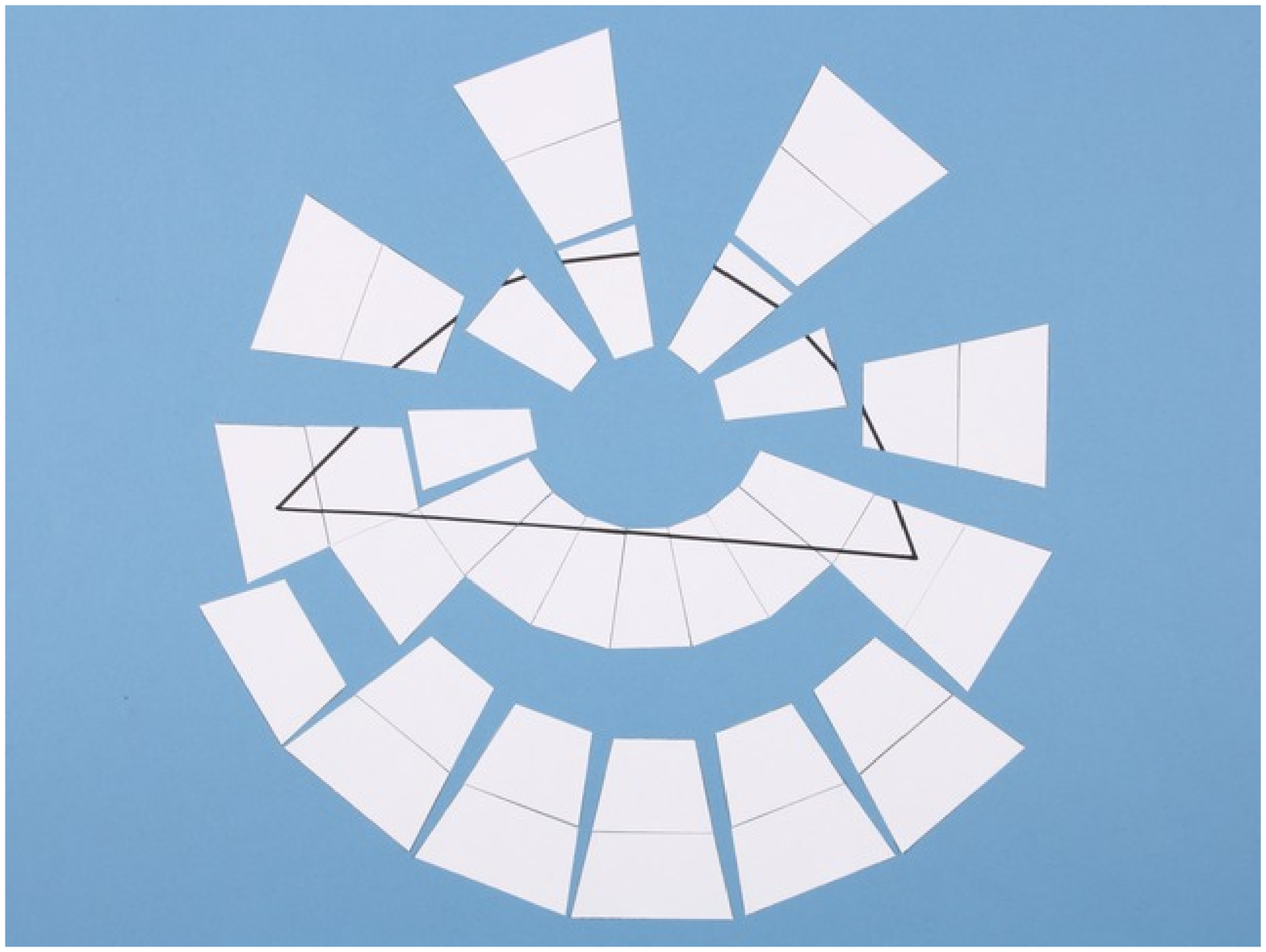}
  }\hfill%
  \subfigure[]{%
    \includegraphics[width=\db]{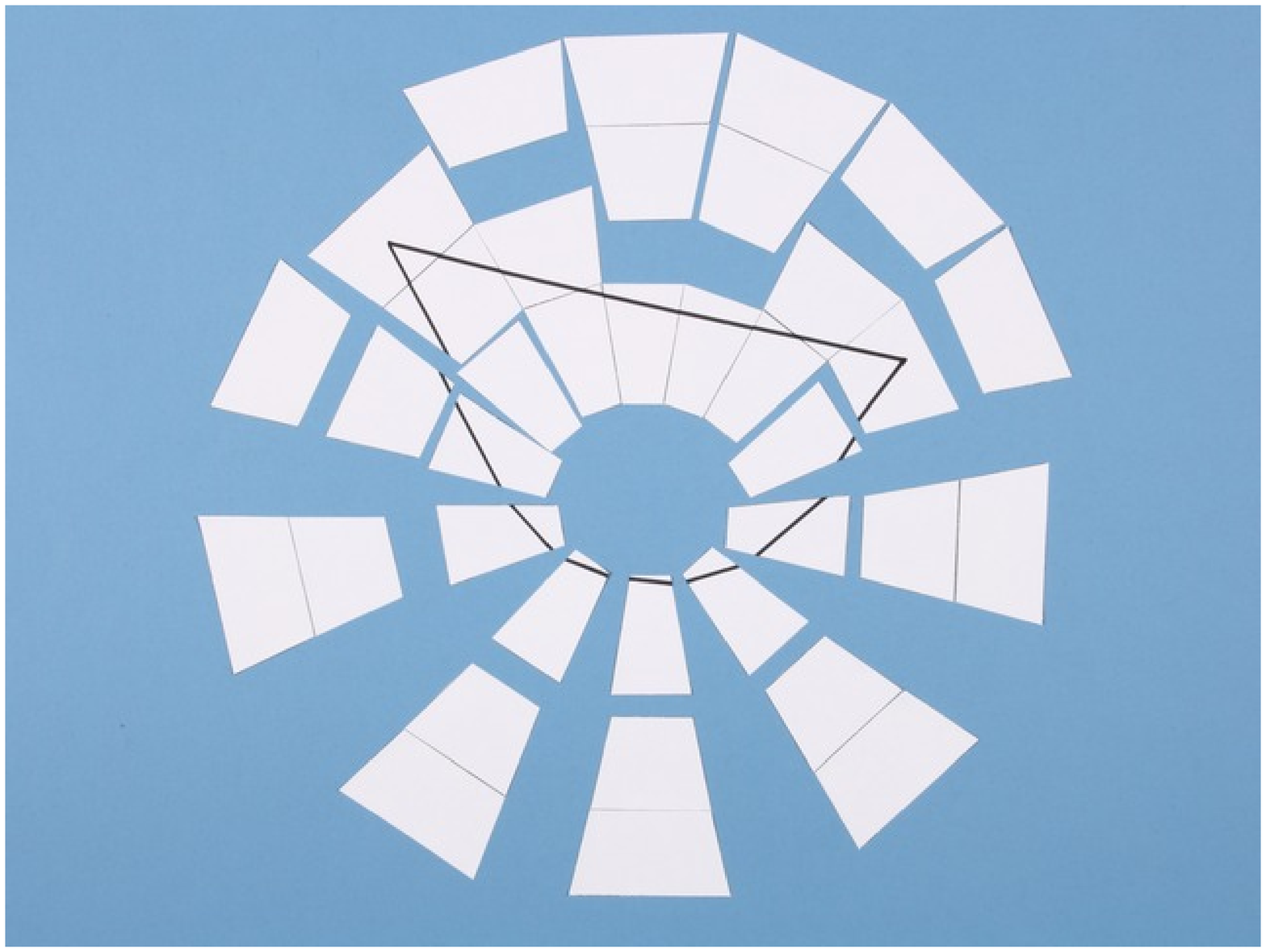}
  }\hfill%
  \subfigure[]{%
    \includegraphics[width=\db]{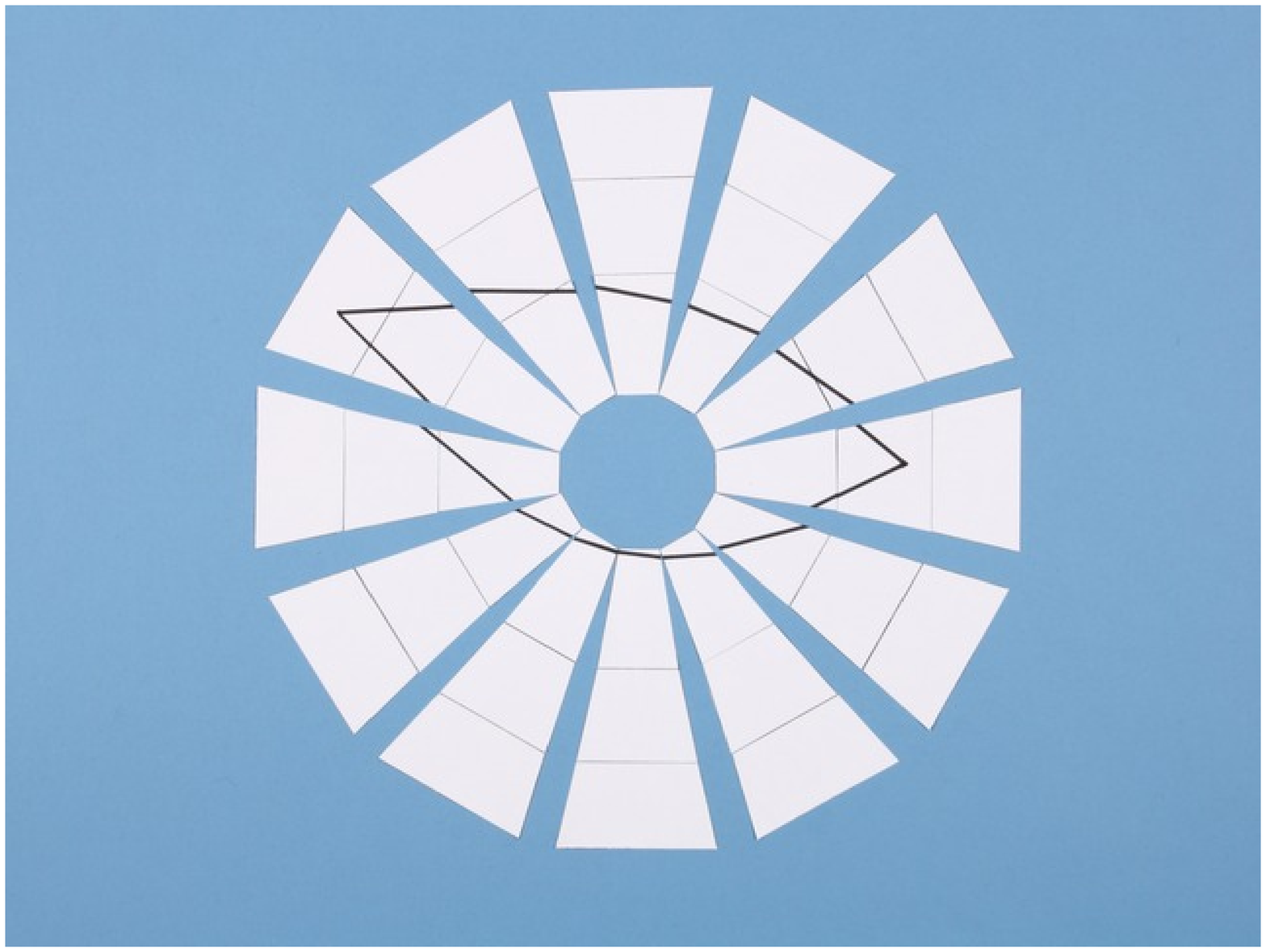}
  }
\caption{\label{fig.zweieck}
Zwei Geodäten bilden ein Zweieck.
Dies illustriert die Entstehung von Doppelbildern
durch Lichtablenkung: Licht einer Quelle erreicht den
Beobachter auf zwei verschiedenen Wegen.
Die Sektoren sind in (a) und (b) jeweils längs einer
der Geodäten aneinandergelegt und in (c) symmetrisch
angeordnet.
}
\end{figure}
Schließlich kann man zwei Geodäten konstruieren,
die vom selben Punkt ausgehend
auf verschiedenen Seiten am Schwarzen Loch vorbeiführen
und sich wieder treffen (Abb.~\ref{fig.zweieck}).
Man kann also aus Geodäten ein Zweieck
bilden.
Übertragen auf Lichtstrahlen zeigt diese Konstruktion,
wie Doppelbilder zustandekommen.

\subsection{Konstruktion von Geodäten mit Transfersektoren}
\label{sec.geodzeichnen}

\begin{figure}
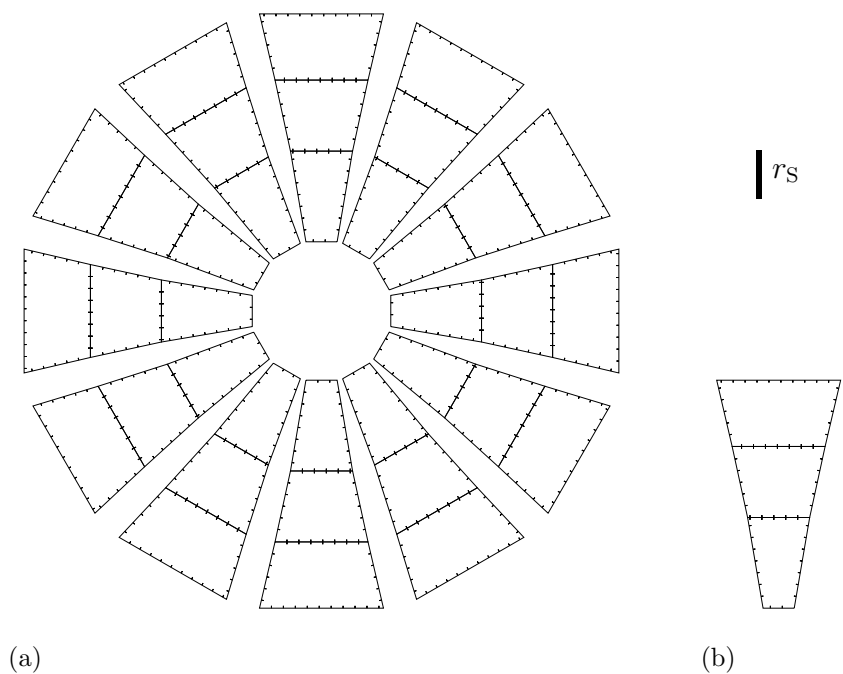

  \centering
  \subfigure[]{%
    \includegraphics[scale=0.3]{\bilder/figure11a.eps}
  }\qquad%
  \subfigure[]{%
    \shortstack{%
      \mbox{\rule{2pt}{18bp} \raisebox{8pt}{$\rs$}}\\
      \vspace*{2cm}\\
      \includegraphics[scale=0.3]{\bilder/figure11b.eps}
    }
  }
  \caption{\label{fig.slmodell}
  Vorlage für die Konstruktion von Geodäten in der Nähe eines
  Schwarzen Lochs. Die Vorlage besteht aus
  dem Sektormodell der Äquatorfläche
  als symmetrisch ausgelegtem Gesamtmodell mit Randmarkierungen (a)
  und einer Spalte von Transfersektoren (b).
  Der Schwarzschildradius $\rs$ des Schwarzen Lochs
  ist als Balken markiert.
}
\end{figure}

\begin{figure}
\centering
  \subfigure[]{%
    \includegraphics[width=\veb]{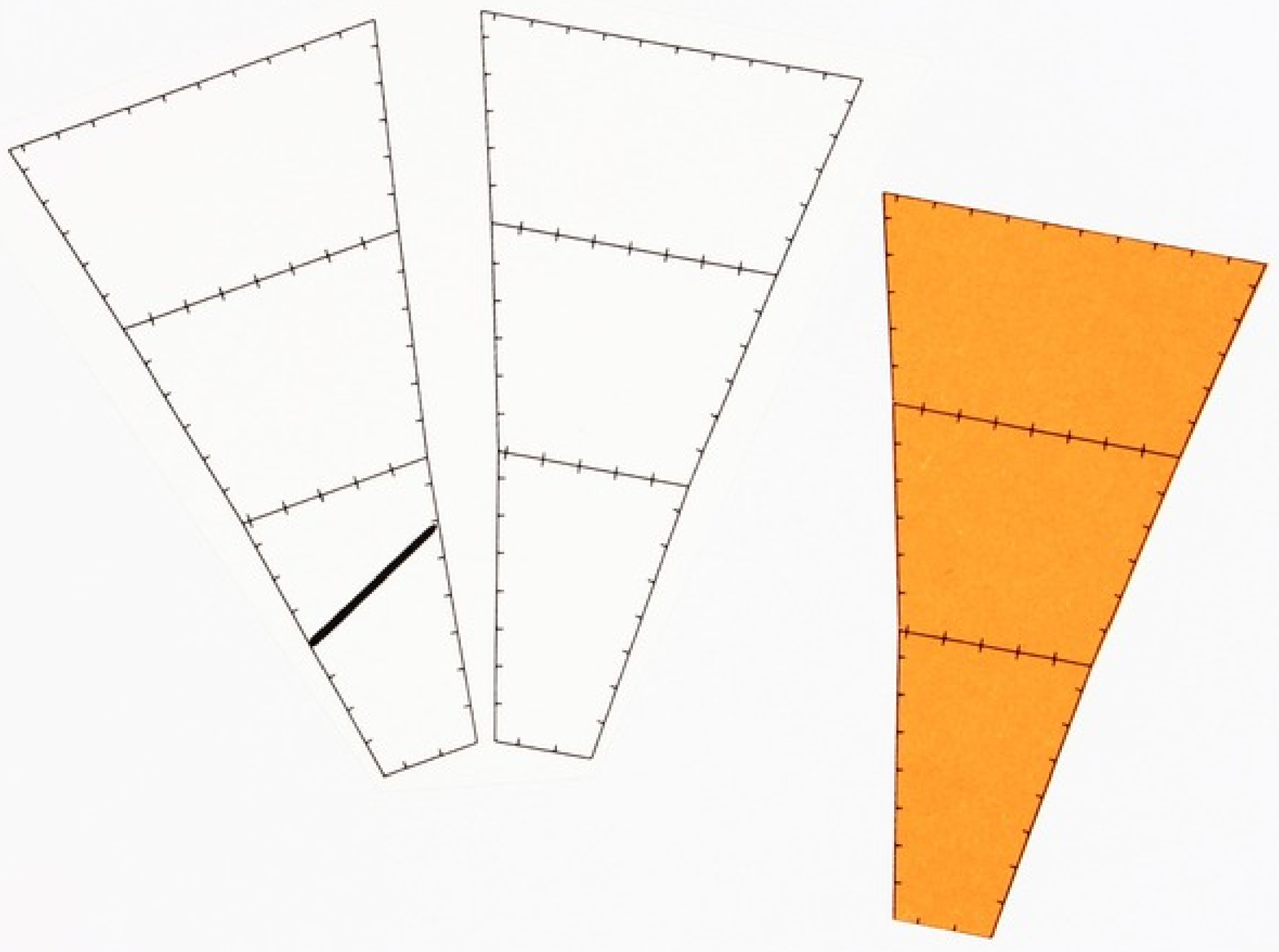}
  }\hfill%
  \subfigure[]{%
    \includegraphics[width=\deb]{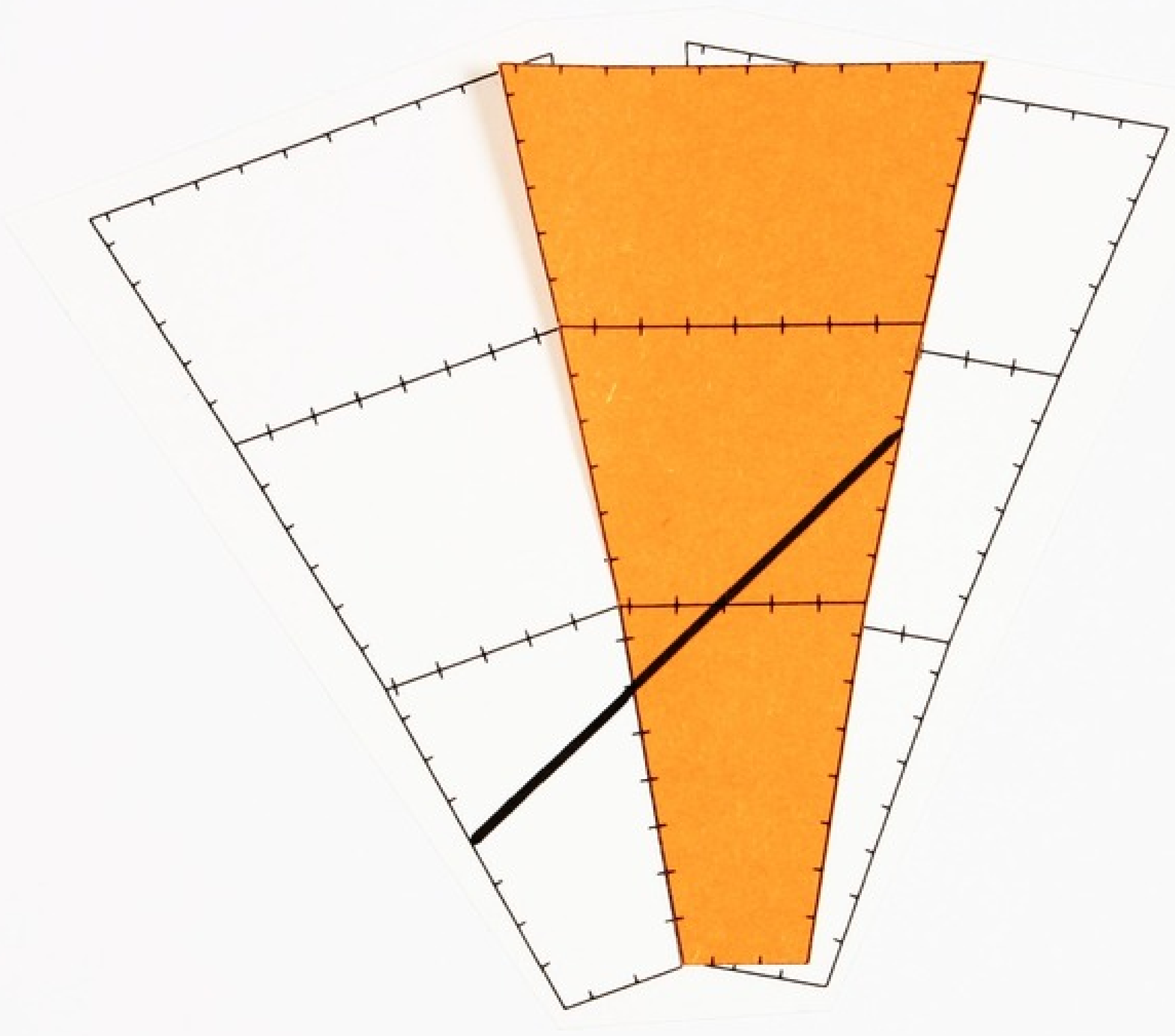}
  }\hfill%
  \subfigure[]{%
    \includegraphics[width=\veb]{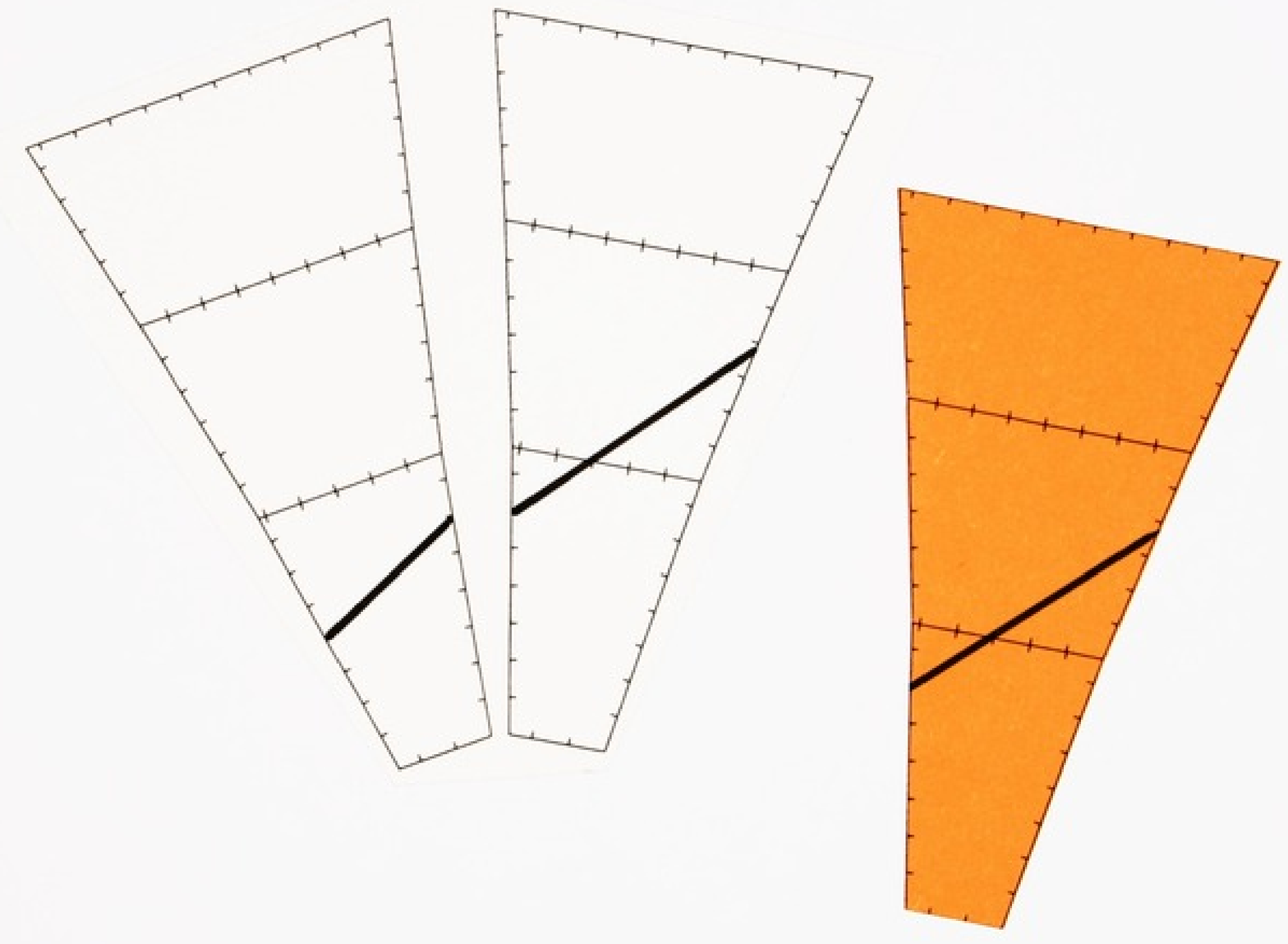}
  }
\caption{
\label{fig.transfer}
Konstruktion von Geodäten mit Transfersektoren (farbig markiert).
Die Geodäte wird bis zum Rand der Spalte gezeichnet (a),
nach Anlegen des passenden Transfersektors
auf der Transferspalte fortgesetzt (b)
und von dort auf die Nachbarspalte übertragen (c).
}
\end{figure}

Für die Abbildungen
\ref{fig.slgeod},
\ref{fig.scharparallel}
und~\ref{fig.zweieck}
wurden die Sektoren ausgeschnitten
und nach Bedarf längs einer Geodäte
oder aber in symmetrischer Anordnung ausgelegt.
Das hat den Vorteil, dass man jede Geodäte für sich
als gerade Linie zeigen kann.
Die Konstruktion der Geodäten
kann man aber einfacher und schneller durchführen,
wenn man auf das Ausschneiden der Sektoren verzichtet.
Dazu wird die
in Abb.~\ref{fig.slmodell}(a)
gezeigte Vorlage mit symmetrischer Anordnung
und Randmarkierungen verwendet.
Ausgeschnitten wird nur eine einzelne zusätzliche
Spalte (Abb.~\ref{fig.slmodell}(b));
dies sind die sogenannten Transfersektoren.
Man beginnt nun mit dem Zeichnen einer Geodäte
auf der symmetrischen Vorlage,
bis man an den Rand einer Spalte gelangt
(Abb.~\ref{fig.transfer}(a)).
Dann wird die Transferspalte so positioniert,
dass an der Kante mit dem Durchstoßpunkt der
passende Transfersektor anliegt,
und die Linie wird geradlinig
über die Spalte der Transfersektoren fortgesetzt
(Abb.~\ref{fig.transfer}(b)).
Von den Transfersektoren wird die Linie auf die Nachbarspalte der Vorlage
übertragen
(Abb.~\ref{fig.transfer}(c)).
Dieses Verfahren wird bis zum gewünschten Endpunkt fortgesetzt.
In dem
in Abb.~\ref{fig.slmodell}(a) gezeigten Sektormodell
sind die Ränder der Sektoren mit äquidistanten Markierungen versehen,
die das Übertragen der Geradenstücke erleichtern.
Die in Abb.~\ref{fig.slmodell} gezeigte Vorlage
ist online verfügbar \citep{zah2018}.

\subsection{Erstellung von Sektormodellen}
\label{sec.modellberechnung}

Ein Workshop kann wie oben beschrieben
mit vorgegebenen Sektormodellen durchgeführt werden.
Dies ist die einfachste und die kürzeste Variante.
Ihr volles Potenzial als Werkzeug zur Untersuchung gekrümmter
Räume entfalten Sektormodelle aber dann,
wenn sie von den Teilnehmer/innen selbst berechnet und konstruiert werden.
Dies versetzt die Teilnehmer/innen in die Lage,
andere gekrümmte Räume selbstständig auf die gleiche Weise zu
untersuchen und beispielsweise zu einer gegebenen Metrik
Geodäten zu bestimmen.
Dabei werden Aufstellen und Lösen der Geodätengleichung
ersetzt durch Erstellung des Sektormodells
und zeichnerische Geodätenkonstruktion.

Im folgenden Abschnitt wird gezeigt,
wie man am Beispiel der Kugeloberfläche
die Berechnung von Sektormodellen einführen kann.
Das Vorgehen wird anschließend
auf die Berechnung des Sektormodells
der Äquatorfläche angewandt.

\begin{figure}[t]
  \centering
  \subfigure[]{%
    \includegraphics{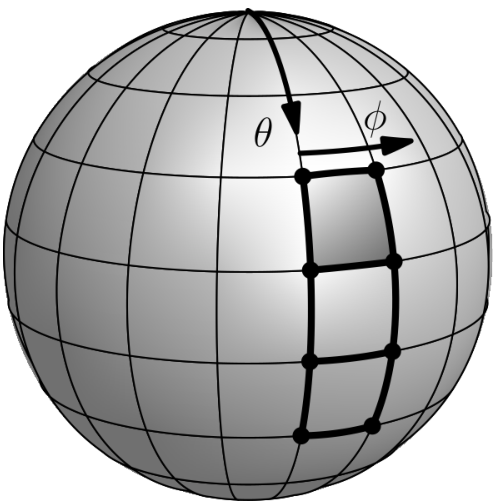}
  }\hfil%
  \subfigure[]{%
    \includegraphics[scale=1]{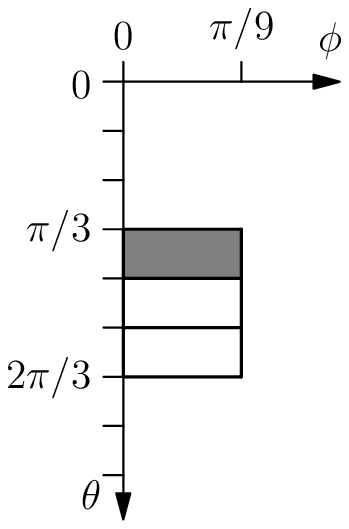}
  }\hfil%
  \subfigure[]{%
    \includegraphics[width=0.11\textwidth]{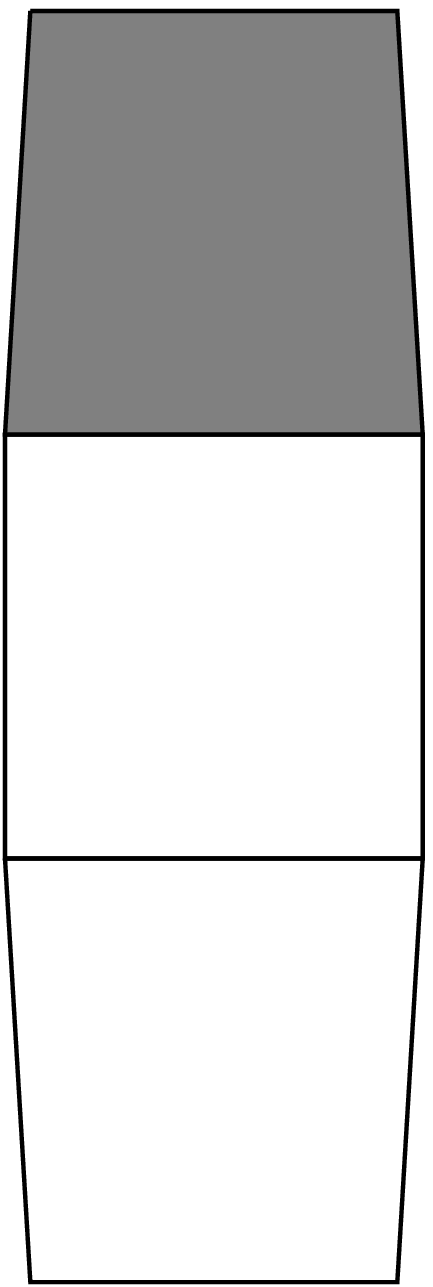}
  }
  \caption{\label{fig.kugelteilung}
   (a) Die Oberfläche einer Kugel
   wird mit Polarkoordinaten $\theta$, $\phi$
   beschrieben und in Flächenstücke mit Winkelausdehnung 20 Grad
   auf 20 Grad
   unterteilt.
   (b) Drei Flächenstücke im $\phi$-$\theta$-Koordinatenraum.
   (c) Die zugehörigen Sektoren.
   Sie bilden eine Spalte des in
   Abb.~\protect\ref{fig.kugelmodell}(b) gezeigten Modells.
   Zur leichteren Zuordnung ist
   in allen drei Teilbildern ein und derselbe Sektor
   grau markiert.
}
\end{figure}

\subsubsection{Erstellung des Sektormodells einer Kugeloberfläche.}

\label{sec.kugelberechnung}

An diesem Beispiel
wird das prinzipielle Vorgehen
bei der Erstellung von Sektormodellen
eingeführt.
Voraussetzung ist der Begriff der Metrik
als einer Funktion,
die den Koordinaten zweier nahe benachbarter Punkte
den Abstand dieser Punkte zuordnet.
Die Metrik kann ausgehend von krummlinigen Koordinaten
auf elementare Weise eingeführt werden
(z.\,B. \citealp{kra2016}; \citealp{har}, S.~21~f; \citealp{nat}, S.~35~f).

Ausgangspunkt der Berechnung ist
die Metrik der Kugeloberfläche
in den üblichen Polarkoordinaten $\theta$, $\phi$
(Abb.~\ref{fig.kugelteilung}(a)):
\begin{equation}
\ud s^2 = R^2\, \ud \theta^2 + R^2 \sin^2\theta \, \ud \phi^2,
\end{equation}
wobei $R$ der Radius der Kugel ist
(für eine elementare Herleitung,
die im Workshop genutzt werden kann, siehe z.\,B. \citealp{har}, S.~23~f;
\citealp{nat}, S. 37 ff).

Die Erstellung des Sektormodells erfolgt in drei Schritten.
Im ersten Schritt
wird
die Kugeloberfläche in Flächenstücke unterteilt,
die durch ihre Eckpunkte definiert sind.
In dem hier betrachteten Beispiel
sind die Flächenstücke Vierecke, deren Eckpunkte
in den Winkelkoordinaten $\theta$ und $\phi$
jeweils Abstände von 20~Grad ($\pi/9$) haben
(Abb.~\ref{fig.kugelteilung}(a), (b)).
Im zweiten Schritt werden die Kantenlängen der Flächenstücke
berechnet.
Dies geschieht näherungsweise, um die Berechnung
einfach zu halten.
Für jede Kante
wird der Abstand zwischen den beiden Endpunkten bestimmt,
indem man
die Endpunkte als nahe benachbarte
Punkte im Sinne der Definition der Metrik behandelt.
Man erhält für die
Kanten zwischen zwei Punkten auf demselben Längenkreis
die Länge
\begin{equation}
\Delta s = R\,\Delta\theta \hspace*{1cm} (\Delta\phi=0),
\label{eq.kugel_laenge}
\end{equation}
im konkreten Beispiel $\Delta s = R\pi/9$.
Für die Kanten zwischen zwei Punkten
auf demselben Breitenkreis erhält man
\begin{equation}
\Delta s = R\,\sin\theta\,\Delta\phi \hspace*{1cm} (\Delta\theta=0),
\label{eq.kugel_breite}
\end{equation}
abhängig vom Winkel $\theta$ des Breitenkreises.
Für die Sektormodelle in den Abbildungen
und Online-Ressourcen werden die Kantenlängen
als Längen der Geodäten zwischen den Eckpunkten
berechnet (s. Teil~I).
Der Unterschied zwischen den genäherten und den exakten
Kantenlängen beträgt bei diesem Beispiel
maximal  $0,13 \%$.

Im dritten Schritt werden
aus den Kantenlängen
ebene Flächenstücke
konstruiert.
In dem hier betrachteten Beispiel
haben die Flächenstücke auf der Kugel
Spiegelsymmetrie;
die ebenen Flächenstücke werden
mit der gleichen Symmetrieeigenschaft
als symmetrische Trapeze
konstruiert
(Abb.~\ref{fig.kugelteilung}(c)).

\subsubsection{Erstellung des Sektormodells der Äquatorfläche
eines Schwarzen Lochs.}
\label{sec.slberechnung}

\begin{figure}
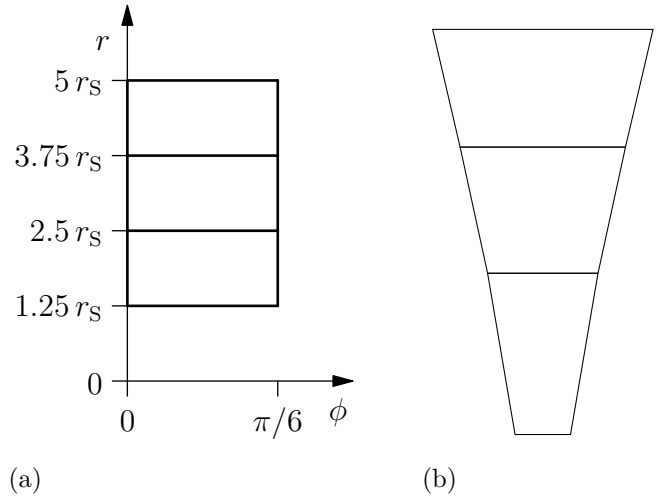

\centering
\subfigure[]{%
  \includegraphics[scale=1]{\bilder/figure14a.eps}
}\qquad%
\subfigure[]{%
  \includegraphics[scale=0.8]{\bilder/figure14b.eps}
}
\caption{\label{fig.sl}
Zur Erstellung des Sektormodells der Äquatorfläche eines Schwarzen Lochs.
(a) Die drei Flächenstücke einer Spalte im $\phi$-$r$-Koordinatenraum.
(b) Die zugehörigen Sektoren.
}
\end{figure}

Ausgangspunkt der Berechnung ist die Metrik der Äquatorebene
eines Schwarzen Lochs
\begin{equation}
\ud s^2 = \frac{1}{1-\rs/r} \ud r^2 + r^2 \ud \phi^2
\label{eq.slmetrik}
\end{equation}
in den üblichen Schwarzschildkoordinaten $r$ und $\phi$.
Hier ist  $\rs = 2 G M / c^2$
der Schwarzschildradius des Schwarzen Lochs mit Masse $M$,
$G$ die Newtonsche Gravitationskonstante
und $c$ die Lichtgeschwindigkeit.
Das Sektormodell stellt einen ringförmigen Ausschnitt der
Äquatorebene dar.
Der Innenrand liegt bei $r=1,25\, \rs$, der Außenrand bei $r=5\, \rs$.
Der Azimutwinkel $\phi$ nimmt Werte zwischen null und $2\pi$ an.

Zunächst wird der Ring in Flächenstücke unterteilt.
Dazu wird
der $\phi$-Bereich in 12 Abschnitte von je $\pi/6$ aufgeteilt.
Da die Metrik nicht von der Koordinate $\phi$ abhängt,
braucht nur einer der 12 Abschnitte berechnet zu werden;
man erhält 12 identische Spalten, die im Kreis angeordnet werden.
Der $r$-Bereich wird
in drei Abschnitte der Koordinatenlänge $1,25\, \rs$ eingeteilt
(Abb.~\ref{fig.sl}(a)).
Im zweiten Schritt werden die Kantenlängen der drei
in Abb.~\ref{fig.sl}(a) gezeigten Vierecke berechnet.
Für den Abstand von
Eckpunkten mit gleicher $r$-Koordinate erhält man
mithilfe der Metrik
\begin{equation}
\Delta s = r \,\Delta\phi \hspace*{1cm} (\Delta r = 0).
\end{equation}
Bei der Abstandsberechnung von
Eckpunkten mit gleicher $\phi$-Koordinate
kommt der erste Term der Metrik zum Tragen.
Er enthält den Metrikkoeffizienten $1\big/(1-\rs/r)$,
der von $r$ abhängt, sich also längs der Kante ändert.
Hier machen wir
eine weitere Näherung und verwenden den
Metrikkoeffizienten
an der mittleren $r$-Koordinate $\rrmi$ der Kante:
\begin{equation}
\Delta s = \sqrt{\frac{1}{(1-\rs/\rrmi)}}\, \Delta r
     \hspace*{1cm} (\Delta\phi = 0),
\end{equation}
wobei $\rrmi = (r_1 + r_2)/2$ mit den Koordinaten
$r_1$ und $r_2$ der zugehörigen Eckpunkte.
Für die Abbildungen und Online-Ressourcen
werden
die Kantenlängen als Längen der Geodäten
zwischen den Eckpunkten numerisch berechnet.
Die Abweichung
der vereinfacht berechneten Kantenlängen
ist am größten für die innerste radiale Kante
und beträgt hier 5,4\%.

Im dritten Schritt werden die
Vierecke konstruiert.
Die Aufteilung des Rings durch radiale Schnitte
erzeugt Flächenelemente mit Spiegelsymmetrie
(Abb.~\ref{fig.gitter2d}(a));
dieser Symmetrie entsprechend werden die Flächenstücke
als symmetrische Trapeze konstruiert.
Abb.~\ref{fig.sl} stellt
die drei Sektoren einer Spalte
den zugehörigen Vierecken
im $\phi$-$r$-Koordinatenraum gegenüber.
Das vollständige Sektormodell mit zwölf Spalten
ist in
Abb.~\ref{fig.slmodell}
dargestellt.

\subsection{Geodäten vs. Krümmung}
\label{sec.ergaenzunggeod}

\begin{figure}[t]
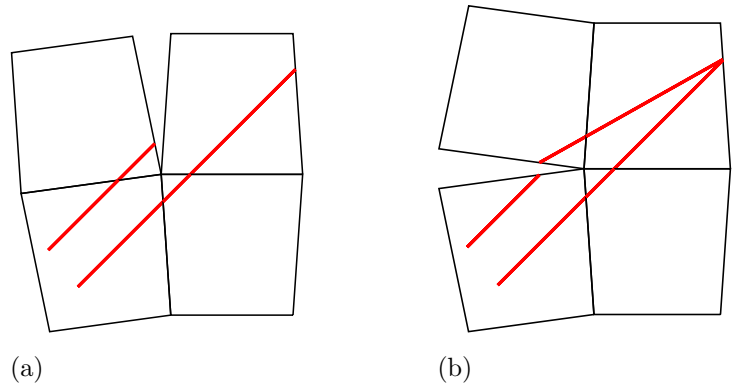

  \centering
  \subfigure[]{%
    \includegraphics[width=\vib]{\bilder/figure15a.eps}
  }\qquad\qquad%
  \subfigure[]{%
    \includegraphics[width=\vib]{\bilder/figure15b.eps}
  }
\caption{\label{fig.vertexgeod}
Zwei zunächst parallele Geodäten (a) führen auf verschiedenen
Seiten an einem Vertex vorbei
und sind anschließend um den Defizitwinkel zueinander geneigt (b).
}
\end{figure}

Den Zusammenhang zwischen
der Krümmung einer Fläche und
dem Verlauf von Geodäten
kann man zum Thema machen,
wenn man zusätzlich zu dem oben geschilderten
Workshop über Geodäten
auch den in Teil~I (Abschnitt~2)
beschriebenen Workshop über Krümmung
durchführt.
In Teil~I werden
die Kugeloberfläche und die Sattelfläche
als Prototypen für Flächen mit positiver
bzw. negativer Krümmung vorgestellt.
Es wird verdeutlicht, dass der Defizitwinkel im Sektormodell
ein Kriterium für die Krümmung ist: Positive Krümmung
wird durch einen positiven Defizitwinkel angezeigt
und vice versa.%
\footnote{%
Der Defizitwinkel ist positiv, wenn
nach Zusammenschieben aller an einen Eckpunkt angrenzenden
Sektoren eine keilförmige Lücke verbleibt
(ein Beispiel gibt Abb.~\ref{fig.vertexgeod}).
Er ist negativ, wenn
nach Zusammenschieben aller an einen Eckpunkt angrenzenden
Sektoren bis auf einen die verbleibende Lücke
zu klein ist für den letzten Sektor.}

Anhand von Sektormodellen kann man aufzeigen,
dass der Verlauf benachbarter Geodäten
ebenfalls ein Kriterium für die Krümmung darstellt.
Abb.~\ref{fig.vertexgeod}
stellt den Verlauf von benachbarten Geodäten
in der Nähe eines
einzelnen Vertex mit positivem Defizitwinkel dar.
Zwei Geodäten, die
vor dem Vertex parallel verlaufen
und auf verschiedenen
Seiten an dem Vertex vorbeiführen
(Abb.~\ref{fig.vertexgeod}(a)),
laufen nach dem Vertex aufeinander zu
(Abb.~\ref{fig.vertexgeod}(b)).
Der Winkel, den die beiden Richtungen nach dem Vertex einschließen,
ist nach Konstruktion der Defizitwinkel.

Parallel startende Geodäten
zeigen also positive Krümmung an,
wenn sie aufeinander zu laufen.
Umgekehrt zeigen sie
negative Krümmung an, wenn sie auseinander laufen.
Zwei Beispiele für dieses Kriterium
gibt
Abschnitt~\ref{sec.kugel}
mit Geodäten auf der Kugeloberfläche
(Abb.~\ref{fig.kugelgeod}(c))
und der Sattelfläche
(Abb.~\ref{fig.sattelgeod}(d)).
Angewandt auf die Äquatorebene eines Schwarzen Lochs
schließt man aus dem Auseinanderlaufen von parallel startenden
Geodäten
(Abb.~\ref{fig.scharparallel})
auf eine negative Krümmung.

Die obige Überlegung ist eine Veranschaulichung für
die Gleichung der geodätischen Abweichung
\begin{equation}
(\nabla_{\bf u}\nabla_{\bf u} {\bf D})^i =
  -  R^i{}_{jkl} u^j D^k u^l
\end{equation}
für zwei Geodäten $x^i(\lambda)$ und
$x^i(\lambda) + D^i(\lambda)$
mit
$u^i = d x^i / d\lambda$
und dem Riemannschen Krümmungstensor
$R^i{}_{jkl}$.
Im Sektormodell werden die Komponenten des Riemannschen
Krümmungstensors durch die Defizitwinkel dargestellt (Teil~I,
Abschnitt~3) und Abb.~\ref{fig.vertexgeod} zeigt bildlich,
wie diese sich auf die Änderung des Abstands benachbarter Geodäten auswirken.

\begin{figure}[p]
  \centering
  \subfigure[]{%
    \includegraphics[width=0.63\textwidth]{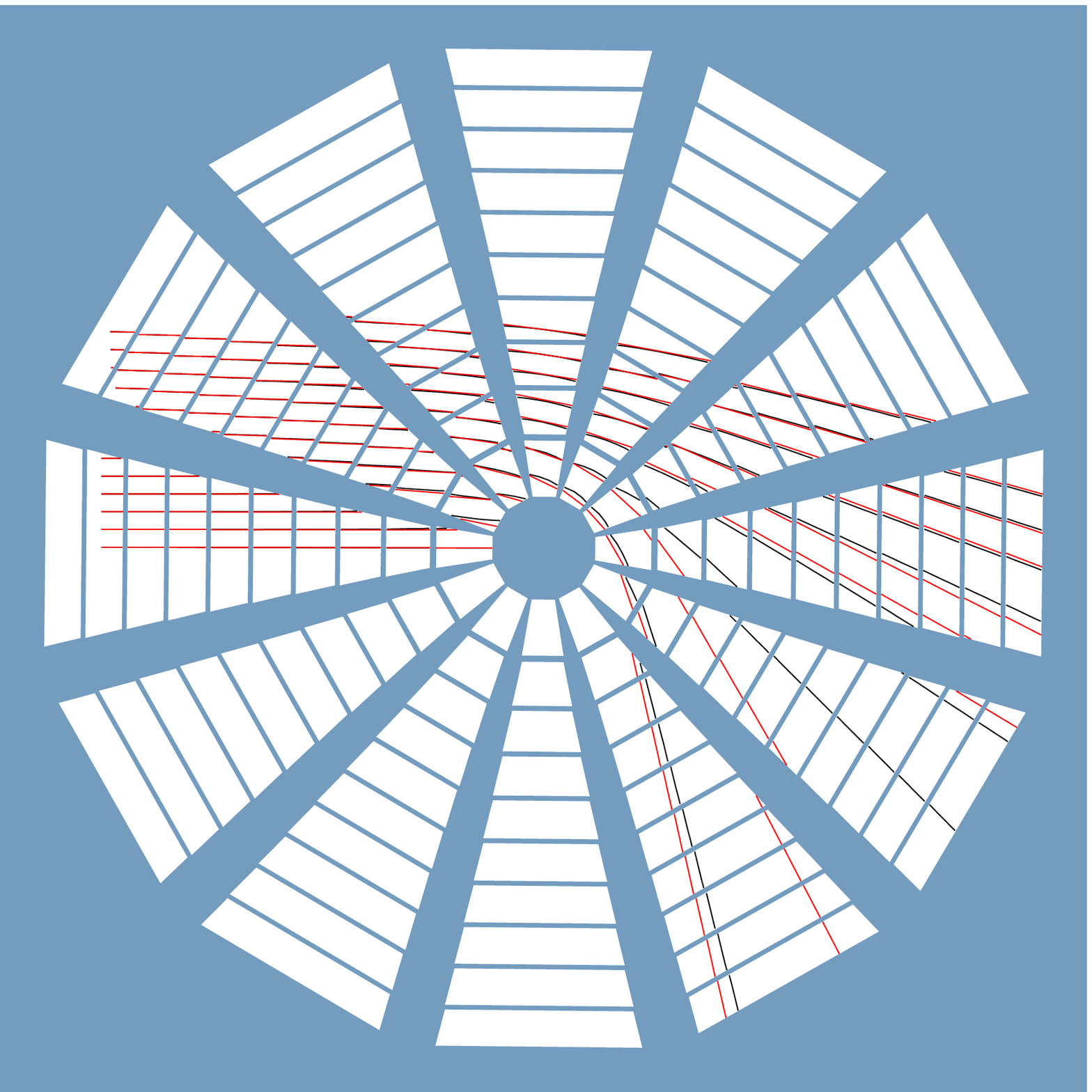}
  }
  \subfigure[]{%
    \includegraphics[width=0.63\textwidth]{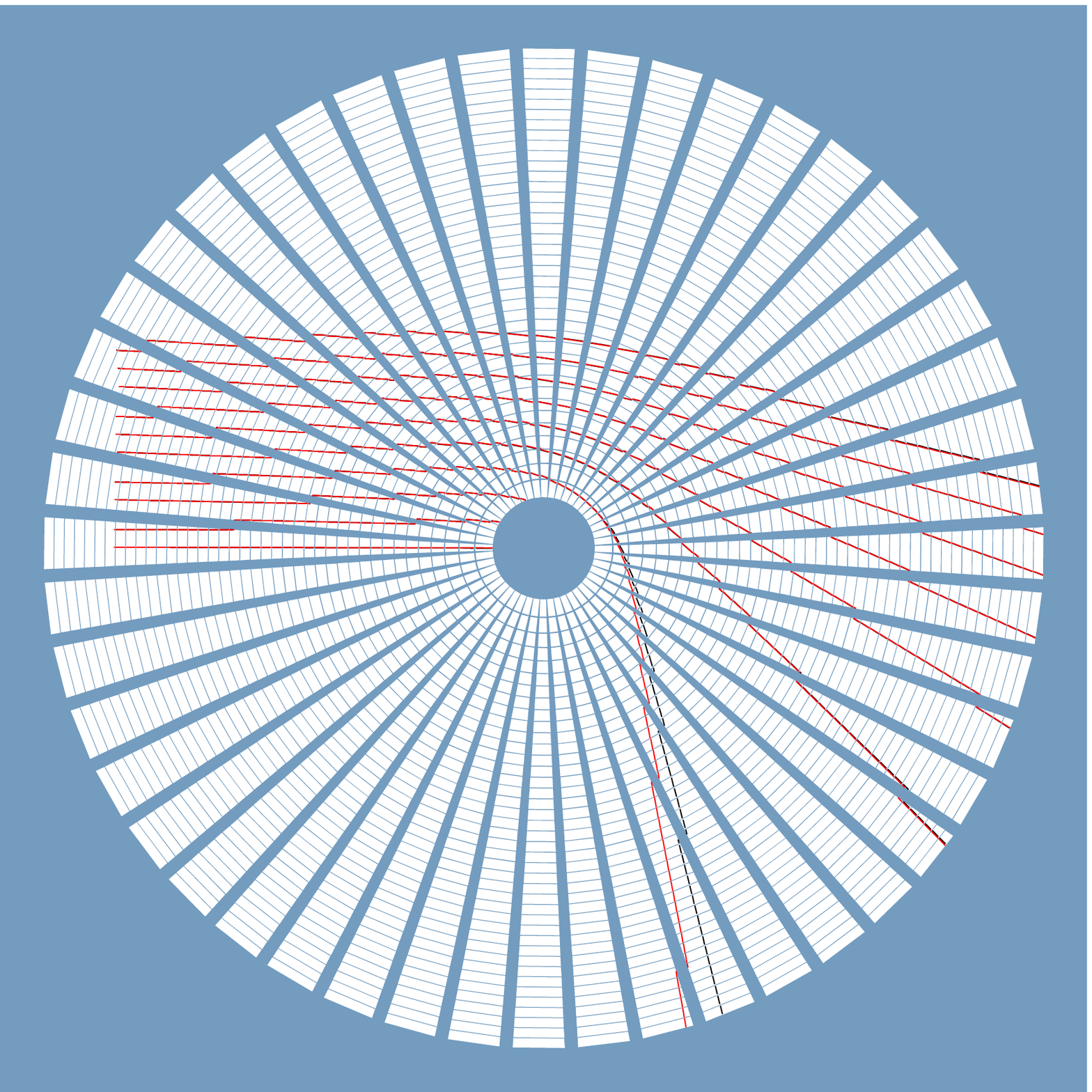}
  }
\caption{\label{fig.geodvergleich}
Auf Sektormodellen konstruierte Geodäten (rote Linien)
im Vergleich mit numerischen Lösungen der Geodätengleichung
(schwarze Linien).
(a) Sektormodell mit der im Workshop verwendeten Auf"|lösung.
(b) Sektormodell mit der vierfachen Auf"|lösung in jeder
Koordinatenrichtung.
}
\end{figure}

\section{Die Genauigkeit von Geodäten auf Sektormodellen}
\label{sec.modell2}

\begin{figure}
  \centering
  \subfigure[]{%
    \includegraphics[width=\hb]{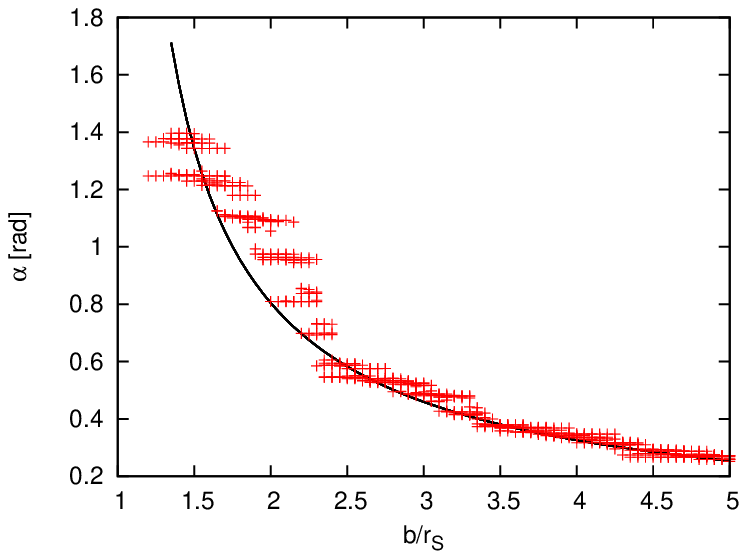}
  }\hfill%
  \subfigure[]{%
    \includegraphics[width=\hb]{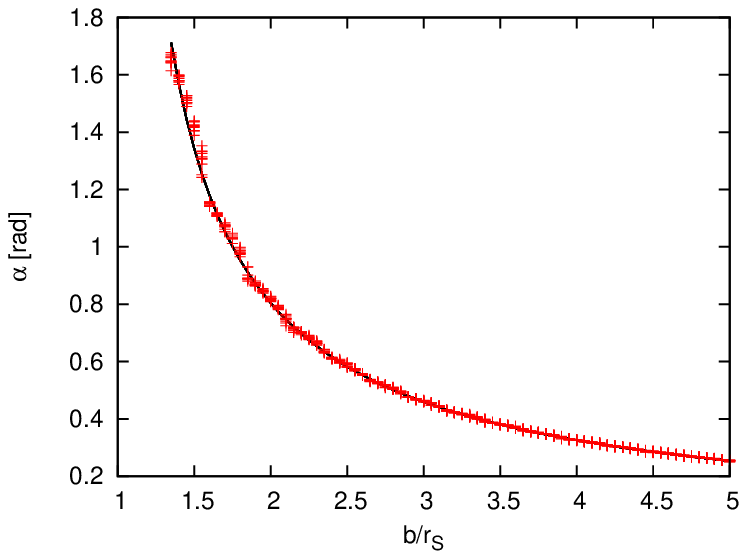}
  }
\caption{\label{fig.ablenkwinkel}
Zusammenhang zwischen dem Ablenkwinkel $\alpha$ und
dem Stoßparameter $b$ einer Geodäte,
berechnet
durch Konstruktion der Geodäten auf einem Sektormodell
(Kreuze)
sowie
durch Integration der Geodätengleichung (Linie).
(a) Sektormodell mit der im Workshop verwendeten
Auf"|lösung.
(b) Sektormodell mit der
vierfachen Auf"|lösung in jeder Koordinatenrichtung.
}
\end{figure}

Im Regge-Kalkül werden Geodäten
als Geradenstücke in den ungekrümmten Sektoren
beschrieben \citep{wil1981,wil1984,bre1993}.
Diese Darstellung wird auf den Sektormodellen
zeichnerisch umgesetzt.
Die so konstruierten Geodäten
sind also grundsätzlich quantitativ
korrekt. Ihre Genauigkeit hängt allerdings
von der Auf\/lösung der Sektormodelle ab.
Für die Workshops
wird die Auf\/lösung bewusst
grob gewählt, damit die Modelle gut handhabbar sind.
In diesem Abschnitt wird durch den Vergleich mit
numerisch berechneten Geodäten untersucht,
welche Genauigkeit mit der zeichnerischen Methode
erreicht wird.
Für den Vergleich werden zwei Sektormodelle
der Äquatorebene eines Schwarzen Lochs
verwendet,
die beide den Bereich zwischen $r=1,25\,\rs$ und
$r=13,75\,\rs$ überdecken.
Das erste hat die Auf"|lösung
des im Workshop verwendeten Sektormodells
($\Delta r = 1,25\,\rs$, $\Delta\phi = \pi/6$);
es besteht aus 10~Ringen zu je 12 Sektoren
(Abb.~\ref{fig.geodvergleich}(a)).
Das zweite hat in jeder Koordinatenrichtung
die vierfache Auf"|lösung
($\Delta r = 0,3125\,\rs$,
$\Delta\phi = \pi/24$)
und besteht somit aus 40 Ringen zu je 48 Sektoren
(Abb.~\ref{fig.geodvergleich}(b)).
In Abb.~\ref{fig.geodvergleich}
sind die
im Regge-Kalkül bestimmten
Geodäten
und die
numerischen Lösungen der Geodätengleichung
im Vergleich
dargestellt.
Dazu werden die berechneten
Bahnen auf dem Sektormodell dargestellt. Die Abbildung der
in Schwarzschildkoordinaten gegebenen Bahnpunkte
auf Punkte im Sektormodell erfolgt durch Interpolation \citep{hor2005}.
Für einen quantitativen Vergleich
wurde auf denselben beiden Sektormodellen
der Ablenkwinkel als Funktion des Stoßparameters
bestimmt
und mit den aus der Geodätengleichung
durch Integration bestimmten Werten verglichen
(Abb. \ref{fig.ablenkwinkel}).
Auf den Sektormodellen wurden dabei für jeden Stoßparameter
zehn Geodäten
konstruiert,
die in $\phi$-Richtung gegeneinander verdreht sind,
so dass sie unterschiedliche Lagen
bzgl. der Sektorgrenzen haben.
Man erkennt, dass für die gröbere Diskretisierung
die Übereinstimmung von Geodäten mit geringer Ablenkung gut ist,
dass bei Geodäten mit starker Ablenkung
aber erhebliche Abweichungen auftreten können
(Abb.~\ref{fig.geodvergleich}(a), \ref{fig.ablenkwinkel}(a)).
Für das Sektormodell mit der höheren Auf"|lösung ist die
Übereinstimmung generell gut
(Abb.~\ref{fig.geodvergleich}(b), \ref{fig.ablenkwinkel}(b)).
Die Genauigkeit auf
dem im Workshop eingesetzten Sektormodell
ist für qualitative Betrachtungen
völlig zufriedenstellend.

\section{Fazit und Ausblick}

\label{sec.fazit2}

\subsection{Zusammenfassung und didaktische Anmerkungen}

Wir haben gezeigt, auf welche Weise man Sektormodelle als Werkzeuge
zur Bestimmung von Geodäten einsetzen kann. Dies vermittelt
einerseits ein anschauliches Verständnis und ermöglicht andererseits
die Ermittlung von Geodäten mit zeichnerischen Mitteln.
Veranschaulicht wird zunächst der Begriff der Geodäte als einer lokal
geraden Linie, indem diese Definition auf einem Sektormodell mit dem
Lineal unmittelbar umgesetzt wird (Abschnitt~\ref{sec.kugel}).
Weiter wird anschaulich klar, dass eine Geodäte nach dem Durchqueren
eines Bereichs gekrümmten Raums eine andere Richtung hat als zuvor
(Abschnitt~\ref{sec.ssmgeod}), wodurch das Zustandekommen der
Lichtablenkung im Schwerefeld verdeutlicht wird.
Da die konstruierten Geodäten Lösungen der Geodätengleichung
darstellen, erhält man quantitativ richtige Ergebnisse. Deren
Genauigkeit ist wegen der relativ groben Auf"|lösung der von Hand
erstellten und eingesetzten Sektormodelle nicht hoch.  Aus
didaktischer Sicht ist eine grobe Auf"|lösung jedoch vorteilhaft.  Die
Defizitwinkel sind dann so groß, dass die Auswirkungen der Krümmung an
einzelnen Vertizes verdeutlicht werden können.  So entsteht ein klares
Bild von dem Zusammenhang zwischen der Krümmung und dem Verlauf
benachbarter Geodäten (Abschnitt~\ref{sec.ergaenzunggeod}).
In diesem Beitrag werden nur raumartige Geodäten betrachtet.
Eine Erweiterung auf Geodäten in der Raumzeit wird in einem
Folgebeitrag beschrieben \citep{teil3}.

Der in diesem Beitrag
vorgestellte Workshop zu Geodäten
und Lichtablenkung
entstand in einem
Wechselspiel zwischen Erprobung und Weiterentwicklung
\citep{kra2005,zah2010,zah2013,kra2016}.
Tests wurden
vorwiegend
mit Schulklassen der Oberstufe sowie mit Lehramtsstudierenden
durchgeführt.

Für das vorgestellte Material
gibt es, je nach Ziel und Zeitbudget,
verschiedene Einsatzmöglichkeiten.
Wenn das Ziel
ein kurzer und direkter Zugang zur gravitativen Lichtablenkung
ist, beispielsweise für einen Astronomiekurs,
kann der vorgestellte Workshop wie in den
Abschnitten~\ref{sec.kugel} bis~\ref{sec.geodzeichnen}
beschrieben mit fertigen Vorlagen der Sektormodelle durchgeführt
werden. Vorkenntnisse zum Begriff der Metrik sind dann nicht
erforderlich;
die zeichnerische Konstruktion ist einfach durchzuführen
und vermittelt ein fachlich befriedigendes
Konzept von Lichtbahnen als Geodäten.
Ist eine Einführung in die Grundkonzepte der Allgemeinen
Relativitätstheorie das Ziel, dann kann man die Sektormodelle
der Kugeloberfläche und der Äquatorebene des Schwarzen Lochs
von den Teilnehmer/innen
selbst berechnen lassen. Der Kurs vermittelt dann die Fähigkeit,
anhand einer vorgegebenen Metrik
die Geometrie einer Fläche
zu untersuchen.
Dabei werden Aufgaben zeichnerisch gelöst, die in einem
Standardkurs rechnerisch bearbeitet würden.
Da Sektormodelle und die zeichnerische Konstruktion von Geodäten
auf direkte Weise der mathematischen Beschreibung
durch die Metrik und die Geodätengleichung entsprechen,
kann dieses Material auch als Ergänzung zu einer
üblichen, mathematisch ausgerichteten Vorlesung eingesetzt werden,
um die geometrische Anschauung zu stärken.

\subsection{Vergleich mit anderen anschaulichen Zugängen}

Im Vergleich mit anderen bildlichen Darstellungen von Geodäten
zeichnen sich Konstruktionen auf Sektormodellen dadurch aus,
dass sie den lokal geraden Verlauf deutlich zum Ausdruck bringen
und dass sie einfach zu erstellen sind.

Zur Erklärung der durch Lichtablenkung bewirkten optischen Phänomene,
z.\,B. der Doppelbilder, werden üblicherweise Grafiken eingesetzt, die
Lichtwege als gebogene Linien darstellen. In diesem Zusammenhang
werden Lichtwege auch häufig als \glqq krumm\grqq\ bezeichnet.
Dass Lichtwege Geodäten, also (lokal) gerade Linien sind,
kommt in diesen Grafiken
und auch in der sprachlichen Beschreibung nicht zum Ausdruck,
was zu Fehlvorstellungen führen kann.
Die Konstruktion auf den Sektormodellen verdeutlicht, dass
der lokal gerade Verlauf der Geodäte und das Auftreten von
Lichtablenkung nicht im Widerspruch stehen (Abb.~\ref{fig.slgeod}).
Es kann auch der Bezug zu den
gekrümmt dargestellten Lichtwegen hergestellt werden:
Auf einer Weltkarte
ist
die Erdoberfläche auf eine Ebene
projiziert
und
Geodäten der Erdkugel erscheinen als gekrümmte Linien.
Dies sind Verzerrungen, die der Projektion geschuldet sind.
Analog werden in einer Projektion,
die das Sektormodell aus Abb.~\ref{fig.slgeod}
auf
einen ebenen Kreisring
abbildet, Verzerrungen auftreten,
welche die Geodäten der Äquatorebene des Schwarzen Lochs als
gekrümmte Linien erscheinen lassen.

Eine häufig verwendete
bildliche Darstellung
zeigt Geodäten auf
der Einbettungsfläche der Äquatorebene eines Sterns oder eines
Schwarzen Lochs mit dem Ziel, die Lichtablenkung zu veranschaulichen
(\citealp{dinv}, S.~209).
Dies ist äquivalent zu den in Abschnitt~\ref{sec.ssmgeod}
konstruierten Geodäten.
Wenn man beim dortigen Sektormodell die Sektoren an den gemeinsamen
Kanten zusammenfügt, erhält man eine Facettenfläche, welche
näherungsweise die Einbettungsfläche darstellt.
Für die Geodäten auf der Einbettungsfläche gilt derselbe Vorbehalt
wie für die Geodäten auf dem Sektormodell:
Der dargestellte Unterraum ist rein räumlich, so dass
die gravitative Lichtablenkung
in Form einer Analogie mit raumartigen Geodäten illustriert wird.
Unsere Erfahrung zeigt, dass das Konzept der Einbettungsfläche
für die Zielgruppe dieser Workshops schwierig ist.
Man muss damit rechnen, dass
die Darstellung als geometrische Form des Schwarzen Lochs
fehlinterpretiert wird \citep{zah2010}.
Im Vergleich zu Einbettungsflächen haben Sektormodelle den Vorteil,
dass ihre Berechnung
einfacher ist, insbesondere wenn
das in Abschnitt~\ref{sec.modellberechnung}
beschriebene vereinfachte Verfahren angewandt
wird.
Sie lassen sich außerdem
auf einfache Weise als Anschauungsmodelle aus Papier
erstellen und gut vervielfältigen,
so dass alle Teilnehmer/innen eines Workshops die Konstruktion
von Geodäten auf eigenen Vorlagen selbst durchführen können.

Eine Beschreibung von Geodäten, die mit der Darstellung auf Sektormodellen
verwandt ist, ist die von \citet{dis1981} entwickelte
Konstruktion auf sogenannten Keilkarten (wedge maps).
Eine Keilkarte entsteht,
wenn die Symmetrieebene einer sphärisch-symmetrischen Raumzeit
durch radiale Schnitte in Streifen zerlegt wird;
diese Streifen werden als ungekrümmt angesehen.
Auf den Streifen werden Geodäten nach dem
Prinzip des Regge-Kalküls auf numerische Weise bestimmt.
Die Konstruktion der Geodäten auf der Keilkarte
folgt also demselben Prinzip wie hier für die Sektormodelle beschrieben.
Das numerische Vorgehen ist anspruchsvoller
als die hier eingesetzte zeichnerische Lösung,
sowohl was die mathematische Beschreibung angeht
als auch wegen der erforderlichen Programmierkenntnisse.

\subsection{Ausblick}

In Teil~I
wurden drei Grundfragen aufgeworfen,
die bei einer Einführung in die Allgemeine Relativitätstheorie
beantwortet werden sollten:
Was ist eine gekrümmte Raumzeit? Wie bewegt sich Materie in einer
gekrümmten Raumzeit? Welcher Zusammenhang besteht zwischen der
Materieverteilung und der Krümmung der Raumzeit?
Der Begriff des gekrümmten Raums bzw.\ der gekrümmten Raumzeit
wurde in Teil~I verdeutlicht.
Zur Beantwortung der zweiten Frage
beschreibt der vorliegende Teil~II Geodäten im Raum
und der Folgebeitrag
Geodäten in der Raumzeit \citep{teil3}.
Ein vierter Teil dieser Serie wird
auf den Zusammenhang zwischen der Krümmung der Raumzeit
und der Materieverteilung eingehen.